\documentclass{aa}

\usepackage{graphicx}
\usepackage{dirtytalk}
\usepackage{txfonts}
\usepackage{hyperref}
\usepackage{xcolor}
\hypersetup{
colorlinks,
linkcolor={red!80!black},
citecolor={blue!80!black},
urlcolor={blue!80!black}
}

\newenvironment{myfont}{\fontfamily{ppl}\selectfont}{\par}
\DeclareTextFontCommand{\textmyfont}{\myfont}

\newcommand{\code}[1]{\texttt{#1}}

\def\nifs{\iso{56}Ni}
\def\nife{\iso{58}Ni}
\def\cofs{\iso{56}Co}
\def\cofn{\iso{59}Co}
\def\fefs{\iso{56}Fe}

\def\cm3{cm$^{-3}$}
\def\kms{\mbox{km~s$^{-1}$}}

\def\msun{$M_{\odot}$}

\def\one{\ts {\,\sc i}}
\def\two{\ts {\,\sc ii}}
\def\three{\ts {\,\sc iii}}

\def\beq{\begin{equation}}
\def\eeq{\end{equation}}

\def\lesssim{\mathrel{\hbox{\rlap{\hbox{\lower4pt\hbox{$\sim$}}}\hbox{$<$}}}}
\def\gtrsim{\mathrel{\hbox{\rlap{\hbox{\lower4pt\hbox{$\sim$}}}\hbox{$>$}}}}

\def\one{{\,\sc i}}
\def\two{{\,\sc ii}}
\def\three{{\,\sc iii}}

\def\v1d{{\code{V1D}}}

\def\kepler{{\code{KEPLER}}}

\def\cmfgen{{\code{CMFGEN}}}

\def\photb{{\code{P-HOTB}}}

\def\ekin{$E_{\rm kin}$}
\def\mej{$M_{\rm ej}$}
\def\mhesh{He-sh}
\def\mosh{O-sh}
\def\msish{Si-sh}
\def\mnish{$^{56}$Ni-sh}
\def\mni{$^{56}$Ni}
\def\avxheinnish{$\bar{X}$($^4$He)$_{Ni}$}
\def\avxcainnish{$\bar{X}$($^{40}$Ca)$_{Ni}$}
\def\avxnistableinnish{$\bar{X}$($^{58}$Ni)$_{Ni}$}
 \def\avxcainosh{$\bar{X}$($^{40}$Ca)$_{O}$}
 \def\avxcainhesh{$\bar{X}$($^{40}$Ca)$_{He}$}

\def\ergs{erg\,s$^{-1}$}

\def\niidoub{[N\two]\,$\lambda\lambda$\,$6548,\,6583$}
\def\niiauroral{[N\two]\,$\lambda$\,$5755$}

\def\kidoub{K\one\,$\lambda\lambda$\,$7665,\,7699$}

\def\caiidoub{[Ca\two]\,$\lambda\lambda$\,$7291,\,7323$}
\def\caiitrip{Ca\two\,$\lambda\lambda\,8498-8662$}

\def\oidoub{[O\one]\,$\lambda\lambda$\,$6300,\,6364$}
\def\oiauroral{[O\one]\,$\lambda$\,$5577$}
\def\oitrip{O\one\,$\lambda\lambda$\,$7771-7775$}

\def\mgi{Mg\one]\,$\lambda\,4571$}
\def\nad{Na\one\,$\lambda\lambda\,5896,5890$}

\newcommand{\iso}[2]{\ensuremath{^{#1}\rm{#2}}}

\begin{document}

   \title{Nebular phase properties of supernova Ibc from He-star explosions}

   \titlerunning{SNe Ibc at nebular times}

\author{
Luc Dessart\inst{\ref{inst1}}
  \and
   D. John Hillier\inst{\ref{inst2}}
   \and
   Tuguldur Sukhbold\inst{\ref{inst3}}
   \and
   S.E. Woosley\inst{\ref{inst4}}
   \and
   H.-T. Janka\inst{\ref{inst5}}
  }

\institute{
Institut d'Astrophysique de Paris, CNRS-Sorbonne Universit\'e, 98 bis boulevard Arago, F-75014 Paris, France.\label{inst1}
\and
    Department of Physics and Astronomy \& Pittsburgh Particle Physics,
    Astrophysics, and Cosmology Center (PITT PACC),  \hfill \\ University of Pittsburgh,
    3941 O'Hara Street, Pittsburgh, PA 15260, USA.\label{inst2}
\and
Department of Astronomy, Ohio State University, Columbus, Ohio, 43210, USA\label{inst3}
\and
Department of Astronomy and Astrophysics, University of California, Santa Cruz, CA 95064, USA\label{inst4}
\and
Max-Planck-Institut  f\"{u}r  Astrophysik,  Postfach  1317,  85741, Garching, Germany \label{inst5}
  }

   \date{}

  \abstract{Following our recent work on Type II supernovae (SNe), we
    present a set of 1D nonlocal thermodynamic equilibrium radiative
    transfer calculations for nebular-phase Type Ibc SNe starting from
    state-of-the-art explosion models with detailed
    nucleosynthesis.  Our grid of progenitor models is derived from He stars
    that were subsequently evolved under the influence of wind mass loss.
    These He stars, which most likely form through binary mass exchange,
    synthesize less oxygen than their single-star counterparts with the same
    zero-age main sequence (ZAMS) mass. This reduction is greater in
    He-star models evolved with an enhanced mass loss rate. We obtain a wide range of
    spectral properties at 200\,d. In models from He stars with an
    initial mass $>$\,6\,\msun, the \oidoub\ is of comparable or
    greater strength than \caiidoub\ -- the strength of
    \oidoub\ increases with He-star initial mass. In contrast, models
    from lower mass He stars exhibit a weak \oidoub, strong \caiidoub,
    but also strong N\two\ lines and Fe\two\ emission below
    5500\,\AA. The ejecta density, modulated by the ejecta mass, the
    explosion energy, and clumping, has a critical impact on the
    gas ionization, line cooling, and the spectral
    properties. Fe\two\ dominates the emission below 5500\,\AA\ and is
    stronger at earlier nebular epochs. It ebbs as the SN ages, while
    the fractional flux in \oidoub\ and \caiidoub\ increases, with a similar rate, as the
    ejecta recombine. Although the results depend on the adopted
      wind mass loss rate and pre-SN mass, we find that He-stars of
    6\,$-$\,8\,\msun\ initially (ZAMS mass of 23\,$-$\,28\,\msun)
    match adequately the properties of standard SNe Ibc. This finding
    agrees with the offset in progenitor masses inferred from
    the environments of SNe Ibc relative to SNe II. Our results for less massive
    He stars are more perplexing, since the predicted spectra are not seen in nature.
    They may be missed by current surveys
    or associated with Type Ibn SNe in which interaction dominates
    over decay power.  }


   \maketitle

%%%%%%%%%%%%%%%%%%%%%%%%%%%%%%%

\section{Introduction}

Recent developments with the nonlocal thermodynamic equilibrium
(nonLTE) radiative transfer code \cmfgen\ \citep{HD12} have led to a
better numerical stability in the steady-state solver and a more
suitable treatment of chemical mixing in core-collapse supernova (SN)
ejecta \citep{DH20_neb,DH20_shuffle}. Following our previous work
\citep{D21_sn2p_neb} on the nebular-phase properties of Type II SNe
arising from the explosion of stars that evolved in isolation at solar
metallicity and died as red supergiants \citep{sukhbold_ccsn_16}, we
here undertake a study of a similar nature based on the He-star
explosion models of \citet{ertl_ibc_20}, with the pre-SN evolution
described in \citet{woosley_he_19}. The underlying assumption is
that such He stars formed initially from the prompt removal of the
H-rich envelope through binary mass exchange, probably as the star
first expanded following the ignition of hydrogen burning in a shell
(case B mass transfer). This scenario is thought to be responsible for
most of the observed Type Ibc SNe and is thus distinct from the
single-star evolution that may produce most of the observed SNe
II. This study thus complements our previous work on SN II
\citep{D21_sn2p_neb} by documenting the nebular-phase properties of a
large grid of Type Ibc models from \citet{ertl_ibc_20}.

The properties of core-collapse SNe at nebular times are complex but offer the potential to constrain some important characteristics of the progenitor star and its explosion (see, for example, \citealt{fransson_chevalier_89}; \citealt{jerkstrand_rev_17}).  These characteristics include the yields from intermediate mass elements (e.g., O or Ca) and iron group elements (for example the initial abundance of \nifs, or  more rarely stable Ni; \citealt{jerkstrand_ni_15}), the geometry of the inner ejecta \citep{mazzali_98bw_01,maeda_neb_06,maeda_neb_08,modjaz_ibc_neb_08,taubenberger_ibc_09,milisavljevic_etal_10}, the formation of dust and molecules \citep{kotak_04dj_05,kotak_05af_06,rho_17eaw_18,rho_20oi_21}
or the late-time source of power for the ejecta (see for example the peculiar nebular phase spectral properties of SN\,2016gkg; \citealt{hanin_16gkg_20}). By combining the analysis of both photospheric-phase and nebular-phase properties, one can build a more consistent picture of core-collapse SNe, which can provide important constraints for massive star evolution and explosion.

Previous modeling of nebular-phase spectra of Type Ibc SNe\footnote{The sample includes Type IIb since they are very similar to SNe Ib, especially at nebular times, and very different from SNe II at all epochs.} has been limited to a few nearby objects for which good quality photometry and spectra could be obtained until late times. The study of the Type IIb SNe 1993J, 2008ax, and 2011dh by \citet{jerkstrand_15_iib} provided a wealth of information on their spectral properties (including nonLTE processes, line formation, molecule formation etc), the progenitor stars, and their explosion physics. The progenitor models for that study were however taken from the single-star explosion models of \citet{WH07} and subsequently trimmed to retain only the innermost layer of the H-rich envelope. This approach is not suitable for SNe Ibc for which the combined effects of binary-mass transfer and Wolf-Rayet wind mass loss appear essential \citep{podsiadlowski_92,eldridge_08_bin,yoon_ibc_10,yoon_wr_17,woosley_he_19,dessart_snibc_20}.

This study is thus devoted to modeling the nebular-phase properties for the large set of explosion models presented by \citet{ertl_ibc_20} and based on the He-star evolution models of \citet{woosley_he_19}. In the next section, we present our numerical setup, summarizing how the pre-SN evolution, the explosion phase, and the radiative transfer calculations are treated. Section~\ref{sect_gen_prop} presents the general properties of the progenitor star and explosion models we study. Section~\ref{sect_std} describes the spectral properties of the He-star explosion models evolved with the nominal mass loss rate. Section~\ref{sect_rep} focuses on three representative models for lower, intermediate, and higher He-star masses. Section~\ref{sect_ii_ib} takes a side step and compares the properties of a SN Ib and a SN II at a post-explosion time when the SN luminosity is comparable. Section~\ref{sect_den} explores the strong impact that ejecta density has on SN radiation properties, in particular as obtained through the introduction of clumping or through variations in explosion energy. Section~\ref{sect_mdot} studies the impact of pre-SN mass loss rate on the resulting SNe Ibc. Section~\ref{sect_positron} relaxes the assumption of local positron trapping to gauge the importance of their contribution to SN Ibc spectra and luminosity. While all simulations so far were done at 200\,d after explosion, Section~\ref{sect_evol} describes the evolution until late times for few of our models. Section~\ref{sect_hei} searches for the signatures of  He\one\ lines in the nebular phase spectra of our He-star explosion models, and in particular how these vary with SN type. Section~\ref{sect_obs} presents a succinct comparison to a few well observed SNe Ibc (including a couple of SNe IIb). Finally, we present our conclusions in Sect.~\ref{sect_conc}. Supplementary tables and figures are provided in the appendix to complement the information given in the main text.

\section{Numerical setup}
\label{sect_setup}

\subsection{He-star models: Progenitor evolution}

The pre-SN evolution of the stars studied here has been discussed in \citet{woosley_he_19} and \citet{ertl_ibc_20} and their models are employed without revision (some remapping is performed before doing the radiative-transfer calculation -- see Sect~\ref{sect_setup_cmfgen}). To summarize, helium stars, consisting of helium plus a concentration of heavier isotopes reflecting the result of hydrogen burning in a massive star with solar metallicity, were evolved, including the effect of mass loss by winds to the pre-SN stage. The initial helium star mass was assumed to reflect the mass of the helium core at the time of central helium ignition in stars with variable main sequence mass. No main sequence evolution or binary mass exchange was calculated. The loss of the envelope was regarded as instantaneous, and the relevant main sequence mass was inferred by inspection of previously existing grids of single star evolution evaluated at helium ignition. Once helium burning ignited, mass loss was included using the prescription of \citet{yoon_wr_17}, which is itself an amalgamation of \citet{hainich_wn_14} and \citet{tramper_wc_16}. Because these rates are uncertain and possibly underestimates, we also considered cases with multipliers of 1.5 and 2 times the base rate of \citet{yoon_wr_17}. Uncertainties in whether the envelope was lost precisely at helium ignition or a lot before or after is absorbed into uncertainties in the mass loss rate. Stars are expected to frequently lose their envelopes at this time because it is at this point that the star expands from a compact main sequence star to a supergiant.

Because of their isolation from an overlying hydrogen envelope, these helium stars did not increase in mass as they burned helium, but instead shrank due to mass loss by winds. The final pre-SN mass was thus substantially smaller than it would have been if the helium core evolved as part of a single isolated star. For example, evolved with the same code and physics, the helium core mass for a 15\,\msun\ single star is 4.3\,\msun\ when the star explodes, but the same star with its envelope removed at helium ignition and subject to mass loss by winds has a pre-SN mass of only 2.4\,\msun\ \citep{woosley_he_19}. These differences in masses translate into different structures at explosion time, as typified for example by their core compactness and different compositions.  The other major difference is that the models were not artificially exploded using pistons. Rather, as explained in the next section, the explosion is triggered by the deposition of neutrino energy, which was treated using a 1D neutrino transport model.

\subsection{He-star models: Explosion phase}

The explosion modeling of the He stars was performed in spherical
  symmetry with the \photb\ code (see
  \citealt{ertl_ccsn_16,ertl_ibc_20}; \citealt{sukhbold_ccsn_16}). In
  order to mimic the basic effects of thermal energy deposition over
  an extended period of several seconds by the neutrino-driven
  mechanism, an approximate, gray treatment of neutrino transport in
  combination with a time-dependent model for the neutrino emission of
  the newly formed neutron star was employed. The parameters of this
  neutrino ``engine'' were calibrated by the requirement that the
  observationally determined explosion energies and \nifs\ yields of
  the Crab SN and of SN\,1987A could be reproduced for progenitors in
  the mass ranges (around 9$-$10\,\msun\ and 15$-$20\,\msun,
  respectively) of these two well-studied SNe.

The application of this neutrino engine led to a progenitor-dependent explosion
behavior with differences in the mass cut, explosion energy, and compact remnant
(neutron star or black hole) mass. The \photb\ simulations were performed with
a nuclear network constrained to a small set of species (the 13 $\alpha$-nuclei,
neutrons, protons, and a ``tracer nucleus'' (hereafter ``Tr'') that represents
neutron-rich nuclei where $Y_{\rm e} < 0.49$). Therefore the SN results from \photb\
were used to guide postprocessing calculations of the explosions with the \kepler\ code,
which employs a large network and thus allows to capture the details of the nucleosynthesis
consistent with the progenitor evolution. The chemical composition of the ejecta could
thus be determined accurately to facilitate the spectral modeling in the study reported here.

\begin{table*}
    \caption{Ejecta model properties from \citet{ertl_ibc_20} -- see also Table~\ref{tab_prog_bis} and \ref{tab_ertliron}.
\label{tab_prog}
}
\begin{center}
\begin{tabular}{l@{\hspace{2mm}}|c@{\hspace{2mm}}c@{\hspace{2mm}}c@{\hspace{2mm}}c@{\hspace{2mm}}c@{\hspace{2mm}}c@{\hspace{2mm}}c@{\hspace{2mm}}c@{\hspace{2mm}}c@{\hspace{2mm}}c@{\hspace{2mm}}c@{\hspace{2mm}}c@{\hspace{2mm}}c@{\hspace{2mm}}c@{\hspace{2mm}}}
\hline
Model  &  $M_{\rm preSN}$  & $M_{\rm ej}$  &     $E_{\rm kin}$   & $V_m$   &  \iso{4}He & \iso{12}C   & \iso{14}N  & \iso{16}O &  \iso{24}Mg &  \iso{28}Si &  \iso{40}Ca &  \iso{56}Ni$_{t=0}$ & $M_{\rm sh}$ & $V_{{\rm max},\nifs}$ \\
       &     [\msun] & [\msun]   &        [foe]    & [\kms]    &    [\msun] & [\msun] & [\msun] & [\msun] & [\msun] & [\msun] & [\msun] & [\msun] & [\msun] & [\kms] \\
\hline
%  Model     MpreSN     Mej        Ekin            Vm           he4            c12           n14           o16          mg24          si28          ca40          ni56    Msh    Vmax_ni56
   he2p6      & 2.15 &   0.79  &     0.13   &      4134   &      0.71    &      0.02  &   4.78(-3)   &   2.28(-2)   &   1.58(-3)   &   3.35(-3)   &   2.40(-4)   &   1.22(-2)  & 0.29 &  2697   \\
   he2p9      & 2.37 &   0.93  &     0.37   &      6336   &      0.77    &      0.04  &   5.15(-3)   &   5.03(-2)   &   3.82(-3)   &   1.01(-2)   &   5.62(-4)   &   2.32(-2)  & 0.38 &  4353   \\
   he3p3      & 2.67 &   1.20  &     0.55   &      6777   &      0.84    &      0.06  &   6.21(-3)   &   1.51(-1)   &   1.75(-2)   &   2.76(-2)   &   1.00(-3)   &   4.00(-2)  & 0.43 &  3712   \\
   he3p5      & 2.81 &   1.27  &     0.41   &      5704   &      0.87    &      0.07  &   6.31(-3)   &   1.72(-1)   &   1.60(-2)   &   2.13(-2)   &   7.34(-4)   &   2.92(-2)  & 0.49 &  3121   \\
   he4p0      & 3.16 &   1.62  &     0.63   &      6272   &      0.92    &      0.10  &   6.46(-3)   &   3.10(-1)   &   2.98(-2)   &   4.70(-2)   &   1.35(-3)   &   4.45(-2)  & 0.82 &  3974   \\
   he4p5L     & 3.49 &   1.89  &     0.54   &      5348   &      0.95    &      0.13  &   6.55(-3)   &   4.29(-1)   &   4.00(-2)   &   4.90(-2)   &   1.73(-3)   &   8.22(-2)  & 1.14 &  3532   \\
   he4p5      & 3.49 &   1.89  &     1.17   &      7884   &      0.95    &      0.13  &   6.52(-3)   &   4.19(-1)   &   3.73(-2)   &   6.14(-2)   &   2.40(-3)   &   8.59(-2)  & 1.14 &  5588   \\
   he4p5H     & 3.49 &   1.89  &     2.44   &     11400   &      0.97    &      0.12  &   6.38(-3)   &   3.96(-1)   &   3.55(-2)   &   7.86(-2)   &   3.33(-3)   &   9.03(-2)  & 1.14 &  8369   \\
   he5p0      & 3.81 &   2.21  &     1.51   &      8286   &      0.97    &      0.15  &   6.60(-3)   &   5.92(-1)   &   5.20(-2)   &   5.55(-2)   &   2.26(-3)   &   9.77(-2)  & 1.43 &  6164   \\
   he6p0      & 4.44 &   2.82  &     1.10   &      6269   &      0.95    &      0.25  &   6.20(-3)   &   9.74(-1)   &   1.01(-1)   &   5.88(-2)   &   2.12(-3)   &   7.04(-2)  & 2.16 &  4990   \\
   he7p0      & 5.04 &   3.33  &     1.38   &      6456   &      0.90    &      0.39  &   5.42(-3)   &   1.29       &   1.07(-1)   &   9.47(-2)   &   3.42(-3)   &   1.02(-1)  & 2.77 &  5356   \\
   he8p0      & 5.63 &   3.95  &     0.71   &      4251   &      0.84    &      0.49  &   5.17(-3)   &   1.71       &   1.10(-1)   &   4.89(-2)   &   2.00(-3)   &   5.46(-2)  & 3.40 &  3435   \\
   he12p0     & 7.24 &   5.32  &     0.81   &      3911   &      0.23    &      1.00  &   1.42(-4)   &   3.03       &   8.73(-2)   &   7.41(-2)   &   3.42(-3)   &   7.90(-2)  & 3.60 &  2531   \\
\hline
   he5p0x1p5    & 3.43  &  1.85  &   1.59   &      9299   &      0.86    &      0.16  &   6.18(-3)   &   4.22(-1)   &   3.27(-2)   &   6.34(-2)   &   2.74(-3)   &   1.11(-1)  & 1.12 &   6758  \\
   he6p0x1p5    & 3.96  &  2.32  &   1.57   &      8237   &      0.83    &      0.27  &   5.68(-3)   &   6.80(-1)   &   4.76(-2)   &   6.31(-2)   &   3.03(-3)   &   1.23(-1)  & 1.25 &   5787  \\
%  he7p0x1p5    & 4.45  &  2.84  &   1.46   &      7203   &      0.76    &      0.42  &   3.48(-3)   &   9.29(-1)   &   5.84(-2)   &   7.44(-2)   &   3.37(-3)   &   1.99(-1)  & 1.69 &         \\
   he8p0x1p5    & 4.92  &  3.26  &   1.49   &      6776   &      0.64    &      0.56  &   6.14(-4)   &   1.28       &   9.73(-2)   &   6.75(-2)   &   2.74(-3)   &   1.06(-1)  & 2.05 &   5096  \\
   he9p0x1p5    & 4.87  &  3.13  &   1.03   &      5743   &      0.26    &      0.62  &   1.55(-5)   &   1.49       &   9.36(-2)   &   8.39(-2)   &   3.58(-3)   &   1.21(-1)  & 2.07 &   4222  \\
   he10p0x1p5   & 4.96  &  3.20  &   1.22   &      6192   &      0.18    &      0.67  &   1.20(-5)   &   1.65       &   9.07(-2)   &   8.89(-2)   &   3.56(-3)   &   9.92(-2)  & 2.00 &   4465  \\
   he11p0x1p5   & 5.19  &  3.44  &   1.11   &      5705   &      0.20    &      0.76  &   4.24(-5)   &   1.78       &   7.16(-2)   &   6.26(-2)   &   2.72(-3)   &   4.26(-2)  & 2.13 &   4015  \\
   he12p0x1p5   & 5.43  &  3.71  &   0.97   &      5132   &      0.23    &      0.79  &   5.00(-5)   &   1.89       &   7.48(-2)   &   7.22(-2)   &   3.25(-3)   &   8.80(-2)  & 2.36 &   3561  \\
   he13p0x1p5   & 5.64  &  3.99  &     1.31   &      5736   &      0.24    &      0.82  &   7.19(-5)   &   2.04       &   7.38(-2)   &   8.26(-2)   &   3.78(-3)   &   1.28(-1)  & 2.60 &   4179  \\
\hline
   he10p0x2p0  &  3.98  & 2.22  &     1.23   &      7446   &      0.15    &      0.57  &   5.53(-6)  &   9.31(-1)    &   5.52(-2)   &   6.92(-2)   &   2.94(-3)   &   1.16(-1)  & 1.29 &  5233 \\
\hline
\end{tabular}
\end{center}
    {\bf Notes:} The table columns correspond to the preSN mass, the ejecta mass, the ejecta kinetic energy (1 foe $\equiv 10^{51}$\,erg), the mean expansion rate $V_{\rm m} \equiv \sqrt{2E_{\rm kin}/M_{\rm ej}}$, the cumulative yields of \iso{4}He, \iso{12}C, \iso{14}N, \iso{16}O, \iso{24}Mg, \iso{28}Si, \iso{40}Ca, and \nifs\ prior to decay. The last two columns give the Lagrangian mass $M_{\rm sh}$ adopted for the shell shuffling (see Sect.~\ref{sect_prep}), as well as the ejecta velocity that bounds 99\% of the total \nifs\ mass in the corresponding model (this value corresponds roughly to the velocity at the  Lagrangian mass $M_{\rm sh}$). All \nifs\ masses used here were taken from the \kepler\ approximation to the \photb\ neutrino simulation, and are between the best guess (\nifs+``Tr'') and generous upper bounds ($3/4 \times$(\nifs\ + ``Tr''+ $\alpha$)) defined by \citet[see also Table~\ref{tab_ertliron}]{ertl_ibc_20}.
\end{table*}

\subsection{Preparation of models for radiative-transfer simulations using \cmfgen}
\label{sect_prep}

In this paper we discuss a large grid of radiative-transfer calculations based on explosion models of solar-metallicity nonrotating He-stars evolved with the nominal pre-SN wind mass loss rate (no suffix; model he2p6 corresponds to the He-star model with an initial mass of 2.6\,\msun, etc). Additional models with the mass loss rate  enhanced by 50\% (suffix x1p5) and 100\% (suffix x2p0) are also considered. For the standard mass loss case, we consider He-star masses of 2.6, 2.9, 3.3, 3.5, 4.0, 4.5, 5.0, 6.0, 7.0, 8.0, and 12.0\,\msun, with fewer models at higher mass because they are in tension with observed properties of SNe Ic (they have residual surface He and a large ejecta mass).

The grid with a 50\% enhanced mass loss rate does not cover the low mass end where the influence of wind mass loss is negligible. Instead, the grid covers with a 1\,\msun\ increment the range of He-star masses between 5 and 13\,\msun. These ejecta models have similar properties to those computed and discussed in the study of \citet{dessart_snibc_20}. In particular, they yield SN spectra that reproduce satisfactorily the dichotomy between SNe Ib and SNe Ic. Finally, for the 100\% enhanced mass loss rate, we only consider a He-star progenitor of 10\,\msun\ (model he10p0x2p0).

The ejecta models described in the previous section (a summary of ejecta properties is provided in Tables~\ref{tab_prog} and \ref{tab_prog_bis}) are evolved until 1\,d after explosion, at a time when homologous expansion is essentially established. To prepare the ejecta models for \cmfgen, we  select only the ejecta layers  with a velocity greater than 50\,\kms\ (this trims any fallback material; however, in most explosion models, the minimum ejecta velocity is several 100\,\kms\ up to $\sim$\,1000\,\kms) and lower than about 20\,000\,\kms\ (the exact value depends on the explosion energy, but it is chosen so that the outer ejecta regions are optically thin in both the continuum and lines).  A full discussion of the approach,  its merits and its weaknesses, is provided in \citealt{DH20_shuffle}).

\begin{figure*}
\centering
\includegraphics[width=0.495\hsize]{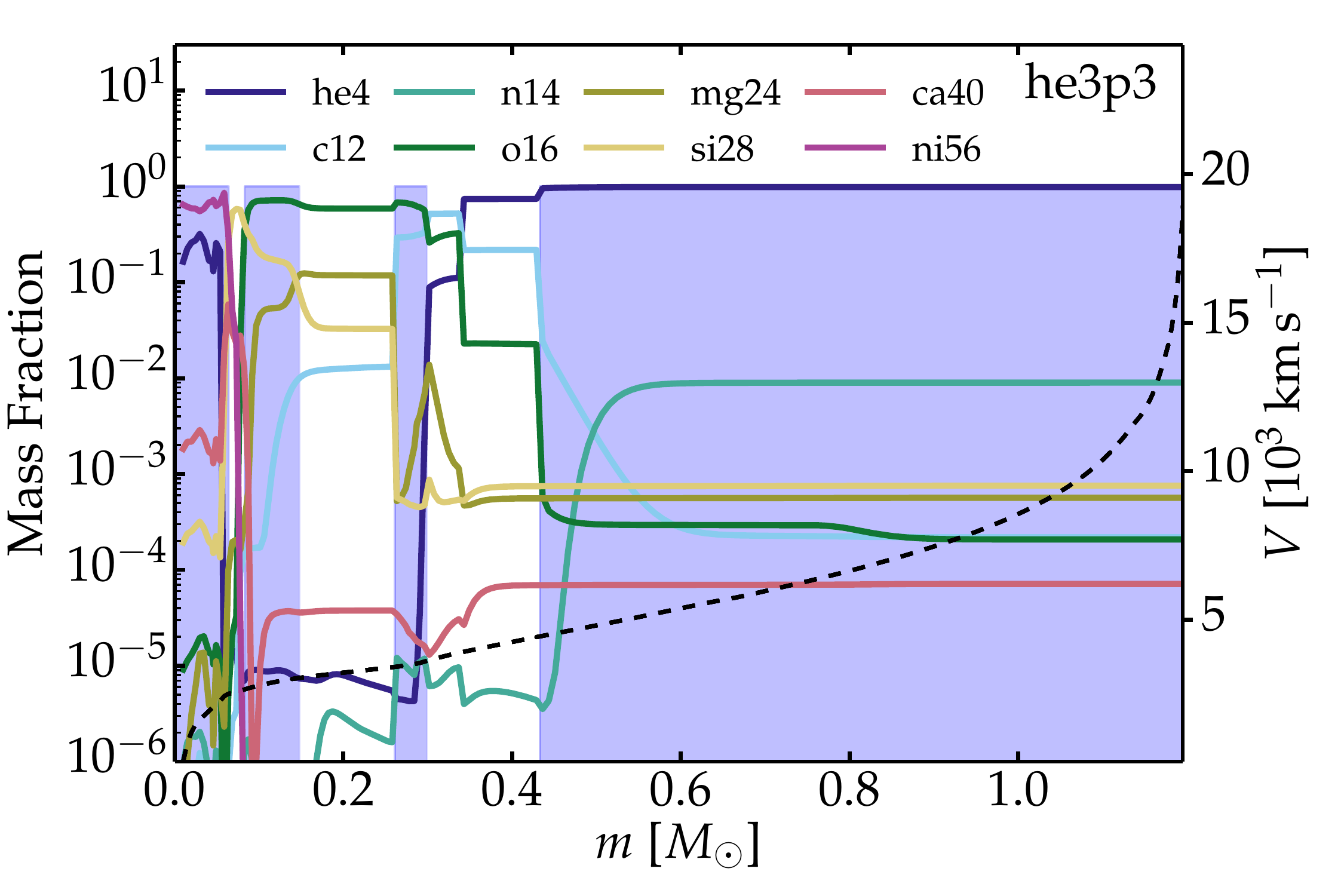}
\includegraphics[width=0.495\hsize]{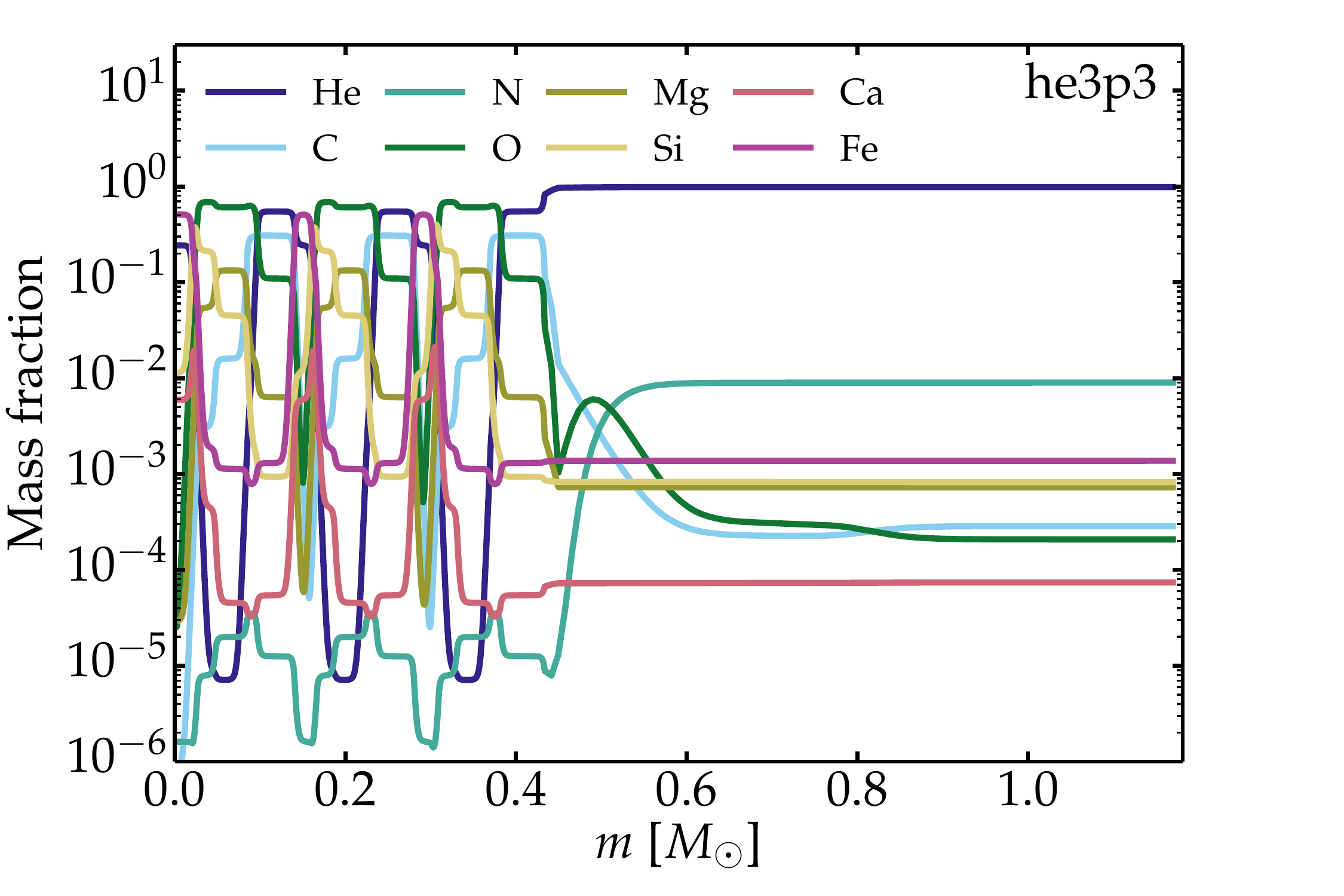}
\includegraphics[width=0.495\hsize]{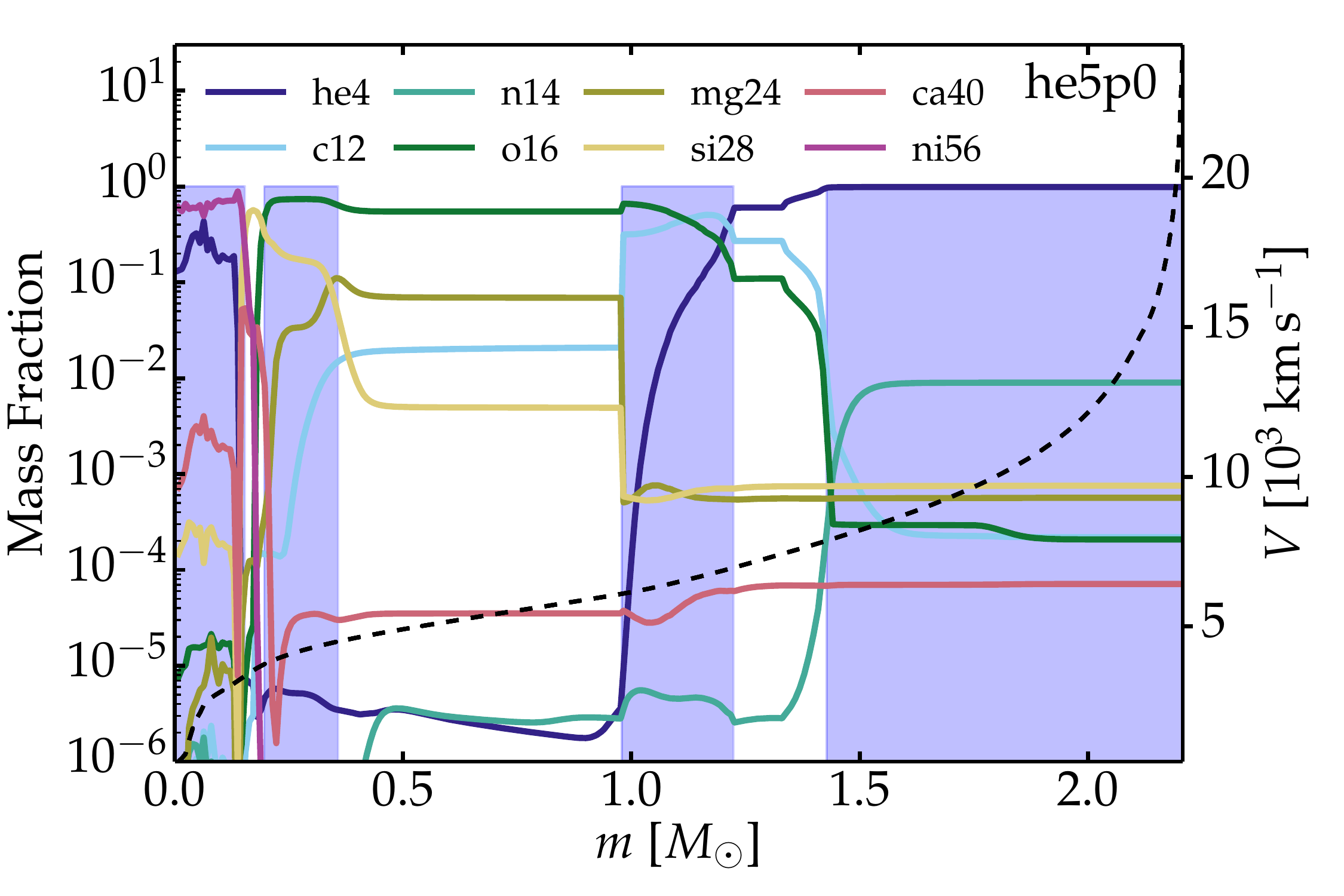}
\includegraphics[width=0.495\hsize]{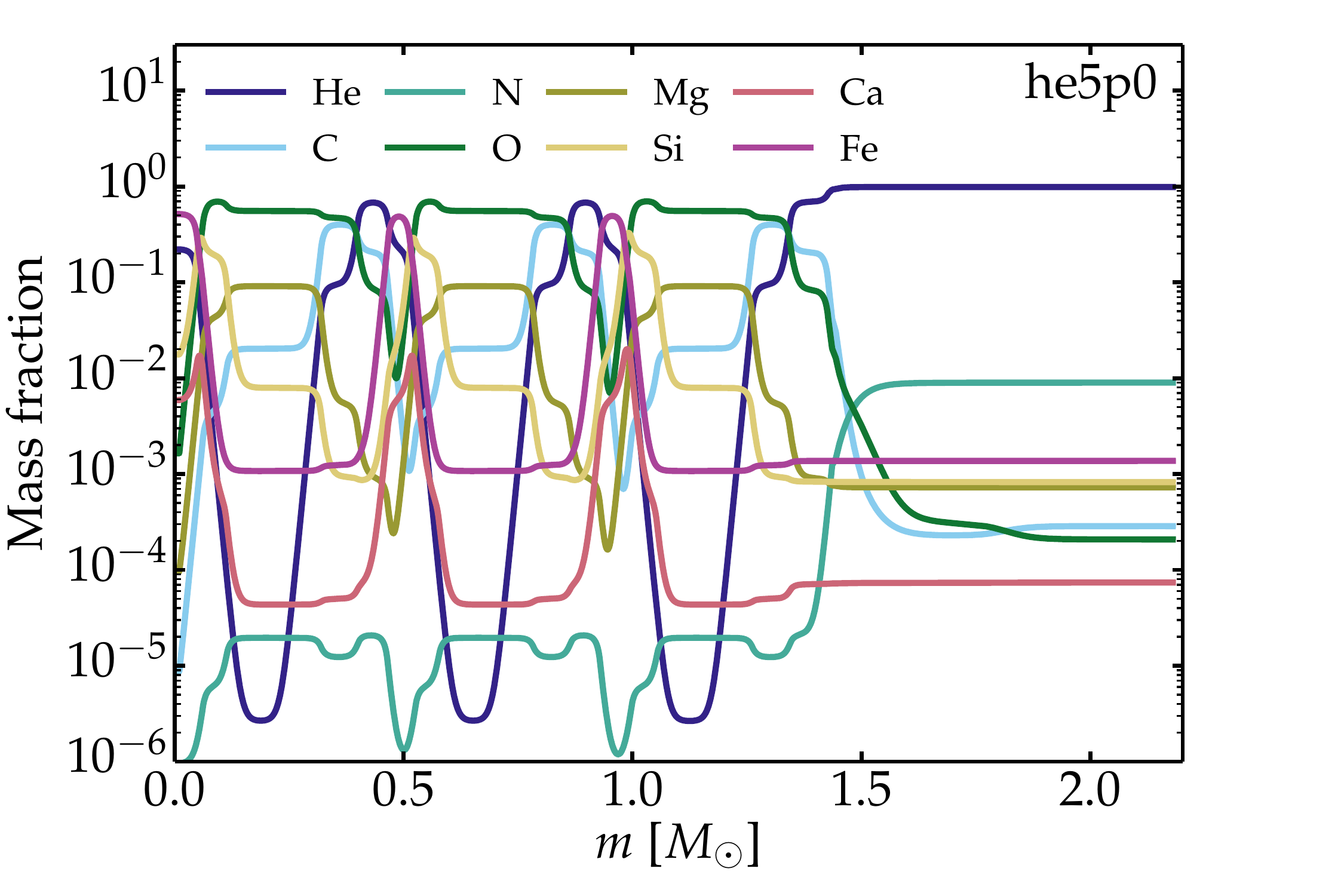}
\includegraphics[width=0.495\hsize]{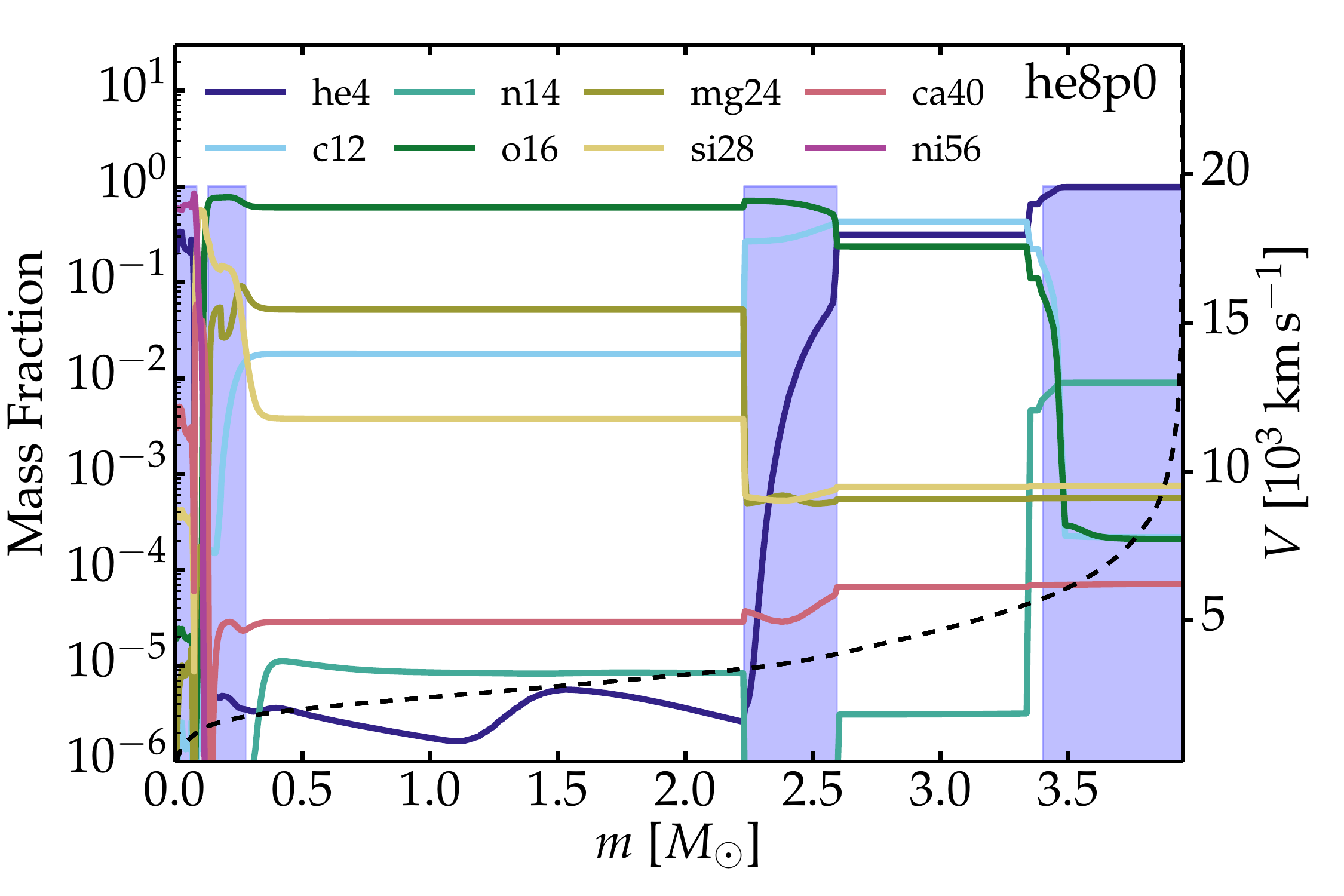}
\includegraphics[width=0.495\hsize]{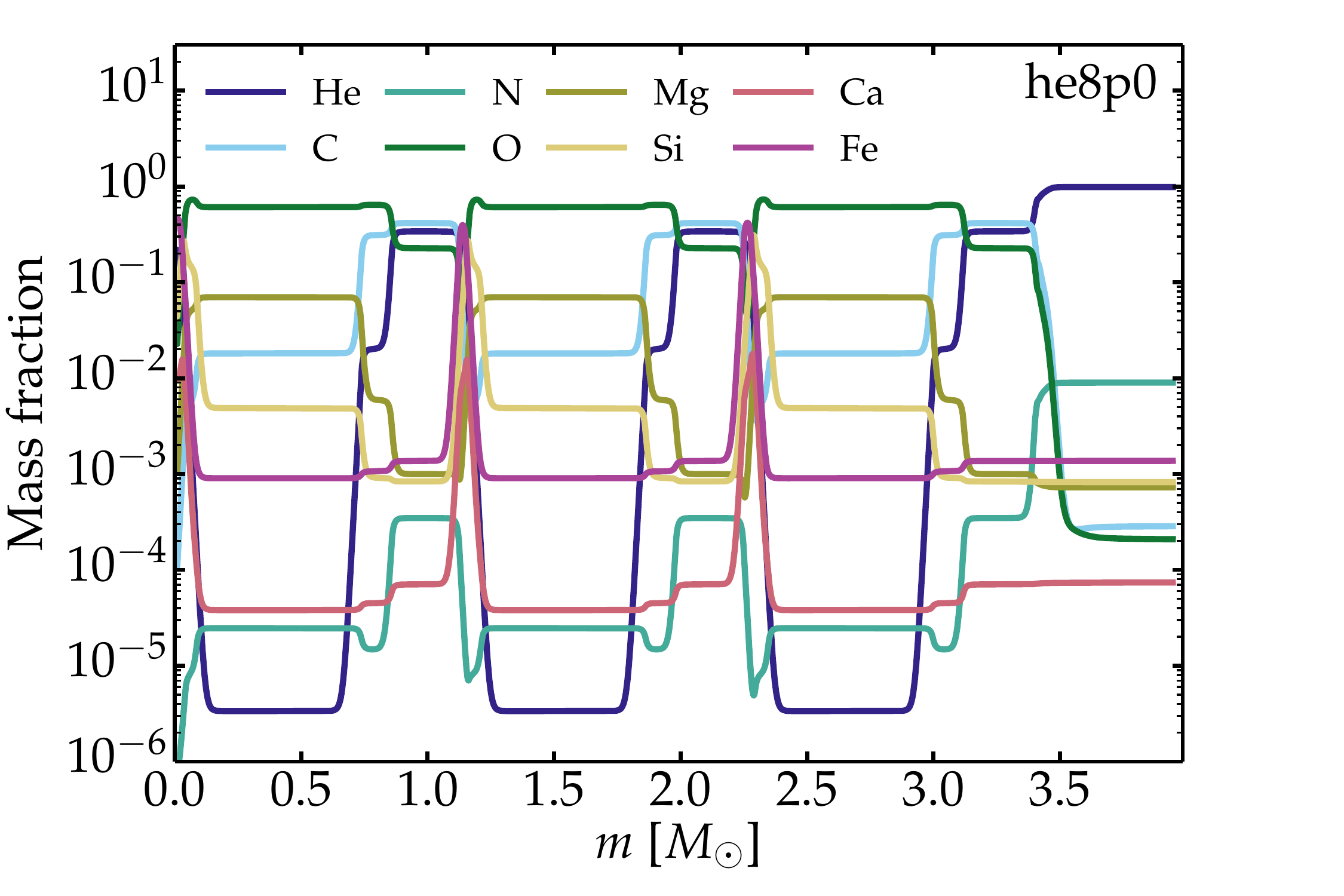}
\vspace{-0.2cm}
\caption{Left column: Chemical composition versus Lagrangian mass in the original, unmixed he3p3, he5p0, and he8p0 ejecta models of  \citet{ertl_ibc_20} at 1\,d (the original mass fraction is shown for \nifs) -- ejecta properties are summarized in Table~\ref{tab_prog}. The shaded areas denote dominant shells, of a given composition, that are used in the shuffling procedure. Right column: corresponding ejecta at 200\,d in which the main progenitor shells have been shuffled in mass space within the Lagrangian mass cut $M_{\rm sh}=\,$0.43, 1.43, and 3.40\,\msun\  (details on the shuffling procedure are given in the appendix of \citealt{DH20_shuffle}). In the right column, the mass fractions shown are obtained by summing over all isotopes associated with a given element (see Sect.~\ref{sect_prep} and Table~\ref{tab_mod_atoms} for details). The x-axis covers the same range in each row and the y-axis covers the same range in all panels.
\label{fig_prog_comp}
}
\end{figure*}

For most models a large nuclear network was used to compute the progenitor evolution and the explosion. Not all models used the same number of isotopes although in all cases the network was much larger than needed to capture the abundance of the elements important for the radiative transfer calculations presented here.  Most models used about 450 isotopes which was sufficient to include all nontrivial nuclear flows up to about $A = 90$. Some models included more isotopes during the explosion.

All \nifs\ masses used here were taken from the \kepler\ approximation to the \photb\ neutrino simulation, and are between the best guess (\nifs+``Tr'') and generous upper bounds ($3/4 \times$(\nifs\ + ``Tr''+ $\alpha$)) defined by \citet{ertl_ibc_20}. Our \nifs\ masses are thus compatible with the neutrino-powered explosions, but perhaps slightly on the high side (see additional details in Table~\ref{tab_ertliron}).

As the radiative transfer is primarily controlled by the most abundant species, and isotopic effects are small, we sum over all isotopes (see Table ~\ref{tab_mod_atoms}) and only include the following species: He, C, N, O, Ne, Na, Mg, Si, S, Ar, K, Ca, Sc, Ti, Cr, Fe, Co, and Ni in our radiative transfer calculations. We allow for the radioactive decay of \cofs\ and \nifs, which  are treated individually. More unstable isotopes could be allowed to decay, but this is either not necessary at one year post explosion (e.g., for \iso{44}Ti; \citealt{jerkstrand_87a_11}), or secondary because of their low abundance (e.g., \iso{57}Ni is about 20 times less abundant than \nifs\ in these explosion models). Treating underabundant unstable isotopes as stable has little impact on the composition at late times.

For each model, we identify all shells that have a distinct composition. There is more than one way to do this since some shells do not have a clear boundary, and some show internal composition gradients. For example, the He mass fraction shows a steady rise as we progress outward in the O/C shell of model he8p0. Some species may be abundant in consecutive shells, such as Ca in the Fe/He and the Si/S shells. The main shells are designated by their dominant species and are typically Fe/He, Si/S, O/Si, O/Ne/Mg, O/C, He/C, and He/N. In some cases, the He/C shell is also O-rich and so one may designate this shell as He/C/O.

The Fe/He and the Si/S shells are the product of explosive nucleosynthesis so their mass depends on the progenitor structure and on the explosion energy, which in nature, is also determined by the progenitor structure.  When the explosion energy is high, the O/Si shell composition may also be altered. In these regions, the composition profile after shock passage is very different from that in the corresponding convective shell of the progenitor. While convection within a shell tends to make the composition uniform, the shock leaves behind a composition gradient, so that \nifs, Si, Ca, etc.\ drop in abundance as we progress outward. Consequently, \nifs\ is abundant in the Fe/He shell but it is also present in the Si/S shell. Similarly, the Si abundance continuously drops through the O/Si shell. By contrast, the O, Ne, and Mg composition profiles are nearly constant in the O/Ne/Mg shell (which is essentially unaffected by explosive nucleosynthesis).

The average mass fractions for the main elements composing each of the main shells are provided in Tables~\ref{tab_he2p60}\,$-$\,\ref{tab_he10Ax2p0}. The typical shell
compositions are as follows:
\begin{description}\itemsep=3pt
\item[He/N shell:]
Composed of He (98\,\%) and N (1\,\%) primarily.
\item[He/C shell:]
Composed of He (30 to 80\,\%), C (10 to 40\,\%), with O present but typically less abundant than 10\,\%.
\item[O/C shell:]
Composed of O (50 to 60\,\%) and C (30 to 40\,\%), with traces of He and Ne.
\item[O/Ne/Mg shell:]
Composed of O (50 to 65\,\%), Ne ($\sim$\,30\,\%), and Mg (5 to 15\,\%), with Na ($\sim$\,0.5\,\%) well above the solar metallicity value. Other abundances also differ from solar -- Co is more than  an order of magnitude above solar, the Ni abundance has been increased by roughly a factor of five, the Ca abundance is nearly a factor of two below solar, and the Fe abundance is also (slightly) below solar.  In contrast to some of the higher-mass progenitors of \citet{sukhbold_ccsn_16}, there is no sign of a merging of the Si shell with the O shell in any of these models.
\item[O/Si shell:]
Composed of O ($\sim$\,70\,\%), Si ($\sim$\,15\,\%), S (few percent), Mg (few percent), and the Ca abundance is well above solar (mass fraction of about 0.0005).
\item[Si/S shell:]
Composed of Si (about 45 to 55\,\%), S ($\sim$\,25\,\%),   with a few percent of Ar and Ca. The Si/S shell also contains some \nifs\ initially and thus is composed of about 10\,\% Fe at late times.
\item[Fe/He shell:]
Composed of Fe (45 to 60\%; the \fefs\ isotope dominates and arises from \nifs\ and \cofs\ decay), \cofs\ (about 10\,\% at 200\,d, as a result of the partial decay of \cofs\ into \fefs), He (15 to 25\,\%), with about 5\,$-$\,10\,\% of \nife\ (i.e., stable nickel), and about 0.5\,\% of Ca.
\end{description}

Having identified the main shells, we can proceed with the shuffling of shells of distinct composition \citep{DH20_shuffle}. We first choose the Lagrangian mass ($M_{\rm sh}$)  that bounds the shuffling.  As  the choice of $M_{\rm sh}$ is uncertain, we have used two options for setting $M_{\rm sh}$. Tests show that they  yield similar spectra (not shown). By default, we shuffle all shells interior to the outermost shell. For SNe Ib progenitors this is the He/N shell, while for  SNe Ic progenitors  it is the He/C, He/C/O, or O/C shell \citep{dessart_snibc_20}. The other option was to limit the shuffling of shells interior to the outermost two shells, which essentially means limiting the shuffling to the Fe/He, Si/S and the O-rich shells.\footnote{We have performed tests for this alternative but find that the impact is small, and mostly limited to lower-mass progenitors with a massive He-rich shell. This exploration is not shown in this paper.} Our approaches mean that macroscopic mixing occurs over 30\,$-$\,85\% of the ejecta mass, leaving the outermost ejecta layers intact.  In reality, the assumption of spherical symmetry is perhaps a greater limitation, in the sense that mixing may be very strong in some directions and weak in others.

With $M_{\rm sh}$ thus specified, we first force a uniform composition in all the main shells within $M_{\rm sh}$. We then split each of the main identified shells in three equal-mass parts and shuffle these truncated parts. This choice places a significant mass of \nifs\ at large velocities. As $\gamma$-rays deposit their energy over an extended ejecta volume, we have found that the specific shell arrangement has little impact on the outcome (see \citealt{DH20_shuffle}). A more profound problem may be the use of 1D smooth ejecta density structure produced in 1D explosion models.  This ignores the action of multidimensional fluid instabilities and the possibility of material compression.

The left column of Fig.~\ref{fig_prog_comp}  and Figs.~\ref{fig_prog_comp_all}\,$-$\,\ref{fig_prog_comp_all_bis} give the unmixed undecayed composition profile versus Lagrangian mass at 1\,d for those explosion models in \citet{ertl_ibc_20} that are based on the He-star progenitors evolved with the nominal mass loss rate \citep{woosley_he_19}.  The right column in Fig.~\ref{fig_prog_comp} shows the shell structure and abundances after we have performed the shell shuffling in models he3p3, he5p0, and he8p0. For completeness, we  provide the composition of each dominant shell in the unmixed ejecta in Tables~\ref{tab_he2p60}\,$-$\,\ref{tab_he12p00} for He-star progenitor models with the nominal mass loss rate, Tables~\ref{tab_he5p00x1p5}\,$-$\,\ref{tab_he13p0x1p5} for He-star progenitor models with a 50\% enhanced mass loss rate, and Table~\ref{tab_he10Ax2p0} for the only model treated here with a 100\% enhanced mass loss.

\subsection{Radiative transfer calculations with \cmfgen}
\label{sect_setup_cmfgen}

Having prepared the shuffled-shell structure for a wide set of He-star explosion models, we remap all models onto a grid with 350 radial points. From the inner ejecta layers up to $M_{\rm sh}$, we use a uniform grid in velocity (or radius) with 300 points. Above $M_{\rm sh}$, we allocate 50 points equally spaced in velocity but on a logarithmic scale.  All simulations are performed at a SN age of 200\,d after explosion unless otherwise stated (see Sect.~\ref{sect_evol}).

The radiative transfer performed with \cmfgen\ \citep{hm98,HD12} is carried out with the approach described in \citet{DH20_neb}. As we assume steady state (except in Sect.~\ref{sect_evol})  a calculation at any epoch can be performed without any knowledge of the previous evolution. This  permits a much broader exploration, allowing various ejecta properties to be varied. We assume that the only source of power in those explosions at nebular times is the decay of \nifs\ and \cofs\ and treat nonthermal processes as per normal \citep{d12_snibc,li_etal_12_nonte}. For simplicity, we compute the nonlocal energy deposition by solving the gray radiative transfer equation for $\gamma$-rays with the gray absorption opacity set to 0.06\,$Y_{\rm e}$\,cm$^2$\,g$^{-1}$, where $Y_{\rm e}$ is the electron fraction. All simulations assume that positrons deposit their power at the site of emission, except for the study presented in Sect.~\ref{sect_positron}. The model atoms included and the number of levels used are listed in Table~\ref{tab_mod_atoms}. The numbers in parentheses correspond to the number of super levels and full levels employed (for details on the treatment of super levels, see \citealt{hm98}).

\begin{table}
\caption{Adopted species and model atoms.}
\label{tab_mod_atoms}
\begin{tabular}{ll}
\hline
Species & Model atoms \\
\hline
He (\iso{3}He and \iso{4}He) & He\one\ (40,51), He\two\ (13,30) \\
C (\iso{10}C to \iso{18}C) &  C\one\ (14,26), C\two\ (14,26) \\
N (\iso{11}N to \iso{21}N) &  N\one\ (44,104), N\two\ (23,41) \\
 O (\iso{13}O to \iso{22}O) & O\one\ (30,77), O\two\ (30,111) \\
 Ne (\iso{16}Ne to \iso{28}Ne) & Ne\one\ (70,139), Ne\two\ (22,91) \\
 Na (\iso{19}Na to \iso{30}Na)  &  Na\one\ (22,71)   \\
 Mg (\iso{20}Mg to \iso{32}Mg)&  Mg\one\ (39,122), Mg\two\ (22,65)  \\
 Si (\iso{23}Si to \iso{37}Si) & Si\one\ (100,187), Si\two\ (31,59) \\
 S (\iso{27}S to \iso{43}S)& S\one\ (106,322), S\two\ (56,324) \\
 Ar (\iso{31}Ar to \iso{49}Ar) & Ar\one\ (56,110), Ar\two\ (134,415) \\
 K (\iso{32}K to \iso{51}K) & K\one\ (25,44) \\
Sc (\iso{36}Sc to \iso{54}Sc)  & Sc\one\ (26,72), Sc\two\ (38,85) \\
Ca (\iso{35}Ca to \iso{52}Ca) &  Ca\one\ (76,98), Ca\two\ (21,77) \\
Ti (\iso{39}Ti to \iso{56}Ti) &  Ti\two\ (37,152), Ti\three\ (33,206) \\
Cr (\iso{43}Cr to \iso{62}Cr) &  Cr\two\ (28,196), Cr\three\ (30,145) \\
Fe (\iso{47}Fe to \iso{67}Fe) & Fe\one\ (413,1142), Fe\two\ (275, 827)\\
                                              & \hbox{ }\hspace{0.3cm} Fe\three\ (83, 698) \\
Co (\iso{48}Co to \iso{70}Co)\tablefootmark{a} &  Co\two\ (44,162), Co\three\ (33,220) \\
Ni (\iso{51}Ni to \iso{73}Ni)\tablefootmark{b}   & Ni\two\ (27,177),  Ni\three\ (20,107)   \\
\hline
\end{tabular}
\tablefoot{The left column gives the range of isotopes considered when determining the abundance for each species treated in our \cmfgen\ calculations. The right column lists the atoms and ions treated in the radiative transfer (the parenthesis gives the number of super-levels followed by the corresponding number of full levels; see \citealt{hm98} for discussion). \\
\tablefoottext{a}\cofs\ is treated as a separate species, but uses the same Co atomic data and model atoms. \\
\tablefoottext{b}\nifs\ is treated as a separate species, but uses the same Ni atomic data and model atoms.
}
\end{table}

\section{General properties}
 \label{sect_gen_prop}

\begin{figure}
\centering
\includegraphics[width=0.99\hsize]{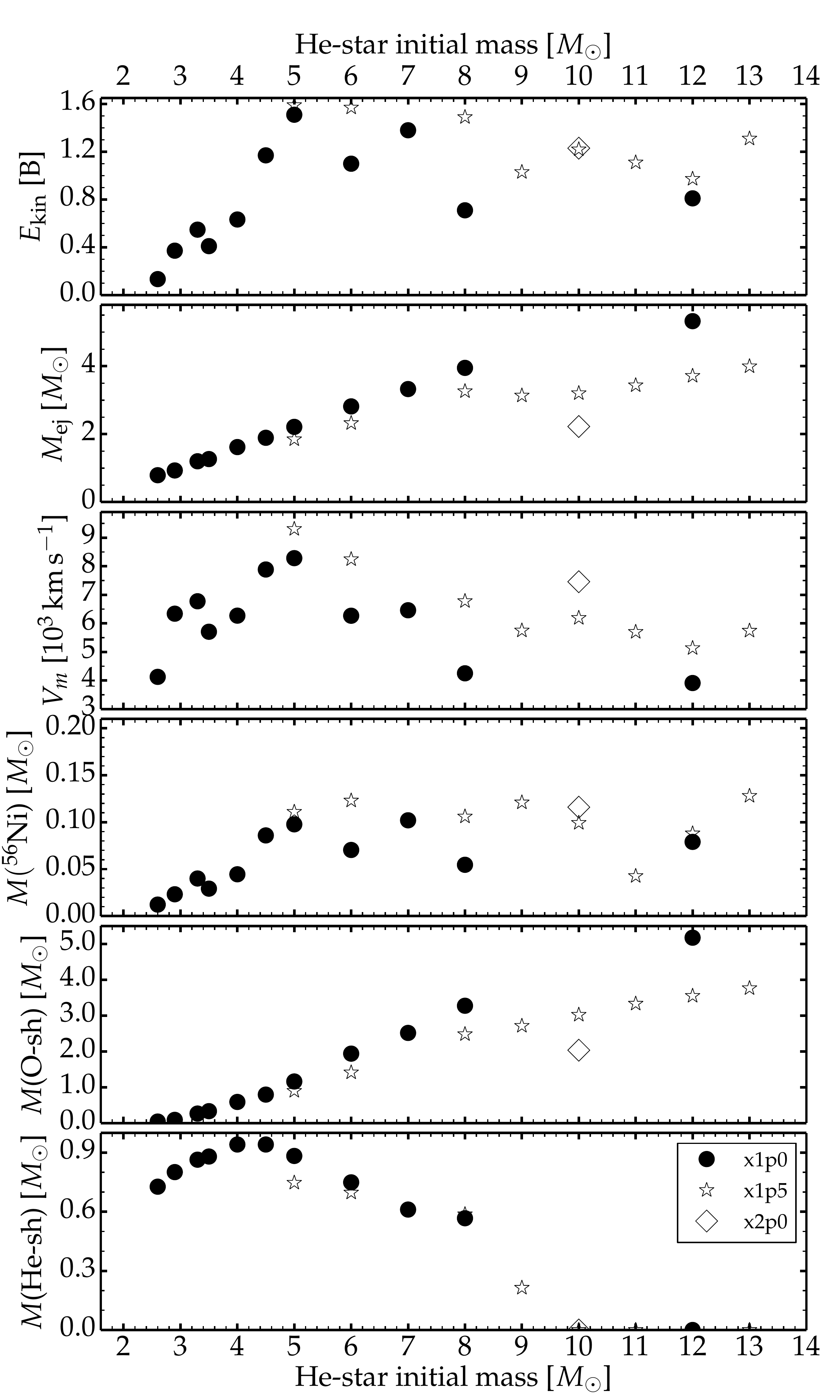}
\caption{Illustration of the ejecta properties for the explosion models based on He stars evolved with the nominal mass loss rate (filled dots), a 50\% enhanced mass loss rate (stars), and a 100\% enhanced mass loss rate (diamond; only model he10p0x2p0 corresponds to this case). From top to bottom, we show  the ejecta kinetic energy $E_{\rm kin}$, the ejecta mass $M_{\rm ej}$, the mean expansion rate $V_{\rm m}$, the \nifs\ mass,  the mass of the O-rich shell (loosely defined by all locations having an O mass fraction greater than 0.1), and the mass of the He-rich shell that accounts for all the mass above the O-rich shell.
\label{fig_prop}
}
\end{figure}

 Fig.~\ref{fig_prop} shows the variation of kinetic energy, ejecta
  mass, mean ejecta expansion rate, \nifs\ mass, O-shell mass, and He-shell mass when the initial
  mass of the He-star is varied. The explosion energy and mass cut
  were obtained for each model using a 1D parametrized
    calculation of neutrino transport as discussed in
  \citet{ertl_ibc_20}. While 1D, the model, which has been
  calibrated against the Crab supernova and SN\,1987A, follows the
  complex interplay between neutrino luminosity from the protoneutron
  star and the infalling material from the progenitor mantle. For the
  lower mass progenitors like model he2p6, calibration to the Crab and
  realistic 10\,\msun\ single star simulations implies a low explosion
  energy of only $\sim$\,10$^{50}$\,erg. However, as the mass of the supernova
  progenitor is increased, the explosion energy increases
  until saturating at a maximum of about $2 \times
  10^{51}$\,erg. Since the binding energy increases monotonically with
  mass, the asymptotic ejecta kinetic energy ultimately reaches a
  maximum of $1.5 \times 10^{51}$\,erg for model he7p0, and then
  decreases for more massive models. The scatter overlaid onto this
  general trend results from the idiosyncrasies of each model.

It should be noted that while results are plotted versus the
  initial He-star mass (which is also contained in the model
  name), the explosion properties and nucleosynthesis are really most
  sensitive to the pre-SN mass. The relation between the two
  depends on the uncertain rate of wind-driven mass loss after
  the envelope is presumably stripped in a binary. Tables and analytic
  expressions relating pre-SN mass to initial He-star mass
  for these models have been given by \citet{woosley_he_19} and
  \citet{ertl_ibc_20}.

For the models evolved with the nominal mass loss rate, the
  ejecta mass ranges from 0.79 to 5.32\,\msun.  He-star models evolved
  with a 50\% boost in the mass loss rate give explosions with an
  ejecta mass 20 to 30\% lower. Model he10p0x2p0 evolved with a 100\%
  boost to the mass loss rate yields an ejecta mass that is an
    additional 30\% lower than its he10p0x1p5 counterpart. These
  reductions in ejecta mass imply a greater expansion rate for the
  same ejecta kinetic energy. They also have implications for
  the SN type since additional wind stripping can lead to the loss of
  the outer He-rich material. Models he8p0x1p5\,$-$\,he9p0x1p5 (i.e.,
  from progenitors evolved with a 50\% boost to the mass loss rate)
  agree better with the transition mass that separates SNe Ib and Ic
  \citep{dessart_snibc_20}. With the nominal mass loss rate, all
  ejecta models retain a He/N shell in their outermost layers, where
  He has a $>$\,90\,\% mass fraction, and may therefore be in tension
  with the observed dichotomy between SNe Ib and SNe Ic (i.e., the
  nominal wind mass loss rate is not as compatible with the inferred
  Ib\,$-$\,Ic transition).

The mean ejecta expansion rate, $V_{\rm m}$, ranges from 4000 to
  9000\,\kms, with some scatter resulting from the scatter in $E_{\rm
    kin}$ and $M_{\rm ej}$. A higher mass loss rate leads to
  progenitors that explode with a higher energy per mass, and
  therefore a higher mean ejecta expansion rate, but the size of the
  changes are far from uniform.  Compare, for example, the model pair
  he12p0/he12p0x1p5 and he8p0/he8p0x1p5 in Fig.~\ref{fig_prop}.

The \nifs\ yield ranges from 0.01 (model he2p6) to
  0.11\,\msun\ (model he8p0) for models with the nominal pre-SN mass
  loss rates. With the 50\% boost in pre-SN mass loss rates, there is
  a systematic shift to higher \nifs\ production, with a maximum of
  0.13\,\msun\ for model he13p0x1p5.

The bottom two panels of Fig.~\ref{fig_prop} show the systematic
  increase (decrease) of the O-rich (He-rich) shell mass with
  increasing initial He-star mass. This variation in O-shell mass is
  qualitatively similar to that seen in Type II SN progenitors (see,
  for example, \citealt{whw02}), while the variation in He-shell mass
  reflects in part the increasing wind stripping for higher He-star
  masses. When the pre-SN mass loss rate is boosted (i.e., for a fixed
  He-star initial mass), the greatest impact is on the O-shell mass
  while the He-shell is weakly affected. However, the composition of
  the He-rich shell also differs. For example, in model he8p0, the
  outer He-rich shell is a 0.5\,\msun\ He/N shell, while in model
  he8p0x1p5, the entire region above the O-rich shell is a He/C shell
  of 0.66\,\msun\ with a sizable O abundance, and N is present only in a thin outer
  layer of 0.15\,\msun. Thus, one needs to interpret this He-shell mass with caution
  since the simplistic criterion for distinguishing one shell from
  another can be ambiguous.

\begin{figure}
\centering
\includegraphics[width=0.99\hsize]{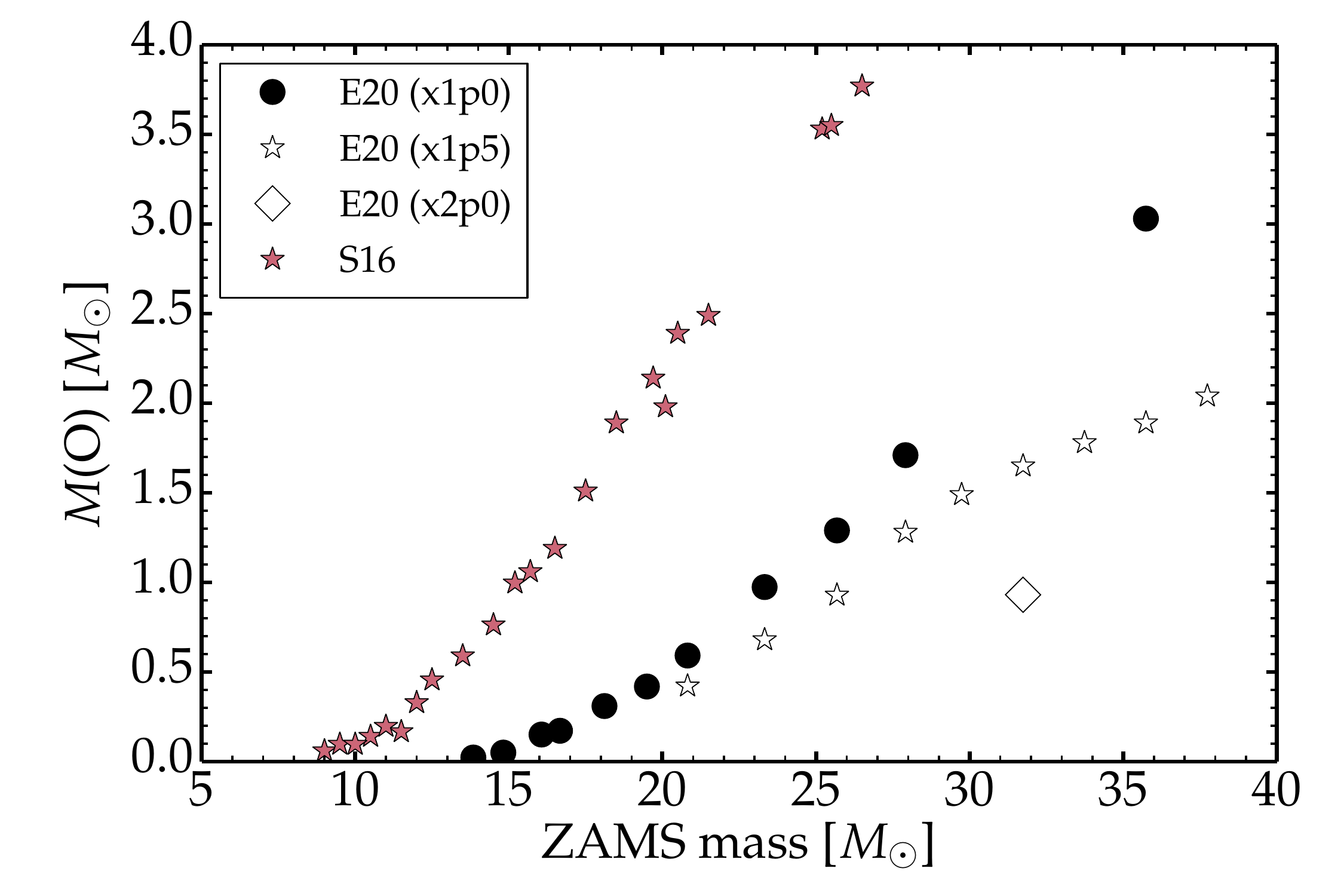}
\includegraphics[width=0.99\hsize]{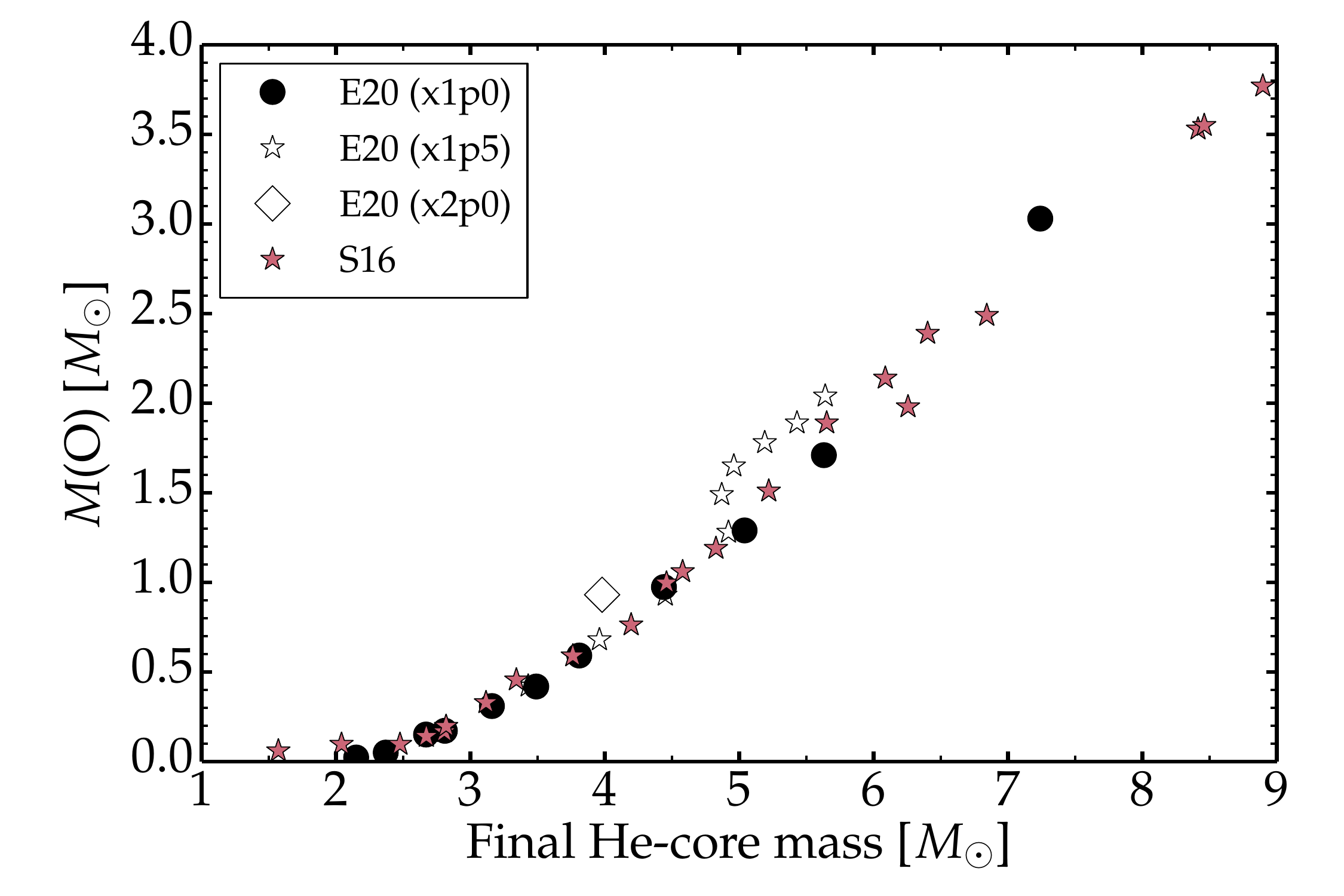}
\caption{Top: Comparison of the O yield for the ejecta arising from
  the He-star progenitor models (which implicitly assume an evolution
  in a binary system; the same set of models as in Fig.~\ref{fig_prop}
  is shown) and for those arising from the single-star models of
  \citet[shown as filled red stars]{sukhbold_ccsn_16}, which all reach
  core collapse with a residual H-rich envelope.  The O yield for a
  given ZAMS mass is much lower for the He-star progenitors of SNe Ibc
  than for the red-supergiant progenitors of SNe II. Bottom: Same
    as top, but now shown versus the pre-SN mass for He-star models and
    the final He-core mass for the single-star models. 
    \label{fig_e20_s16}
}
\end{figure}

An interesting property of He-star models is the moderate O yield
  even for high initial mass. This happens because mass loss by winds
  shrinks the He-star mass even after the envelope has been removed
  while in single stars the He core continues to grow due to hydrogen
  shell burning.  To compare with the O yield of single stars that die
  as red supergiants, we use Eqs.~4 and 5 of \citet{woosley_he_19} to
  estimate the ZAMS mass for all He-star models in this study (see
  also Table~1 in \citealt{ertl_ibc_20}). The top panel of
  Fig.~\ref{fig_e20_s16} shows the O yield versus ZAMS mass for the
  He-star progenitor models of \citet{woosley_he_19} used in this
  study, which produce SNe Ibc, and for the single-star progenitor
  models of \citet{sukhbold_ccsn_16}, which produce SNe
  II-Plateau. Evidently, there is a large offset, which is greater in
  He-star models evolved with a greater mass loss rate. He-star model
  he2p9, which corresponds to a ZAMS mass of about 15\,\msun\ produces
  only 0.05\,\msun\ of O while its single star counterpart of
  15\,\msun\ on the ZAMS produces about 0.76\,$-$\,1.0\,\msun\ (models
  s14.5 and s15.2). Similarly, He-star model he8p0, which corresponds
  to a ZAMS mass of about 28\,\msun, produces 1.71\,\msun\ (nominal
  mass loss rate) or 1.28\,\msun\ \ (50\% enhanced mass loss rate),
  while a single-star of 28\,\msun\ on the ZAMS produces about
  4\,\msun\ of O. This implies that a given O yield inferred from a
  nebular spectrum in a SN Ibc translates into a much larger ZAMS mass
  than when that same O yield is inferred in a SN II (assuming it
  evolved as a single star). This is also a warning for the reader:
  when we discuss a He-star model, say he8p0, that model should not be
  evaluated relative to the properties of a single star that died with
  a He core of 8\,\msun.

To make the comparison clearer and more physically consistent, we
  show in the bottom panel of Fig.~\ref{fig_e20_s16} the same model
  results plotted instead as a function of pre-SN mass (for the He-star
  models) or He-core mass at core collapse (for single stars). 
   That the diverse models now collapse into almost a single line shows that
  the spread in the top panel results mainly from pre-SN
      mass loss and not explosion physics, which is similar for stars
      with the same progenitor mass.
  The mass of oxygen that can be ejected is clearly
  limited by the mass of the ejecta. Plotting this way greatly
  reduces the dispersion between SN Ibc and SN II-P as shown in the
  top panel of Fig.~\ref{fig_e20_s16}.

For clarity and completeness detailed information on the models
  is given in the appendix. In Table \ref{tab_prog}, we provide the
  model composition, the pre-SN mass, the ejecta mass, the explosion
  energy, and the mean expansion velocity and the ejecta velocity that
  bounds 99\% of the \nifs\ mass for each model.  The mass of each
  shell is provided in Table~\ref{tab_prog_bis} while
  Tables~\ref{tab_he2p60}\,$-$\,\,\ref{tab_he10Ax2p0} give the
  elemental composition of all main shells in all progenitor models.

\section{Spectral Properties of the He-star explosions with the nominal pre-SN mass loss rate}
\label{sect_std}

\begin{figure*}
\centering
\includegraphics[width=0.75\hsize]{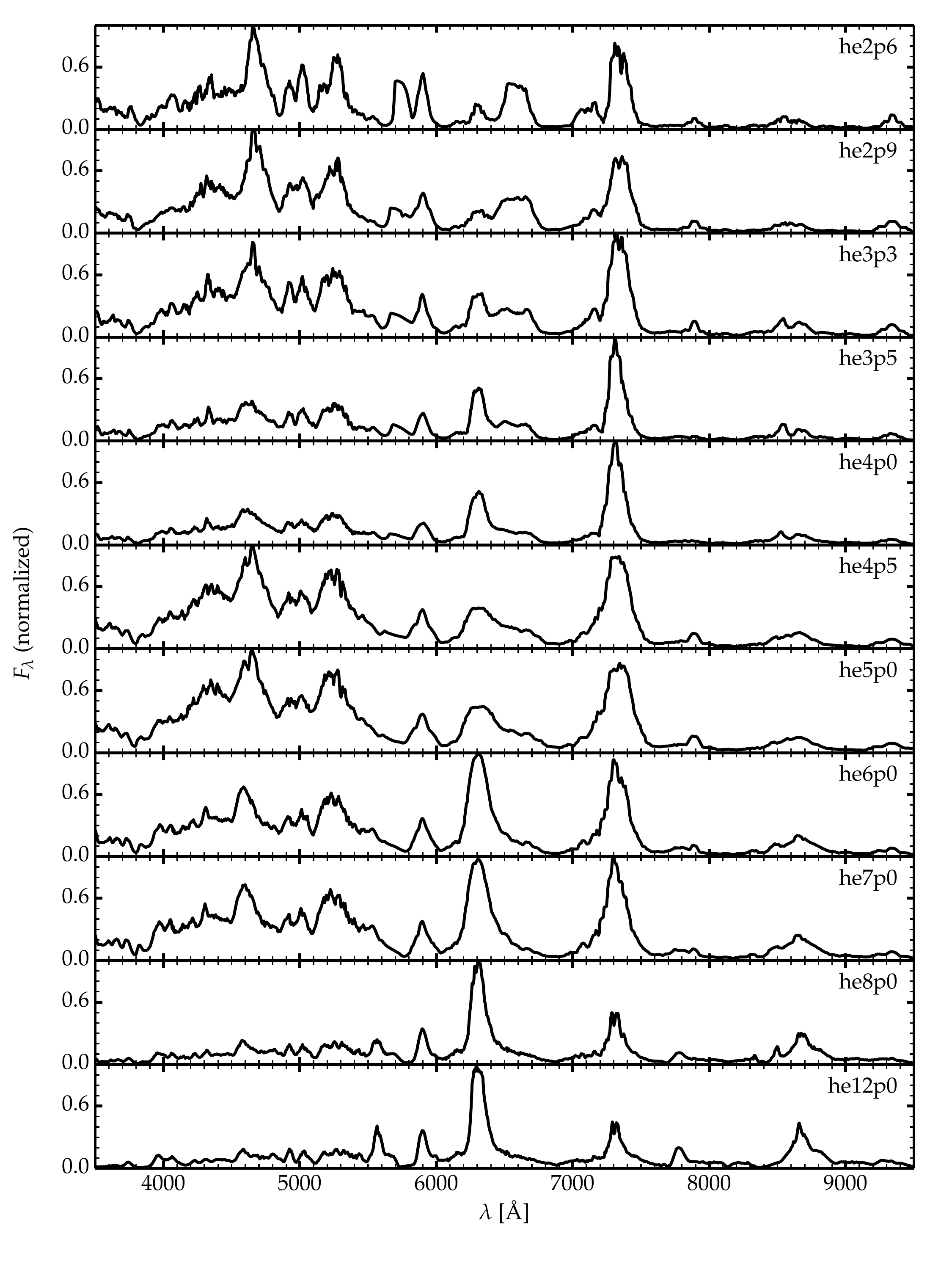}
\vspace{-0.5cm}
\caption{
Montage of spectra for the set of explosion models at 200\,d after explosion (each spectrum is normalized so that the maximum flux is unity). There are strong variations in the iron forest between 4000 and 5500\,\AA, and in the strength of \niidoub, \niiauroral, \oidoub, \caiidoub\ and \caiitrip. The models are based on the He-star progenitors evolved with the nominal mass loss rate  \citep{woosley_he_19,ertl_ibc_20}.
\label{fig_montage_opt}
}
\end{figure*}

\begin{figure*}
\centering
\includegraphics[width=0.75\hsize]{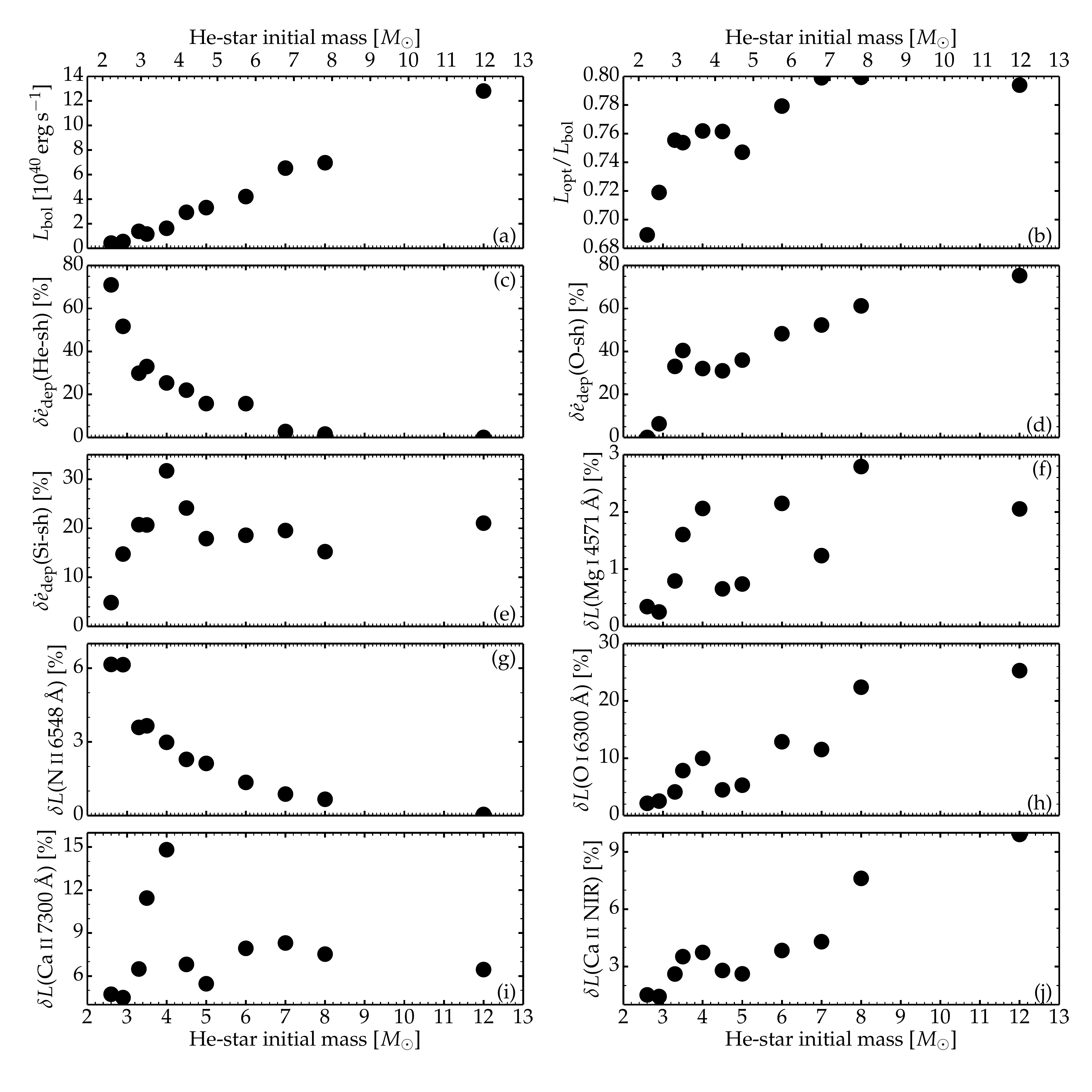}
\vspace{-0.3cm}
\caption{Line fluxes, powers, and their ratios for our radiative-transfer calculations based on the ejecta models arising from He-star progenitors evolved with the nominal mass loss rate \citep{woosley_he_19,ertl_ibc_20}. Quantities $\delta \dot{e}$ are fractional powers relative to the total decay power absorbed in the ejecta, while $\delta L$ corresponds to the fractional luminosity relative to the total optical luminosity. Both are given as a percentage.
\label{fig_line_fluxes}
}
\end{figure*}

Figure~\ref{fig_montage_opt} shows the optical spectra for the whole set of He-star explosions with the nominal  pre-SN mass loss. Particularly evident in the figure is the changing line widths (arising  from differences in the ejecta masses and explosion energies), the varying importance of the Fe forest shortward of 5500\,\AA, and the changing fluxes (and relative fluxes) in the \oidoub\ and \caiidoub\ forbidden lines. Measurements of the various powers and line fluxes are shown in Fig.~\ref{fig_line_fluxes}. Table~\ref{tab_edep} gives the fraction of the total decay power emitted that is absorbed, and the fractional power absorbed by each shell for all models. The photometric properties of the models are provided in Tables~\ref{tab_mag}.

For this model set the bolometric luminosity increases with initial mass (panel (a) of Fig.~\ref{fig_line_fluxes}) because of two factors. The faintness at the low mass end is caused by the low \nifs\ mass (e.g., it is only 0.012\,\msun\ in model he2p6 compared to 0.10\,\msun\ in model he7p0). The other critical factor is the variation in $\gamma$-ray trapping efficiency, which is only 10\,$-$\,30\,\% for models he2p6 to he7p0, and rises to 55\,\% (76\,\%) in model he8p0 (he12p0).

The optical luminosity, here defined as the luminosity falling between 3500 and 9500\,\AA, represents about $78 \pm 2$\,\% of the bolometric luminosity in all models more massive initially than he3p3. In lower mass models, this fraction is lower by up to 10\,\%. This constancy occurs despite the large disparity in optical spectra, especially in the strength of Fe emission in the blue part of the spectrum. At the low mass end, models tend to be bluer, with non-negligible emission down to 3000\,\AA. By contrast, the cooler, more recombined ejecta model he8p0 has little flux shortward of  3500\,\AA.

The eleven models from the ab~initio explosions may be grouped in different ways. We place them into three groups:
\begin{enumerate}
\item
he3p3-like: These have a  broad \niidoub\ which forms in the He-shell, and is only conspicuous in the He-star explosion models he2p6 to he5p0 -- it is present in he6p0 and he7p0, but very weak. These models also have significant emission shortward of 5500\,\AA, mainly due to Fe.  Identifying the \niidoub\ line at nebular times is unambiguous evidence that the progenitor was He rich. Failure to observe that line, however, is not  proof that the progenitor is He deficient (none of our models are He free), since the progenitor may have a He/C shell (with O present at various levels).
\item
he5p0-like: The broad \niidoub\ doublet is not readily identified, and \caiidoub\ line strength is greater than, or similar to, that of  the \oidoub, which is significantly blended with \niiauroral. A significant fraction of the flux is emitted in the Fe forest shortward of 5500\,\AA.
\item
he8p0-like: The \oidoub\ doublet is the strongest spectral feature, and the iron forest has weakened (in a relative sense). The flux in \oidoub\ shows a clear trend of increasing strength with increasing He-star initial mass.  It dominates the he8p0 and he12p0 spectra because of the large O-rich shell mass and the low O ionization of the O-rich material.
\end{enumerate}
In  Sect.~\ref{sect_rep} we examine various properties, including line identification and formation, shell cooling, and ionization effects, for the three representative classes using models he3p3, he5p0, and he8p0.

In contrast to  \oidoub,  \caiidoub\ is not a very useful diagnostic of the progenitor or explosion properties (compare panels (h) and (i) in Fig.~\ref{fig_line_fluxes}). This line forms in the Fe/He and Si/S shells, which are both formed during the explosion and share very similar properties (mass, composition, ionization) across the whole set of models. This line is also badly blended with [Ni\two]\,$\lambda$\,7378 which contributes a significant part of its strength in several models (see Figs.~\ref{fig_fill_all}).

We do find, however,  that the Ca\two\ near-infrared triplet follows a gradual strengthening from model he2p6 to he12p0 (panel (j) in Fig.~\ref{fig_line_fluxes}). This triplet cannot form in the He-rich layers because Ca is Ca$^{2+}$ in these regions. Instead, it forms in the Fe/He, Si/S and O-rich shells. Since the He-rich shell dominates the ejecta mass in lower mass models, the Ca\two\ triplet is weak in these models. As the initial mass is increased,  the triplet can form over an increasing fraction of the ejecta. It is the strongest in models he8p0 and he12p0 because of the lower ionization of the O-rich material, where Ca is a mixture of Ca$^+$ and Ca$^{2+}$. This Ca\two\ near-infrared triplet emission occurs despite the factor of $\sim$\,1000 difference in the Ca mass fraction between the O-rich material and the Fe/He and Si/S shells (see, for example, Table~\ref{tab_he8p00} for model he8p0). As illustrated in Sect.~\ref{sect_obs}, our models have no difficulty in explaining the observed triplet fluxes, and hence triplet emission from the He-shell is not needed.

As readily apparent from Fig.~\ref{fig_montage_opt}, Fe emission extending from around 4000 to 5500\,\AA\ is a characteristic of all the models at 200\,d. Typically, Fe emission arises from all shells (see  Section~\ref{sect_rep}). It is an important coolant for He-rich material because He is a weak coolant, and also because of the relatively high temperature and ionization in the He-rich zones -- in the low mass model Fe can be particularly important because a larger fraction of the power is absorbed in the He shell. Further, because of its rich line spectrum in the UV,  Fe can also reprocess radiation from deeper layers. For example, C\two\ is an important coolant in shells rich in C but, due to Fe absorption, emitted line photons do not escape the ejecta.

Iron can also be an important coolant for the O-rich material owing to the partial ionization of O.
The stark difference in Fe emission between models he4p0 and he4p5, which have very similar yields, is caused by the partial ionization of O (O is nearly O$^+$) in the O-rich shell. This higher ionization results from the greater ejecta expansion rate of model he4p5 relative to he4p0 (this also holds for models he4p5/he5p0 relative to he3p5/he4p0).

%%%%%%%%%%%%%%%%%%%%%%%%%%%%%
%%%%%%%%%%%%%%%%%%%%%%%%%%%%%
%%%%%%%%%%%%%%%%%%%%%%%%%%%%%

\section{Results for representative models for lower, intermediate, and higher He-star masses}
\label{sect_rep}

The progenitor mass is probably the most  important property one wishes to infer from observation, and in this section we discuss its influence on nebular spectra. Nebular spectra are strongly influenced by variations in explosion energy, ejecta mass, chemical yields, and nonLTE processes, and these are all affected by progenitor mass. As most spectral changes occur slowly with mass, and for the
reasons outlined in the previous section, we have chosen three models (he3p3, he5p0, he8p0) to discuss the causes and physics underlying the changes. A brief summary of these models is provided in Table~\ref{model_info}.

\begin{table}
\caption{Salient properties of the reference models discussed in Sect.~\ref{sect_rep}.}
\label{model_info}
\begin{tabular}{l|c|ccc}
\hline
                    &  &    \multicolumn{3}{c}{Model} \\
                    \hline
Quantity  & Unit &  he3p3 &  he5p0 &  he8p0 \\
\hline
$M_{\rm ej}$ & (\msun) & 1.2  &  2.21 & 3.95  \\
$E_{\rm kin}$ & ($10^{51}$\,erg)  & 0.55 & 1.51 & 0.71\\
$V_{\rm m}$ & (\kms) & 6777  & 8286 & 4251 \\
$M$(He) & (\msun)   & 0.84 & 0.97 & 0.84  \\
$M$(O) & (\msun)  &  0.15 & 0.59 & 1.71 \\
$M$(\nifs) & (\msun)  & 0.04  & 0.098  & 0.055  \\
$f$(He-sh) &  & 30\% & 16\% & 2\% \\
$f$(O-sh) & & 32\% &  47\% & 80\% \\
$f$(Si-sh) & & 20\% &  8\% & 6\% \\
$f$(Fe-sh)  & & 19\% & 29\% & 12\% \\
$f_{\rm positron}$  & & 21\% & 21\% & 6\% \\
$f_{\rm abs}$  & & 15\% &  15\% & 55\% \\
$L_{\rm bol}$ &  ($10^{40}$\,\ergs)  & 1.38 & 3.32 & 6.97 \\
$V$  & (mag)  & -11.88 &  $-12.81$ & $-13.56$ \\
\hline
\end{tabular}
\end{table}

\begin{figure}
\centering
\includegraphics[width=0.90\hsize]{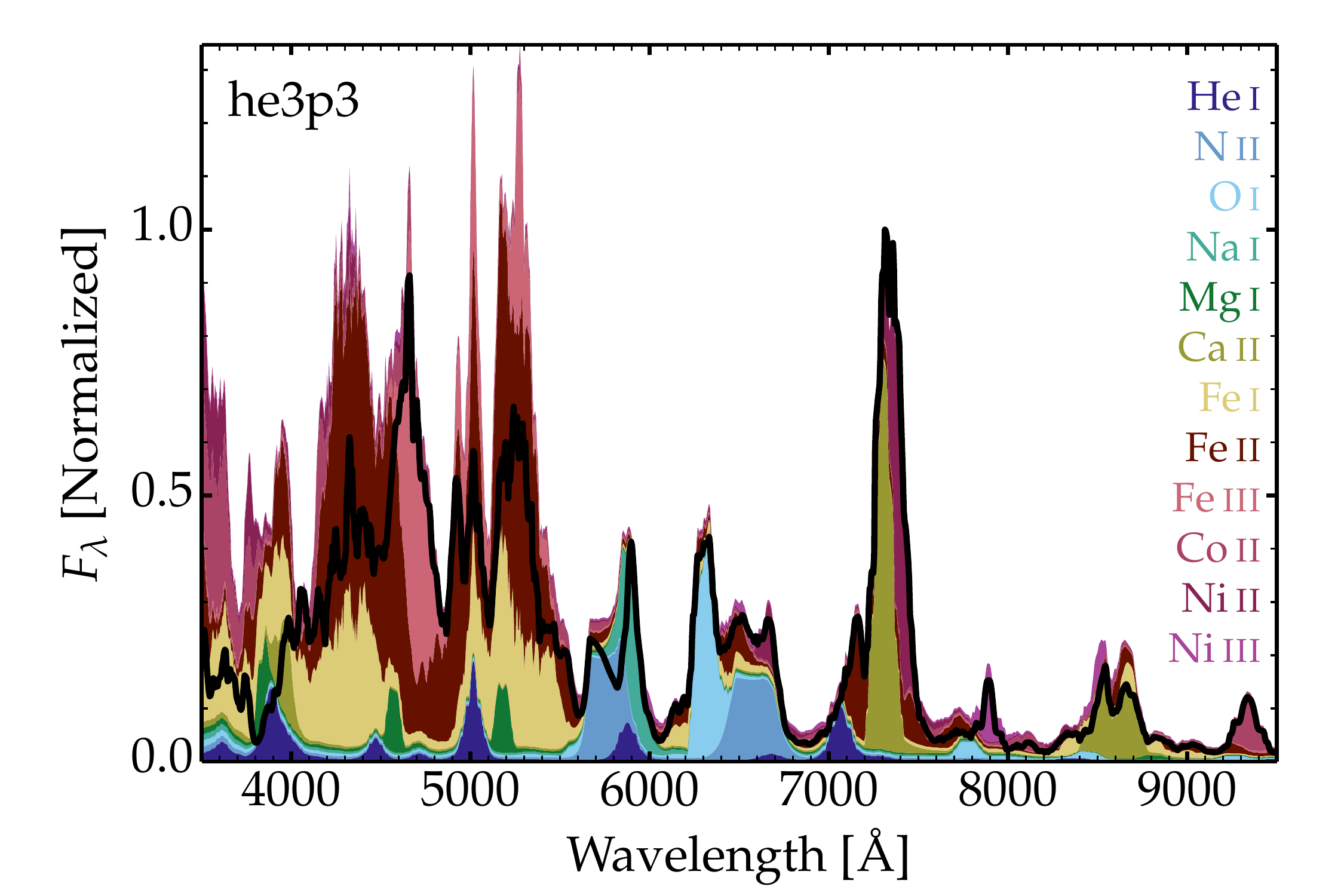}
\includegraphics[width=0.90\hsize]{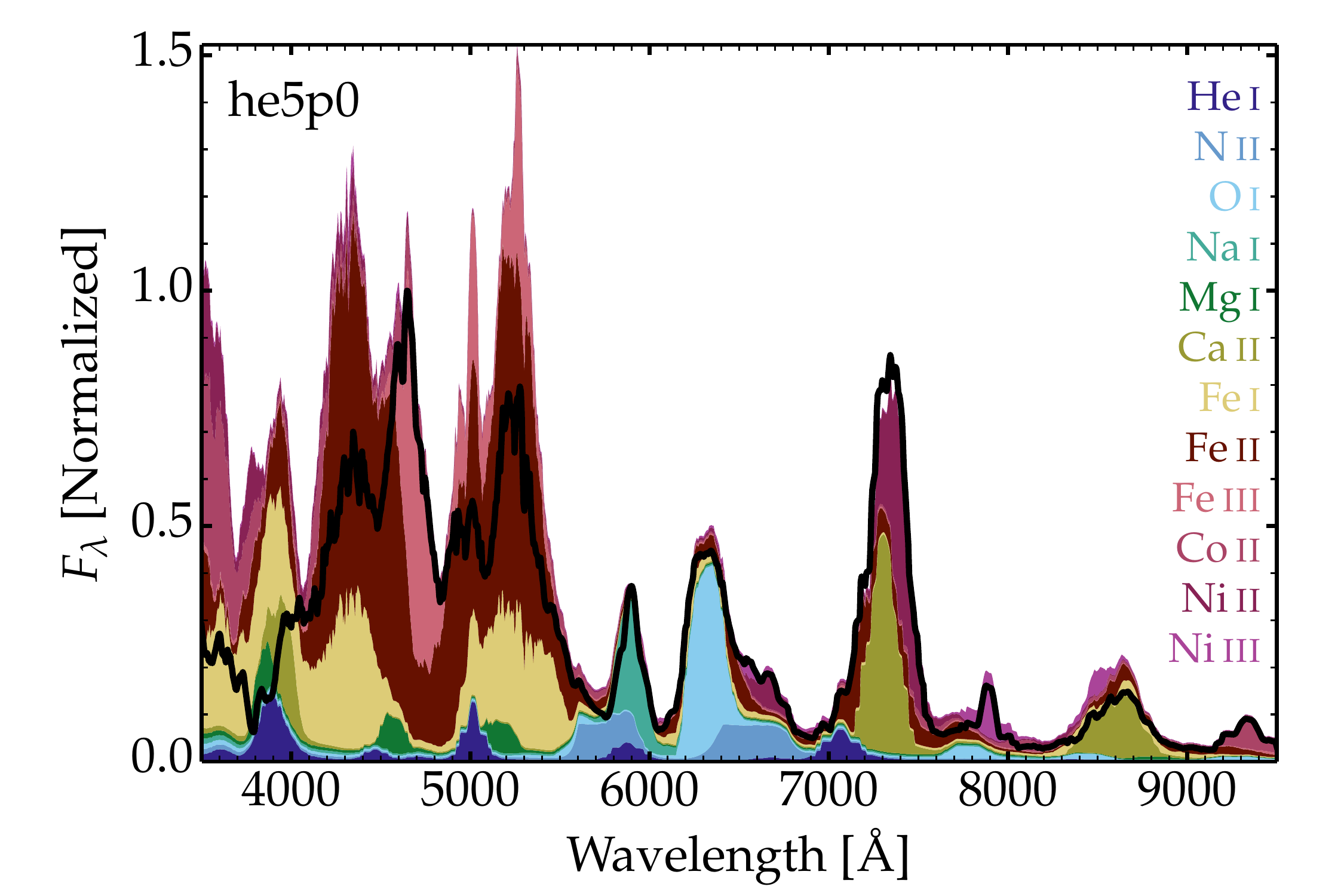}
\includegraphics[width=0.90\hsize]{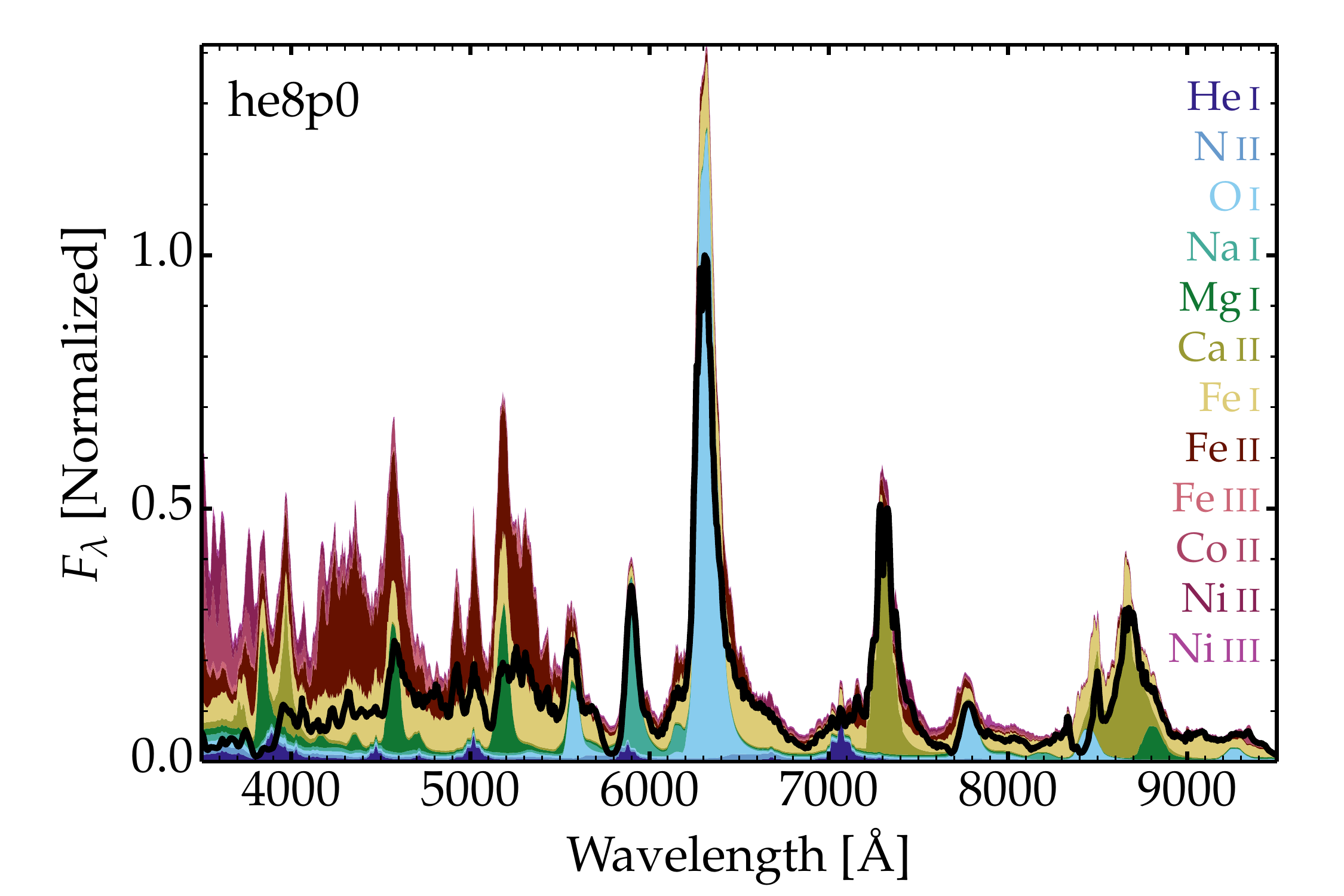}
\caption{Illustration of the flux contribution of individual ions to the total flux (thick black line) for models he3p3, he5p0, and he8p0. The individual contributions account for all bound-bound transitions of a specific ion but ignores the influence of line transitions in other ions. Thus, it accounts only for optical-depth effects intrinsic to that ion. For the plotting, we avoid overlap of these flux contributions by showing the cumulative flux contributions, starting with those due to He\one, then to N\two, etc. This stacking implies that at a given wavelength, the contribution is given by the height separating consecutive colors. This also helps to distinguish similar colors. In numerous spectral regions, the total flux lies below the cumulative flux from the depicted ions, which indicates that photons emitted by one species are being absorbed by another species. \label{fig_fill_all}
}
\end{figure}

\begin{figure}
\centering
\includegraphics[width=\hsize]{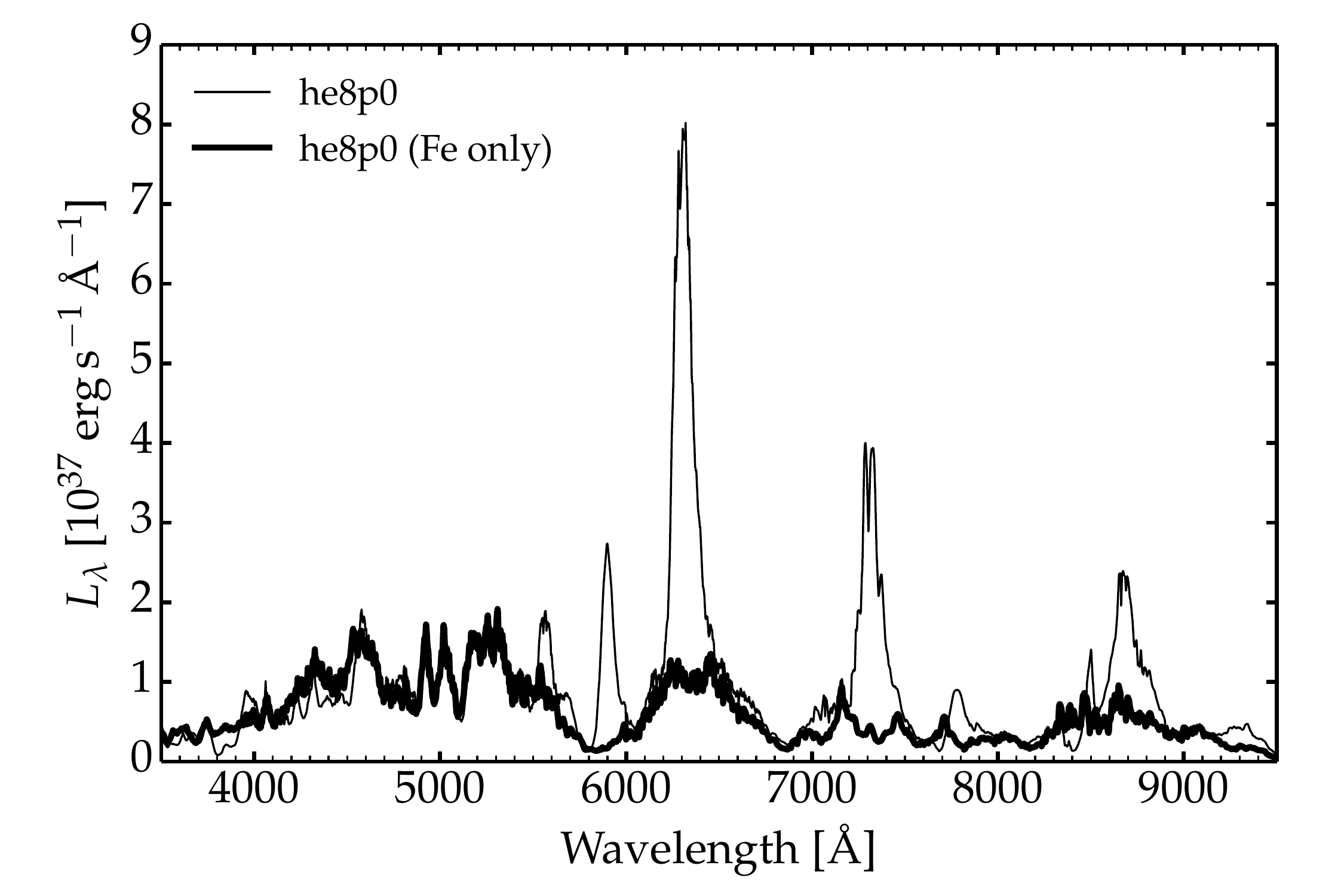}
\caption{Illustration of the contribution from Fe (primarily Fe\one\ and Fe\two; thick line) in the total spectrum (thin line) of model he8p0 at 200\,d after explosion.
 \label{fig_he8_Fe}
}
\end{figure}

At 200\,d, the fraction of $\gamma$-rays absorbed plays a crucial role in setting the luminosity of the ejecta. Of these three models, he8p0 is the most luminous, and this occurs primarily because it has the largest ejecta mass, and consequently it absorbs a larger fraction (55\%) of the decay energy. In the other two models, only 15\% of the decay energy is absorbed. Model he8p0 also has the lowest average expansion velocity, and this gives rise to narrower line widths during the nebular phase.

From  Table~\ref{model_info} we see that a given shell (e.g., the He shell) absorbs different amounts of energy, and this will be reflected in the nebular spectrum. An obvious factor influencing the energy absorbed is the shell mass, but equally important is the location of the shell relative to the \nifs\ (and other shells). In model he3p3, for example, He represents 70\% of the ejecta mass and receives 30\% of the total decay power absorbed. By contrast, in model he8p0, He is 15\% of the ejecta  mass but only accounts for 2\% of the absorbed energy -- in this model most of the energy is absorbed in the O shell.

In our models we assume that the positron energy is deposited locally, and hence in the Fe/He and Si/S shells. Due to $\gamma$-ray escape, positron deposition is more important in lower mass ejecta, and more important at later times. In principle, the increasing importance of the power absorbed in the Fe/He and Si/S shells, relative to other shells, should allow the assumption of local deposition to be tested. Alternatively, one should see a change in the luminosity decline rate as positrons become the dominant energy source. However ionization changes, and their influence on line emission and on the fraction of energy emitted in the optical, will make such a test difficult (see Sect.~\ref{sect_positron}).

Since the mass of O is a strong function of the progenitor mass (see Table~\ref{tab_prog}, Fig.~\ref{fig_prop}, and Fig.~\ref{fig_e20_s16}) it has long been realized that \oidoub\ is potentially an important diagnostic of the progenitor mass. As expected, as the O mass increases, so does the power absorbed in the O shell. In going from model he3p3 to he8p0 the mass of the O shell increases by a factor of 11, while the fraction of power absorbed in the shell increases by a factor of 2.5. More importantly, we see a dramatic change in the ratio of power absorbed in the oxygen shell relative to other shells. For the same models, the ratio  $f$(O-sh)/$f$(Fe-sh) increases by a factor of four while the ratio  $f$(O-sh)/$f$(He-sh)  increases by almost a factor of 40. Thus if  \oidoub\ is the dominant coolant, its strength relative to other features can be used to diagnose the O yield and hence progenitor mass.  Unfortunately, the \oidoub\ is not always the dominant coolant in the O shell (see Tables \ref{tab_cool_he3p3}\,$-$\,\ref{tab_cool_he8p0}), which affects the robustness of this diagnostic for inferring the progenitor mass.

Figure~\ref{fig_fill_all} shows the spectrum (for our three representative models) arising from a given species with each curve corresponding to the species contribution summed with the last curve previously plotted (i.e., a cumulative spectrum). Thus the contribution of each ion as a function of wavelength is given by the height separation between consecutive curves. In the absence of optical depth effects\footnote{Optical depth effects for each species are taken into account when we compute its spectrum. However, bound-bound processes in other species are not taken into account.}, the last curve plotted would correspond to the total flux.  While there are similarities between the three sets of spectra, there are also important differences.

The most striking feature of the illustrated spectra is their complexity below 5500\,\AA. While the emission in this region is primarily due to Fe, three ionization stages (\one, \two, and \three) contribute. He\one, Mg\one\ and Ca\two\ can also make contributions, although only in he8p0 can a feature be identified as having significant contributions from  either Mg\one\ and Ca\two.  Mg\one\ emission becomes more prominent as the SN ages (Sect.~\ref{sect_evol}). As the black line lies below the cumulative spectrum, blanketing effects (i.e., the absorption of radiation by lines and its subsequent emission at other wavelengths) are very important in this spectral region. While blending is less severe longward of 5500\,\AA, it cannot be ignored.

In all three models, the strongest O\one\ line is due to \oidoub, and its prominence increases with increasing ejecta mass. In the two lowest models it is severely contaminated on the red side, primarily by \niidoub\ and Fe emission. Two other O\one\ lines are seen in the plots (\oiauroral\ and \oitrip), but even in model he8p0, the lines have other significant contributors.  This blending limits the usefulness of the \oiauroral\ line as a temperature diagnostic.

[N\two] contributes in the he3p3 and he5p0 models at 5755 and 6548\,$-$\,6583\,\AA. It is
more prevalent in the he3p3 model since a higher fraction of the decay power emitted is absorbed in the He shell. Conversely, [N\two] is essentially absent in the he8p0 model because very little energy is absorbed by the He shell. We identify moderate contributions from He\one\ to the spectrum (e.g., at 5875 and 7065\,\AA) although all are blended.  Similar to N\two, He\one\ features are very weak in the he8p0 model.  The strongest  He\one\ feature is at 10830\,\AA\  (Sect.~\ref{sect_hei}).

The models also predict the presence of \nad\ -- indeed it is a relatively clean spectral feature in all three models. Ca\two\ lines are limited in the optical to \caiidoub\ and the near-infrared triplet. We can also identify [Fe\three]\,$\lambda$\,4658, [Co\two]\,$\lambda$\,9342, [Ni\two]\,$\lambda$\,7378 and [Ni\three]\,$\lambda$\,7889. Fe\one\ is responsible for three broad emission bumps, one of which overlaps with \oidoub\ and mimics the presence of \niidoub.

% \clearpage

As readily apparent from Fig.~\ref{fig_fill_all} the Fe emission, relative to that of \oidoub\ and the \caiidoub\ doublet is much less prominent in model he8p0. We also see that in he8p0, [Fe\one] is responsible for the broad weak components at the base of \oidoub\ and \caiidoub.

In the he8p0 model it seems that most of the Fe\two\ emission is absorbed by Fe\one\ line blanketing since the emergent spectrum closely follows the Fe\one-only spectrum. If we recompute the spectrum by accounting only for Fe\one\ and Fe\two\ bound-bound transitions, we obtain a spectrum that follows closely the total model spectrum except for missing a few strong features associated with O\one, Na\one, and Ca\two\ (Fig.~\ref{fig_he8_Fe}). Furthermore, this Fe-only spectrum is very different from the sum of the individual contributions from Fe\one\ and Fe\two\ (Fig.~\ref{fig_fill_all}), which indicates optical depth effects. Some of the blanketed radiation in the blue escapes beyond 6000\,\AA.

In addition to the amount of energy absorbed, two other key factors influencing the line emission from a  given shell are its composition, and its ionization state.  As  an example we primarily discuss the he3p3 model. Similar discussions can be provided for the other models although the details differ. They change because the densities of the shells are different,  and because of the complex and nonlinear processes that depend on composition, heating rate, and  temperature. The dominant coolants, and how they change, is also highly dependent on the atomic properties.

\begin{figure*}
\centering
\includegraphics[width=\hsize]{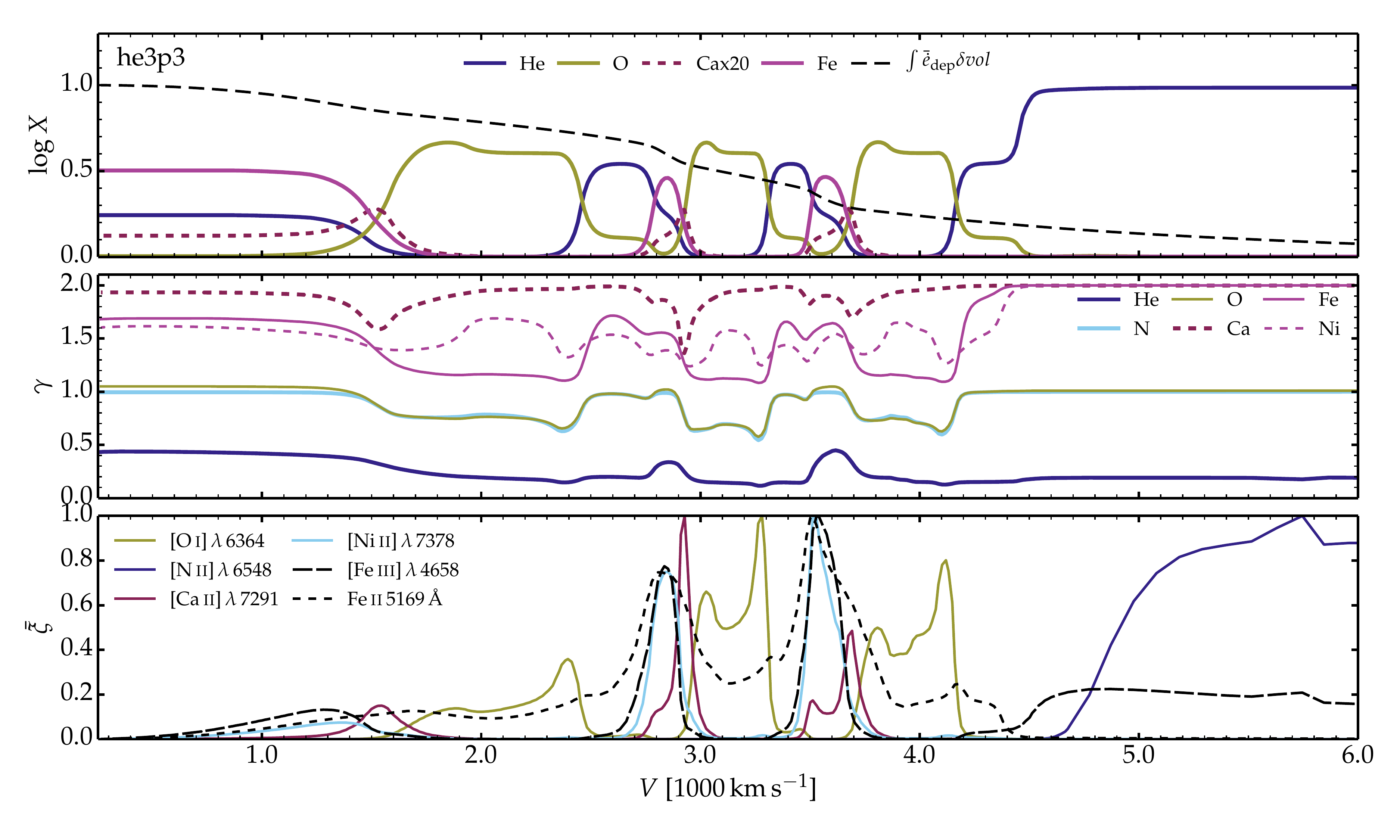}
\includegraphics[width=\hsize]{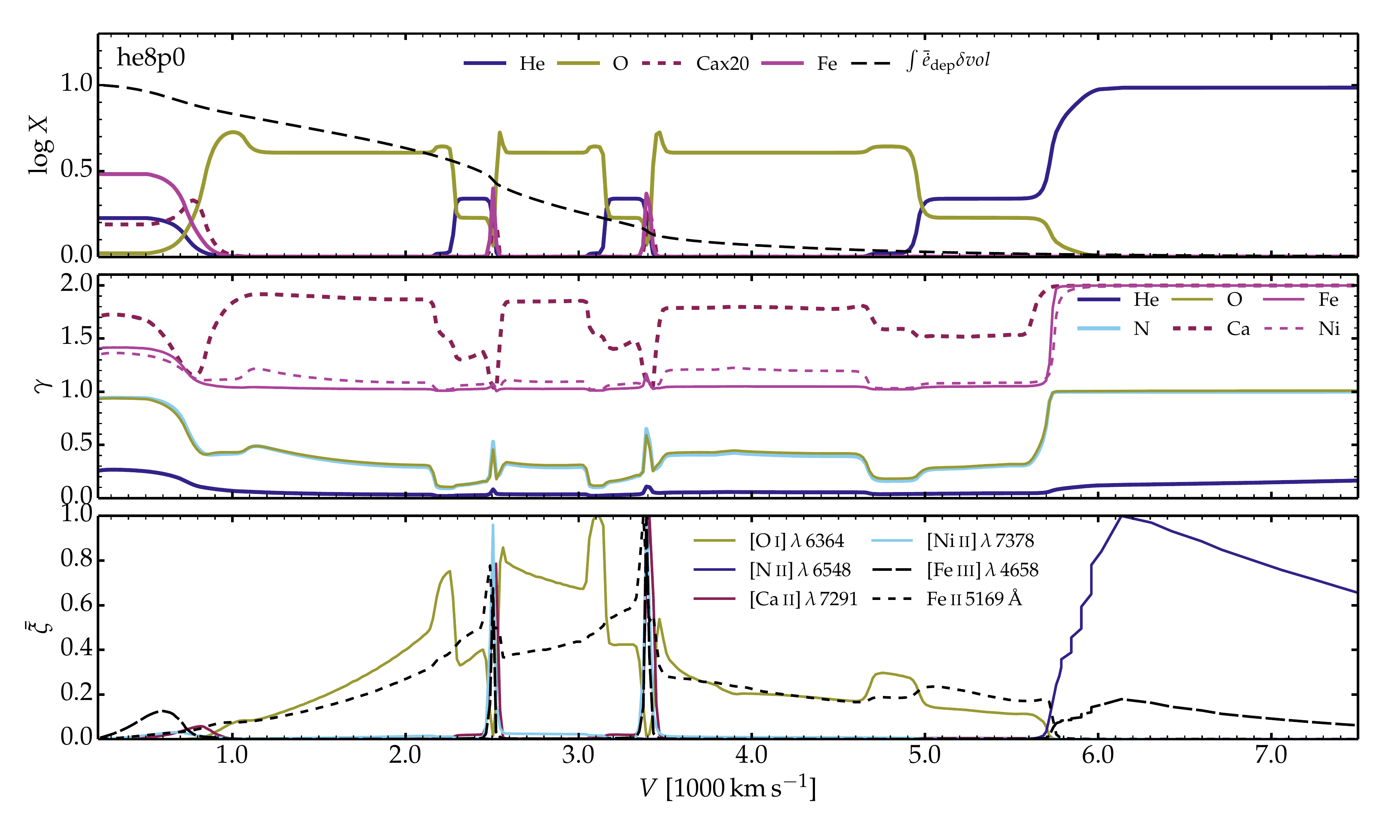}
\vspace{-0.5cm}
\caption{Illustration of ejecta and radiation properties for the shuffled-shell models he3p3 and he8p0. From the upper to the lower panels, we show the profiles versus velocity for the mass fraction of He, O, Ca, and Fe (including the volume integral of the decay power absorbed normalized to unity; dashed line), their ionization state (zero for neutral, one for once ionized, etc.; we also include N and Ni), and the formation regions for [O\,{\sc i}]\,$\lambda$\,6364, [N\,{\sc ii}]\,$\lambda$\,6548, [Ca\,{\sc ii}]\,$\lambda$\,7291, [Ni\,{\sc ii}]\,$\lambda$\,7378, [Fe\,{\sc iii}]\,$\lambda$\,4658, and Fe\,{\sc ii}\,$\lambda$\,5169. The latter shows how the emission varies with velocity in a selection of lines. For better visibility, this emission $\xi$ is normalized so that its maximum is set to unity (and the quantity is then named $\bar \xi$). Quantitatively, the line equivalent width $EW$ is related to $\xi$ through the relation $EW =  \int \xi \, d\log V $.
}
\label{fig_ew_he3p3}
\end{figure*}

Figure~\ref{fig_ew_he3p3} illustrates how the ionization state and line emission (for a few selected lines) vary with velocity, and how these are connected to composition.  Throughout the outer regions (He/N and He/C shells), He is neutral, N is once ionized, and Fe is twice ionized. In the O-rich regions, O is primarily O$^+$ (with some O$^{0+}$) while Fe is primarily Fe$^+$ (with some Fe$^{2+}$). In Fe-rich regions, Fe (and similarly for Ni) is a mix of Fe$^+$ and Fe$^{2+}$ (Ni$^+$ and Ni$^{2+}$). Throughout the ejecta, Ca is mostly Ca$^{2+}$ apart from the Si/S shell locations where it is a mix of Ca$^+$ and Ca$^{2+}$.

The high ionization in the He-rich shell (i.e. He/N and He/C) explains the strong N\two\ and Fe\three\ cooling and the negligible contribution from  Ca\two. Indeed, [N\two]\,$\lambda$\,6548 and [Fe\three]\,$\lambda$\,4658 form in those regions. The latter forms also in the Fe/He shell, together with [Ni\two]\,$\lambda$\,7378. Because Ca$^+$ is only present in the Si/S region, [Ca\two]\,$\lambda$\,7291 only forms in that region. We find that [O\one]\,$\lambda$\,6364 forms throughout the O-rich shell, despite the partial ionization of O. Finally, Fe\two\,$\lambda$\,5169 forms throughout the Fe/He, Si/S, and O-rich shells where Fe$^+$ is abundant, but not in the outer He-rich shells where Fe$^{2+}$ dominates. Hence, Fe\two\ cooling operates in all metal-rich regions (i.e., non He-rich) and is therefore not tied exclusively to the \fefs\ produced by the decay of \nifs\ and \cofs.

\begin{figure*}
\centering
\includegraphics[width=\hsize]{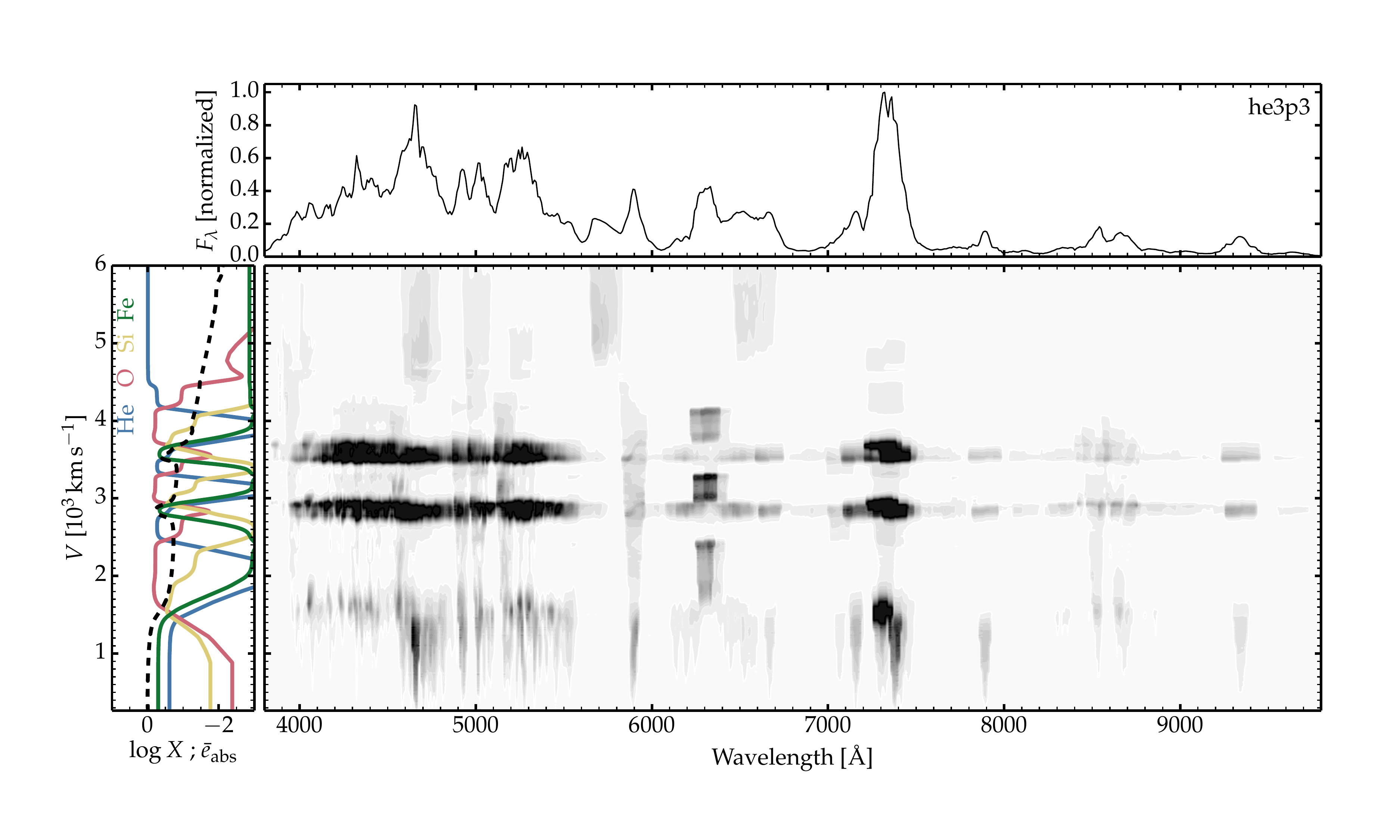}
\vspace{-1cm}
\caption{Illustration of the spatial regions (here shown in velocity space) contributing to the emergent flux in the he3p3 model. The grayscale image shows the observer's frame flux contribution $\partial F_{\lambda,V} / \partial V$ (the map maximum is saturated at 20\,\% of the true maximum to bias against the strong emission lines and better reveal the origin of the weaker emission) versus wavelength and ejecta velocity.  The panel at left connects these emission contributions to ejecta shells and the decay-power deposition profile (dashed line) on the same velocity scale. The upper panel shows the corresponding (scaled) flux $F_{\lambda}$ integrated over all ejecta velocities.
\label{fig_dfr_he3p3}
}
\end{figure*}

The origin of the observed line emission for model he3p3 is shown in Fig.~\ref{fig_dfr_he3p3}. As the
plot indicates the last interaction point of an escaping photon, the plots are affected by scattering
(due to electron scattering and line scattering) and as such do not show the original emission site of all escaping photons. From the figure it is apparent  that only a few lines form in the O-rich shell (primarily \nad\ and \oidoub) and the He-rich shell (primarily \niidoub\ and [Fe\three]\,$\lambda$\,4658), while many lines from Fe, Co, and Ni arise from the Fe/He and Si/S shell. An additional strong line from the Si/S shell is the \caiidoub\ doublet, which also forms at a low level in the O-rich shell and in the He/C shell.

To further understand the origin of the line emission we selected one location in each of the main shells identified in the unmixed ejecta, namely the He/N, He/C, He/C/O (similar to the He/C shell but with some O), O/C, O/Ne/Mg, O/Si, Si/S, and Fe/He shells. At these locations we determined the five main cooling processes that balance decay heating, and these are provided for the three models in Tables~\ref{tab_cool_he3p3}, \ref{tab_cool_he5p0}, and \ref{tab_cool_he8p0}.  Because the composition (other factors also intervene) varies drastically between shells, the coolants are shell-dependent. The coolants are also model dependent -- primarily because of
density and ionization differences. The common theme though is that the cooling processes are nearly exclusively collisional excitation (followed by radiative de-excitation) and nonthermal excitation (followed by photoionization and recombination, or followed by radiative de-excitation).

The tables in the Appendix provide the dominant coolants only at one epoch -- in general the dominant coolant will be epoch-dependent. We further stress that the coolants can be very dependent on impurity species. As extensively discussed by \cite{DH20_neb} (see also \citealt{fransson_chevalier_89} and \citealt{D21_sn2p_neb}), a relatively small amount of Ca mixed into O-rich regions can dramatically weaken the cooling by \oidoub.

Rather than discuss each shell separately we provide only a very broad, and generalized, overview below. The  tables  provided in the appendix should be consulted for greater detail. In all three models, the He/N shell, with a representative temperature of  $\sim$\,10000\,K, cools through N\two\ collisional processes, He\one\ nonthermal excitation, or Fe\three\ collisional processes. N and Fe are once and twice ionized, respectively (see Fig.~\ref{fig_ew_he3p3}). The He/C shell, with a representative temperature of  $\sim$\,6500\,K, cools through C\two\ collisional excitation as well as He\one\ nonthermal excitation. The He/C/O shell is similar, with an additional contribution from Mg\two\ collisional excitation.

In the O/C, O/Ne/Mg, and O/Si shells,  Mg\two\ collisional excitation  tends to be the dominant coolant with O\one\ collisional excitation ranking second. The O-rich shells are cooler than the He-rich shells, with a temperature of $\sim$\,5000\,K in he3p3, and  $\sim$\,4700\,K in he8p0.  Because of the lower ionization of O in he8p0,  O\one\ is the dominant coolant in the O/C and O/Ne/Mg shells.

The Si/S and the Fe/He shells cool primarily through Fe\two, Ca\two, Co\two, and Ni\two\ collisional excitation, with the dominant process being shell- and model-dependent. For example, Fe\two\ is the dominant coolant for he3p3 while Ca\two\ is dominant for he8p0. The stronger coolants in these shells lead to  temperatures less than 4000\,K. The similarity of the coolants in these last two shells arises from their similar composition, mostly because the composition mixture in both results from explosive nucleosynthesis (the Fe/He is more biased toward Fe group elements, but Ca is abundant in the Fe/He shell and Fe is abundant in the Si/S shell).

\section{Salient differences between nebular spectra of Type II and Type Ibc supernovae}
\label{sect_ii_ib}

Since this study on SNe Ibc follows a recent study on SNe II (with a very similar modeling approach for the pre-SN evolution, the explosion phase, and the radiative transfer at nebular epochs), it is instructive to compare the ejecta and radiative properties resulting from a red-supergiant star explosion and an He-star explosion. For this comparison, we select the Type II SN model s15p2 at 350\,d \citep{D21_sn2p_neb} and the Type Ib model he6p0 at 200\,d. The models have a similar $E_{\rm kin}$, \nifs\ mass and metal yields. However, they have very different ejecta masses because the s15p2 progenitor has a massive H-rich shell while the he6p0 model arises from a H-free, He-star progenitor. Specifically, models s15p2 (he6p0) are characterized by $E_{\rm kin} = 0.8 \times 10^{51}\,(1.1 \times 10^{51})$\,erg, $M_{\rm ej} = 10.95\,(2.82)$\,\msun, $M({\rm H}) = 5.24 \,(0.0)$\,\msun, $M({\rm O}) = 1.0\,(0.97)$\,\msun, and  $M(^{56}{\rm Ni}) = 0.063\, (0.070)$\,\msun. Despite the similar O yield, the He-star model corresponds to a ZAMS mass of 23.3\,\msun,  much greater than  the SN II ZAMS mass of 15.2\,\msun\ (see also discussion in Sect.~\ref{sect_gen_prop} and Fig.~\ref{fig_e20_s16}). This offset in ZAMS mass for the same O yield has important implications, for example when the explosion sites of SNe II and Ibc are compared and the SN progenitor masses are inferred.

\begin{figure*}
\centering
\includegraphics[width=0.99\hsize]{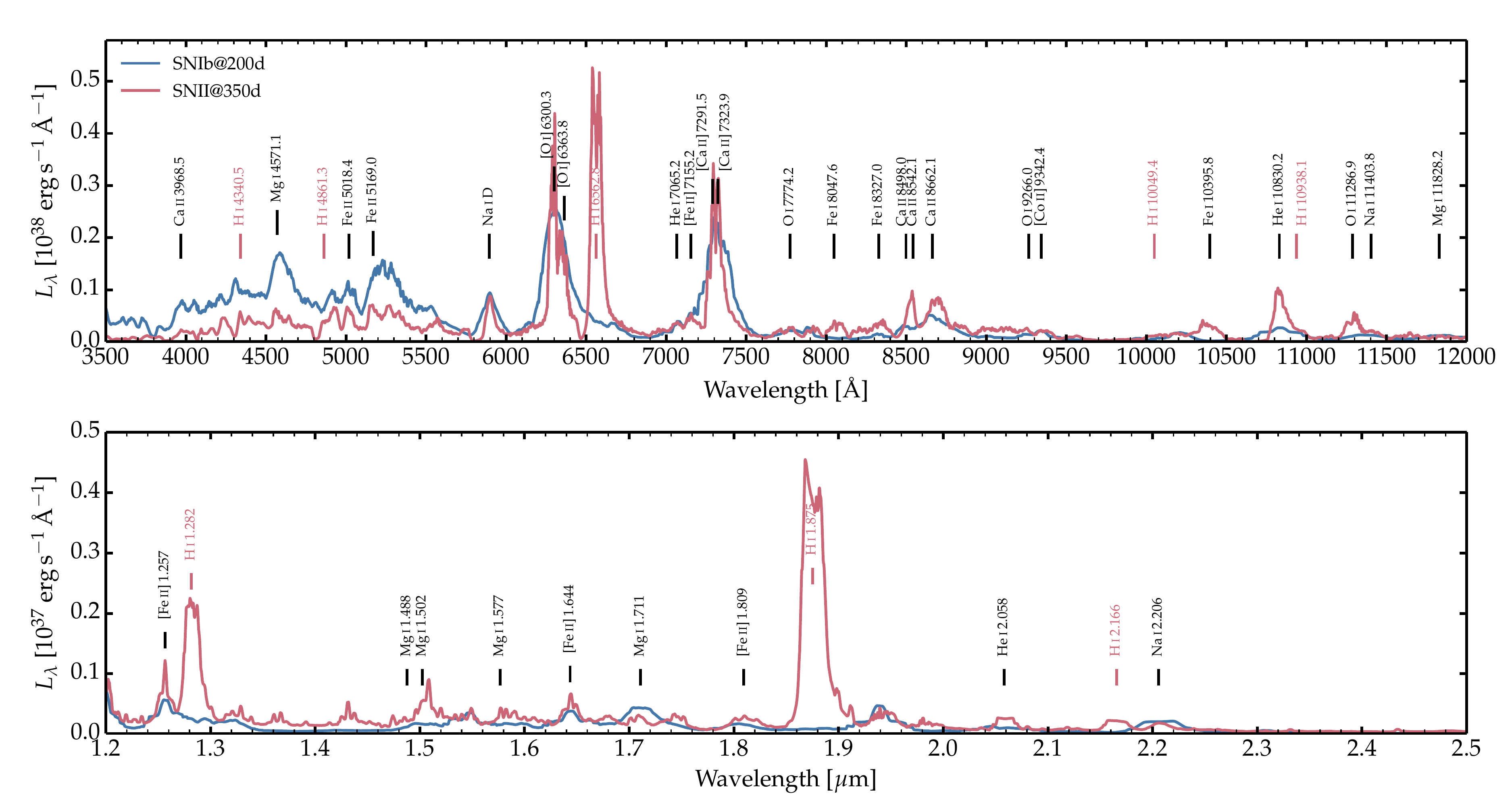}
\caption{Comparison of optical and near-infrared spectra of the Type II SN model s15p2 at 350\,d \citep{D21_sn2p_neb} and the Type Ib model he6p0 at 200\,d (present work). Some line identifications are indicated but should be interpreted with caution since apart from the strongest features (which are often line multiplets), most spectral features are blends of many transitions. The strongest Balmer and Paschen lines are present only in the SN II model and are labeled in red.
\label{fig_comp_ii_ib}
}
\end{figure*}

Having the same $E_{\rm kin}$ within 25\% but a factor of about four difference in $M_{\rm ej}$ means that the SN Ib ejecta expands twice as fast as the SN II counterpart. Although both models have a similar \nifs, the SN Ib model is much fainter at nebular epochs because of the enhanced $\gamma$-ray escape. To compensate for this in the spectral comparison we choose an age of 350\,d  for the Type II SN (12\% of $\gamma$-rays escape; total power absorbed of $3.5 \times 10^{40}$\,\ergs) and 200\,d for the Ibc (74\% of $\gamma$-rays escape; total power absorbed of $4.2 \times 10^{40}$\,\ergs).

 Apart from the H-lines emission in the Type II spectrum, the spectra are quite similar (Fig.~\ref{fig_comp_ii_ib}). In this comparison the Fe emission from 4000 to 5500\,\AA\ is stronger in the SN Ib but this, in part, is caused by the adopted comparison epochs. Because we chose epochs when the absorbed decay power is the same more energy is absorbed in the Fe/C/He layers of the SN Ib model (since, it does not have a H layer) than in the same layers of the SN II model. The  [O\one] or [Ca\two] are also stronger in the SN Ib. This is more difficult to see in Fig.~\ref{fig_comp_ii_ib} because the O\one\ and Ca\two\ profiles are broader.

The flux ratio of the \oidoub\ and \caiidoub\ doublets is comparable in both models, which may be coincidental, although both models have a similar O-rich shell mass (set by pre-SN composition) and a similar Si-rich shell mass (set by explosive nucleosynthesis). \nad\ is present in both models, but it is much broader in the faster expanding SN Ib model.

This illustration helps to see that SN Ib (an by extension SN Ic) and SN II nebular spectra share many similarities, and the physics controlling nebular-phase spectra applies to both. We thus anticipate that the main differences between SNe Ibc and SNe II, apart from the presence/absence of H\,\one\ lines, will primarily be the result of the faster ejecta expansion in Ibc's which leads to lower densities, enhanced $\gamma$-ray escape, and broader spectral lines. In Type II SNe,  the radiation emitted by deeper, metal-rich shells is also strongly reprocessed by the outer H-rich layers (see for example Sect.~4 of \citealt{DH20_neb}) -- this reprocessing in the outer ejecta (for example in the He-rich shells) is either weak or absent in SNe Ibc.

%%%%%%%%%%%%%%%

\section{Sensitivity to the ejecta density structure}
\label{sect_den}

 In this section, we perform exploratory calculations to assess the sensitivity of our results to variations in ejecta density, either caused by clumping or modulations in ejecta kinetic energy. We either assume that clumping is uniform throughout the ejecta  (Sect.~\ref{sect_fvol}) or present exclusively within the O-rich shell (Sect.~\ref{sect_O_fvol}). We also check the impact of variations in explosion energy in otherwise smooth ejecta (Sect.~\ref{sect_ekin}). To go beyond such explorations will require using more realistic 3D explosion calculations as initial conditions for the radiative-transfer calculations (e.g., those of \citealt{gabler_3dsn_21}).

\subsection{Uniform clumping throughout the ejecta}
    \label{sect_fvol}

 Following the method of \citet{d18_fcl} and \citet{D21_sn2p_neb}, we test the impact of clumping on the predicted radiation and gas properties of He-star explosion models. Starting from a converged model with a smooth density structure, we recalculate the radiative transfer solution by adopting a uniform volume filling factor for the material of 50\,\% (corresponding to a density compression of two) or 20\,\%  (a density compression of five).

\begin{figure*}
\centering
\includegraphics[width=0.75\hsize]{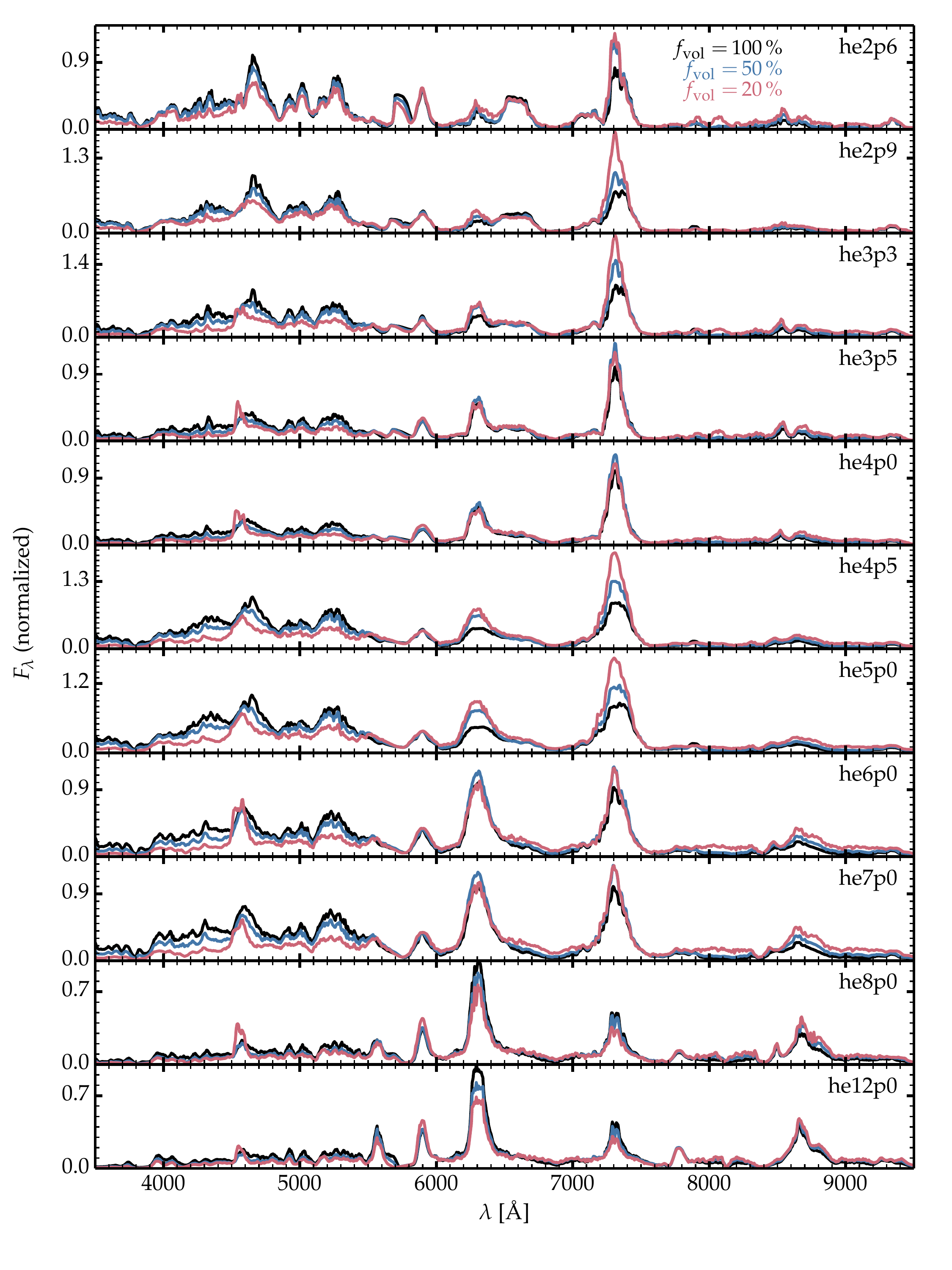}
\vspace{-0.5cm}
\caption{We show the same, unclumped models as in Fig.~\ref{fig_montage_opt} (black), and overplot the clumped counterparts that use a  50\,\% (blue) and 20\,\% (red) volume filling factor. Each triad of models uses the same normalization, chosen so that the first model in the triad (shown in black) has a maximum flux of unity within the wavelength range depicted. [See discussion in Section~\ref{sect_fvol}.]
\label{fig_montage_fvol}
}
\end{figure*}

       Figure~\ref{fig_montage_fvol} is a duplication of Fig.~\ref{fig_montage_opt} but now overplotting the two clumped counterparts for each model between he2p6 and he12p0. In all cases, the introduction of this uniform clumping leads to a reduction of the Fe line emission below about 5500\,\AA. The reduction is greatest in models that were originally (i.e., without clumping) more ionized (for example, model he5p0 relative to he8p0). In the low ionization models he8p0 and he12p0, clumping has the effect of boosting the \nad\ and the \mgi\ lines, partially at the expense of \oidoub. This shift in line strength is driven in part by the reduction of C\two\ and Mg\two\ cooling, which dominate in the O-rich zones in the smooth ejecta models (see Tables~\ref{tab_cool_he3p3} to \ref{tab_cool_he8fvol0p2}). The ionization potentials of Na and Mg are lower than that of O, and in models he8p0 and he12p0 they remain singly ionized even when O is dominantly neutral.\footnote{The ionization energies of Na, Mg, and O are 5.14\,eV, 7.65\,eV, and 13.62\,eV respectively.} Thus the boosts in density, which enhances recombination, boosts their neutral abundance. In the other, lighter models, enhanced clumping generally leads to a strengthening of \oidoub\ as well as \caiidoub.

      Since our treatment of clumping does not alter the total decay power absorbed nor its spatial distribution, the total emergent flux is preserved but its distribution in wavelength is altered. Although the decay power absorbed in a given shell sets the limit for the maximum flux in a line, the relative strengths of nebular lines emitted from a given shell is dependent on the ionization state and temperature. Density-squared processes are boosted with the introduction of clumping, leading to enhanced recombination, and hence a decrease in the ionization. This can favor the appearance of collisionally excited lines  such as \mgi\ while weakening others (e.g, those due to Fe\three).  Clumping will have the strongest influence on emitted line fluxes in those shells where a species contributing to the cooling is in a mixed ionization state (e.g., when N(O$^{0+}$)/N(O$^{1+}$) $\sim 1$).

\begin{figure}
\centering
\includegraphics[width=0.95\hsize]{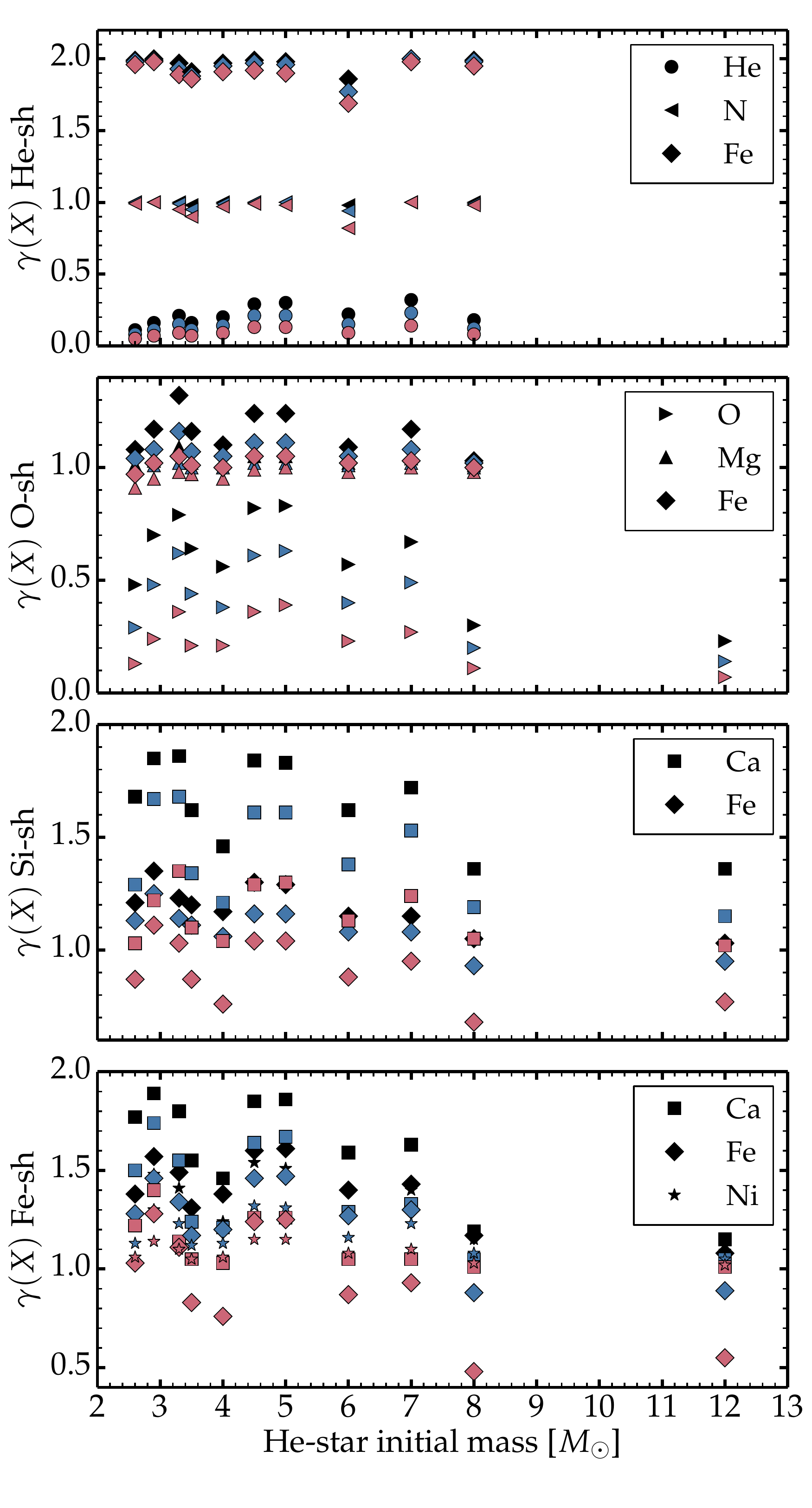}
\vspace{-0.4cm}
\caption{Illustration of the ionization of important elements in the He-rich, O-rich, Si-rich, and Fe-rich shells for the models shown in Fig.~\ref{fig_montage_fvol} (the same color coding is used to differentiate the volume filling factors: black for 100\,\%, blue for 50\%, and red for 20\%). Different symbols correspond to different elements. The smaller the volume filling factor, the smaller the species ionization. In some cases, clumping causes only a marginal reduction (as for the He shell), but in others the effect is strong (for example, for O in the O-rich shell).
\label{fig_shell_ionization_fvol}
}
\end{figure}

Figure~\ref{fig_shell_ionization_fvol} illustrates the ionization stage of the main coolants in the He-rich, O-rich, Si-rich, and Fe-rich shells for the smooth density case and for the two clumped cases. As is readily apparent, in the O-rich shell, clumping shifts the dominant ionization stage from O$^{1+}$ to O$^{0+}$, potentially boosting the importance of \oidoub\ as a coolant. We see similar ionization shifts in the Si-rich and Fe-rich shells. In both these shells,  for example,  the dominant ionization state of Ca shifts from Ca$^{2+}$ to Ca$^+$, and this will, in general, boost the emission in the Ca\two\ lines. However the boost is very model-dependent -- the increase in the \caiidoub\ flux is small in he4p0, and much larger in he4p5 and he5p0. In model he4p0, unlike models he4p5 and he5p0, Ca$^+$ was already (or very close to being) the dominant  ionization stage in the smooth ejecta.

\subsection{Uniform clumping limited to O-rich material}
    \label{sect_O_fvol}

A uniform clumping throughout the ejecta is not expected on theoretical grounds. In the outer regions, unless the ejecta interacts with some dense material (which occurs in interacting SNe), there should not be any physical process to compress the gas. So, clumping in the outer ejecta is unlikely. In addition, clumping of the Fe/He and Si/S shells may not be justified because these regions should be heated by radioactive decay (more so for the Fe/He shell). This is the so-called bubble effect, which should lead to a rarefaction of that material \citep{woosley_87A_late_88,herant_87A_2D_92,basko_56ni_94}. Hence, one would expect the clumping in SNe Ibc to be present, if at all, in the O-rich shell.

We therefore repeated the radiative transfer calculations based on the explosion of He-star progenitors evolved with the nominal pre-SN mass loss rate (omitting he8p0 and he12p0, which hardly changed with the introduction of a uniform clumping; see previous section) but with clumping limited to regions rich in O (i.e., all ejecta locations with an O mass fraction greater than 0.3 and an He mass fraction less than 0.1).

% \clearpage

\begin{figure*}
\centering
\includegraphics[width=0.75\hsize]{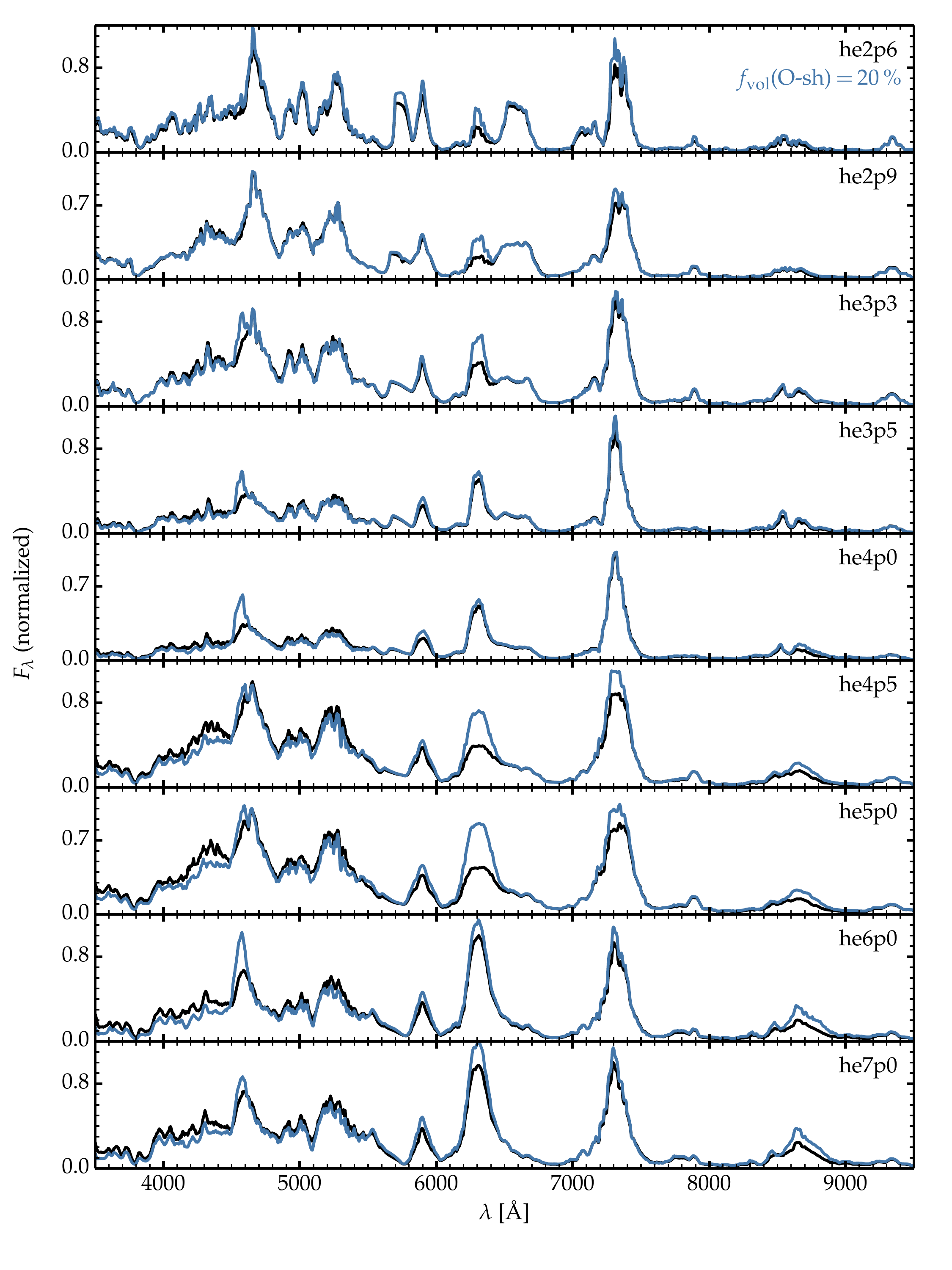}
\vspace{-0.8cm}
\caption{A spectral comparison between the  standard unclumped model  set, and the clumped counterpart in which we enforce a uniform volume filling factor of 20\% only in the ejecta layers that are O rich (i.e., those layers corresponding to the O/Si, O/Ne/Mg, and O/C shells in the progenitor). The significance of the changes varies from model to model, but typically we see enhancements in \oidoub, \mgi, \caiitrip, and to a lesser extent in \caiidoub\ and \nad, and a small reduction in the Fe flux. Models he8p0 and he12p0 are omitted.
\label{fig_montage_O_fvol0p2}
}
\end{figure*}

Clumping limited to the O-rich shell has a smaller impact than a uniform clumping (Fig.~\ref{fig_montage_O_fvol0p2}). This emphasizes the importance of the Fe/He and Si/S shells  in shaping the optical spectrum of these models. With O-shell clumping only, the influence in this model set is to enhance  \oidoub\ (and in a few models, \mgi\ too), which occurs at the expense of the extended Fe\two\ emission that falls between 4000 and 5500\,\AA. So, a flux offset is clearly seen in \oidoub\ but hardly discerned below 5500\,\AA. The underlying change is really the reduction in Mg\two\ and C\two\ cooling for the O-rich material (see Tables~\ref{tab_cool_he3p3} to \ref{tab_cool_he8fvol0p2}).

\subsection{Variations in ejecta kinetic energy}
    \label{sect_ekin}

\subsubsection{Simple velocity scaling}

    Variations in explosion energy for a given ejecta mass and progenitor composition will lead to a diversity in ejecta density. A larger explosion energy implies a larger expansion rate which, in turn, implies a lower mean density of the ejected material. In reality, the shift in density is not just a global scaling, but rather a change in the whole density profile. Furthermore, variations in energy will modulate the explosive nucleosynthesis.

     To help distinguish between the various effects we first consider models in which the total decay power emitted and the composition remain unchanged, delaying a more consistent analysis to the next section. We recalculated the he2p6 to he12p0 model grid (nominal pre-SN mass loss rate) but scaled the velocity up or down by a factor of two. In practice, we scaled the energy up (down) when the model had a lower (higher) ejecta kinetic energy than the ``adjacent'' models of comparable mass. To increase $E_{\rm kin}$ by a factor of 2, we scaled the velocity by a factor of $\sqrt{2}$. Since we did not want to change the SN age, we updated the radius as given by homologous expansion. Finally, we scale the density so that $\rho R^3$, or the total mass, is unchanged.

While a decrease in ejecta kinetic energy leads to a general increase in ejecta density, the impact is more complex than obtained with clumping, which leaves the radial column density unchanged. Clumping does not alter the deposition profile of $\gamma$-rays, and for constant ionization, it does not alter the electron-scattering optical depth of the ejecta. By contrast, varying the ejecta kinetic energy varies the radial column density, and hence modifies the ejecta optical depth for both $\gamma$-rays and low energy photons. Hence, in addition to a flux  redistribution in wavelength, we expect a change in luminosity.

\begin{figure*}
\centering
\includegraphics[width=0.75\hsize]{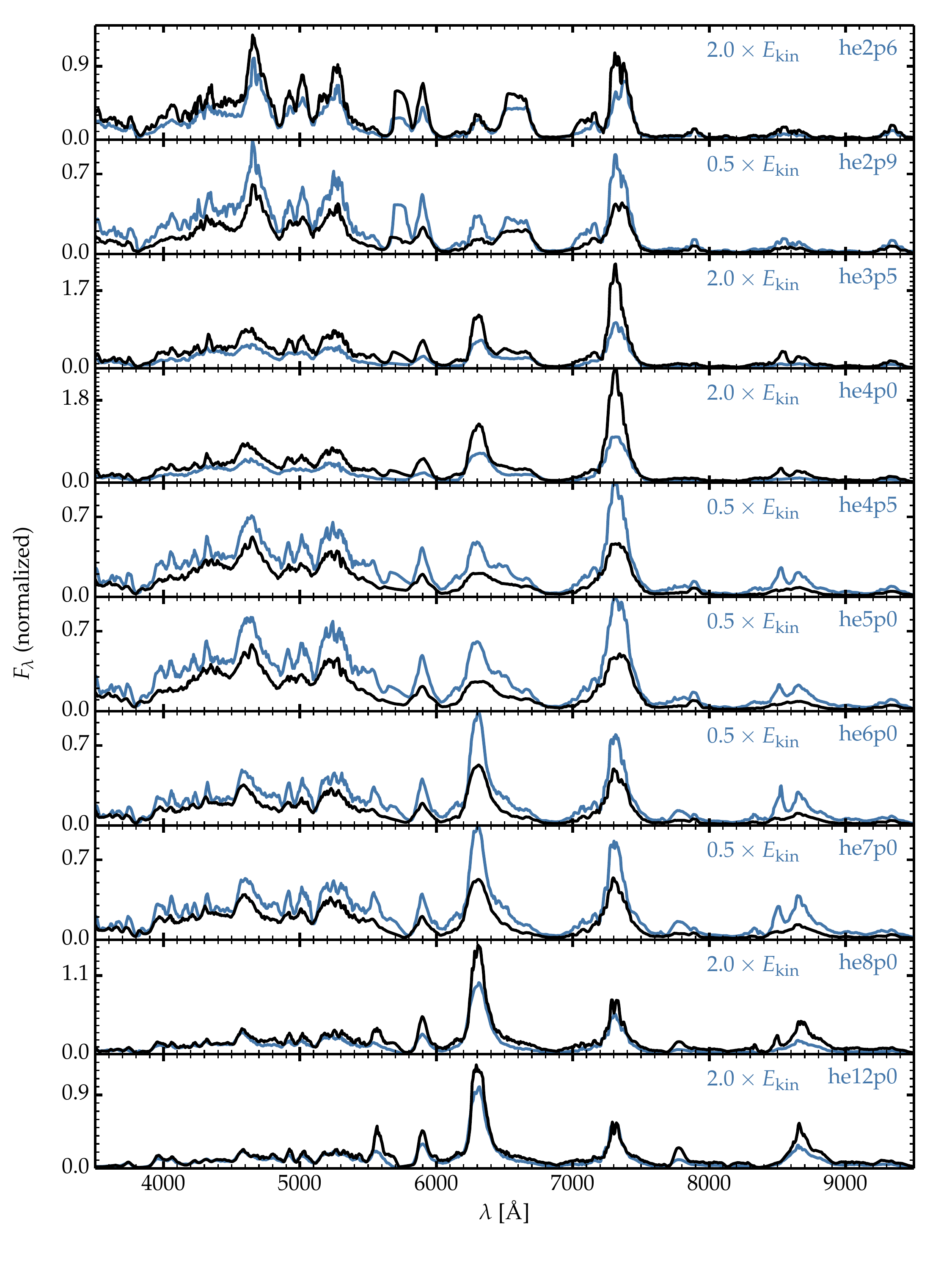}
\vspace{-0.5cm}
\caption{Same as Fig.~\ref{fig_montage_opt}, but now showing the unclumped models (black) and a counterpart at twice or half the ejecta kinetic energy (blue). Since the \nifs\ mass is the same for each pair of models, and since each pair of models has been normalized by the same factor (its value is such that the maximum flux of the first model in each pair is unity), the offset in absolute fluxes reflects the difference in the energy absorbed by the ejecta.
\label{fig_montage_ekin}
}
\end{figure*}

Figure~\ref{fig_montage_ekin} shows a comparison of optical spectra for the He-star explosion models with the nominal pre-SN mass loss rate, and their counterpart at twice or half the ejecta kinetic energy. In all cases, the higher energy ejecta is fainter because of the enhanced $\gamma$-ray escape. The spectral changes, which reflect complicated processes, are both wavelength and model-dependent.  While greater power deposited in a given shell will lead to a general enhancement in line strengths, some individual lines may weaken because of changes in ionization (and temperature).

For the most massive ejecta, belonging to he12p0, increasing $E_{\rm kin}$  quenches the lines that form  in the O-rich shell. This change is driven by the reduced $\gamma$-ray trapping (56 rather than 76\,\%) and a small rise in ionization. The same holds for model he8p0. In he12p0   \caiidoub\  does not change, while Ca\two\ near-infrared triplet shows substantial changes. The latter is stronger in the model with the lower kinetic energy, and this is true for all the test models.

For models he4p5 to he7p0, the reduction in $E_{\rm kin}$ increases the model luminosity (greater $\gamma$-ray trapping), with a slight reduction in line widths (e.g., clearly seen for \oidoub\ in the scaled variant of model he5p0). There are also numerous alterations to the line strengths; just a few
will be discussed. The \oidoub\ strength relative to \caiidoub\ may increase (e.g., model he7p0; more power goes to the O-rich shell while the enhanced density decreases slightly the O ionization despite the enhanced heating). The Ca\two\ near-infrared triplet is also boosted, probably because of an increase in optical depth. Finally, for the lighter models, the variations in $E_{\rm kin}$ in a given model are comparable to differences between models (the scaled he2p9 looks similar to the original, unscaled he2p6 model).

Overall, this exercise suggests that the ejecta kinetic energy can strongly alter the radiative properties of He-star explosions, in contrast to red-supergiant explosions where it was found to have little impact (see \citealt{D21_sn2p_neb}). This sensitivity arises because of the change in the $\gamma$-ray mean free path, which modulates the decay power absorbed (and thus the luminosity), as well as the change in ionization.

\subsubsection{Physically consistent models with half and twice the $E_{\rm kin}$ of model he4p5}

In this section, we analyse the influence of a different explosion energy on the radiation and ejecta properties for a He-star progenitor. Using the progenitor model corresponding to he4p5, we recalculate the explosion with \kepler\ and prescribe an explosion energy that yields either half (he4p5L) or twice (he4p5H) the nominal value for model he4p5. The explosion phase, including nuclear burning, is treated consistently. All three models are therefore physically consistent, unlike the scaled models discussed in the preceding section. The model characteristics are given in Table~\ref{tab_prog}.

Going from model he4p5L, to he4p5, and he4p5H the $E_{\rm kin}$ increases from 0.54, to 1.17, and $2.44 \times 10^{51}$\,erg. This increase in energy is associated with an increase in yields from explosive nucleosynthesis (i.e., an increase in the mass of the explosively produced Fe/He and Si/S shells). In the same order, the \nifs\ mass increases from 0.0822, to 0.0859, to 0.0903\,\msun, and the mass of Si increases from 0.049, to 0.0614, to  0.0786\,\msun. Such small variations in yields will be hard to discern. More significant is the drop in the fractional decay power absorbed from 28.0, to 14.5, and 8.4\,\% from model he4p5L, to he4p5, and he4p5H.

\begin{figure*}
\centering
\includegraphics[width=\hsize]{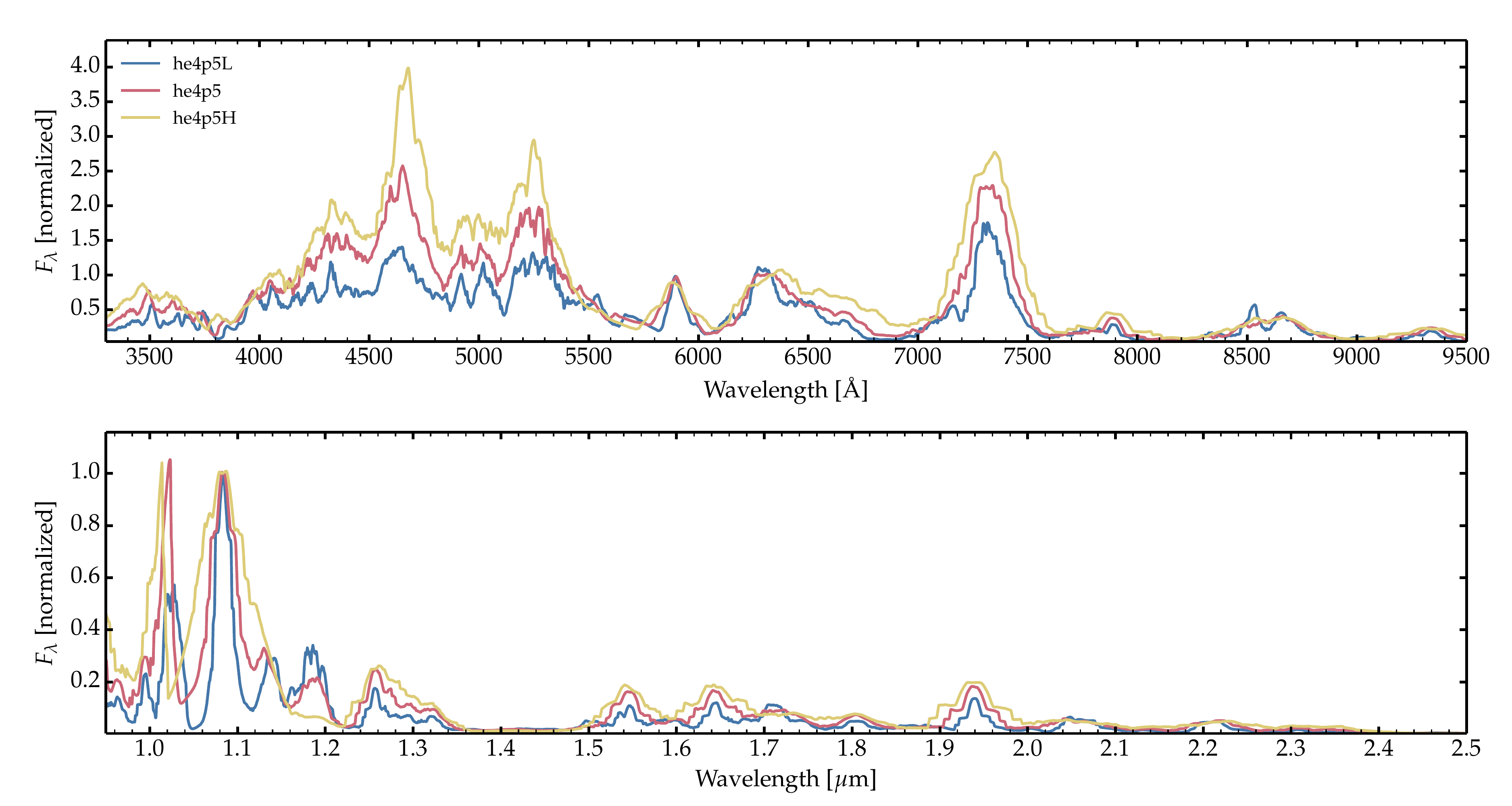}
\vspace{-0.2cm}
\caption{Comparison of normalized optical and near-infrared spectra for model he4p5 and two variants with a lower (he4p5L) and a higher (he4p5H) ejecta kinetic energy. To cancel the offset in model luminosity, the spectra are normalized at 6330\,\AA\ in the optical and at 1.083$\mu$m in the near infrared. As readily apparent, the change in ejecta kinetic energy has  caused a quantitative change in the spectra  --- a change in kinetic energy does not simply alter the luminosity and line widths.  All three models correspond to a SN age of 200\,d.
\label{fig_he4p5_ekin}
}
\end{figure*}

Figure~\ref{fig_he4p5_ekin} compares the optical and near-infrared spectra for models he4p5L, he4p5, and he4p5H. The spectral morphology is analogous in all three models, with  enhanced Doppler broadening and line overlap in the higher energy models. Ejecta ionization increases for O and Mg as the energy is increased, leading to a weakening of the O\one\ and Mg\one\ lines. Some lines, such as Mg\one\,1.183\,$\mu$m  even disappear  in model he4p5H. Because of the complicated variations in ionization, the ratio of line fluxes like \oidoub\ and \caiidoub\ are not preserved when varying the $E_{\rm kin}$, despite the very similar metal yields. This emphasizes the importance of constraining the density structure in velocity space when inferring the yields from SNe Ibc. Because of the complicated 3D structure of core-collapse SNe, this is difficult to achieve and suggests some inherent uncertainty in abundance estimates from SNe Ibc.

\section{Impact of pre-SN mass loss}
\label{sect_mdot}

    In this section, we discuss the impact of the adopted wind mass loss of He-star progenitors. So far in this paper, we have focused on results for the progenitors from \citet{ertl_ibc_20} evolved with the nominal mass loss rate. One potential tension of these models with observations is that, over the whole mass range considered, some residual He (i.e., He associated with the He/N shell where it has a $>$\,90\,\% mass fraction) remains in the outer ejecta, even in model he12p0. This is problematic for explaining SNe Ic since significant mixing of \nifs\ should lead to the production of He\one\ lines and the classification as Type Ib. The He-star progenitors of \citet{dessart_snibc_20}, which were produced with a greater wind mass loss, reproduce the dichotomy between Type Ib and Type Ic SNe. In these models, the classification as Type Ib is robustly associated with the survival of the He/N shell where the He mass fraction exceeds 95\,\%. The dichotomy between SNe Ib and Ic occurs for He-star initial masses around 9\,\msun. The models from \citet{dessart_snibc_20} match closely the properties of the models from \citet{woosley_he_19} that were evolved with a 50\,\% enhancement to the nominal wind mass loss rate. It is therefore interesting to consider the nebular-phase properties of such models (x1p5 series).

Since pre-SN wind mass loss has only a weak effect in lower mass He stars, we consider He-star models with a 50\,\% enhanced mass loss rate with initial masses between 5 and 13\,\msun, but now spaced every 1\,\msun\ (model he7p0x1p5 was discarded due to convergence issues with the radiative transfer). Models up to he9p0x1p5 (which resembles closely model ``he9'' from \citealt{dessart_snibc_20}) are of Type Ib, and those with a higher initial mass are of Type Ic. Compared to models calculated with a nominal mass loss rate and the same initial mass (i.e. he9p0x1p5 vs. he9p0), these ejecta are typically 10\,$-$\,20\% more energetic, contain more \nifs, but their mass  is 15 to 30\,\% lower. Progenitors initially more massive than 9\,\msun\ no longer possess a He/N shell;  they have a He/C or He/C/O outer shell. These objects would be classified as carbon-rich Wolf-Rayet stars \citep{pac_wr_07}. They also have a lower O abundance for the same initial mass (e.g., 1.89\,\msun\ in model he12p0x1p5 compared to 3.03\,\msun\ in model he12p0). A summary of ejecta properties is provided in Table~\ref{tab_prog} and illustrated in Fig.~\ref{fig_prop}.

\begin{figure*}
\centering
\includegraphics[width=0.75\hsize]{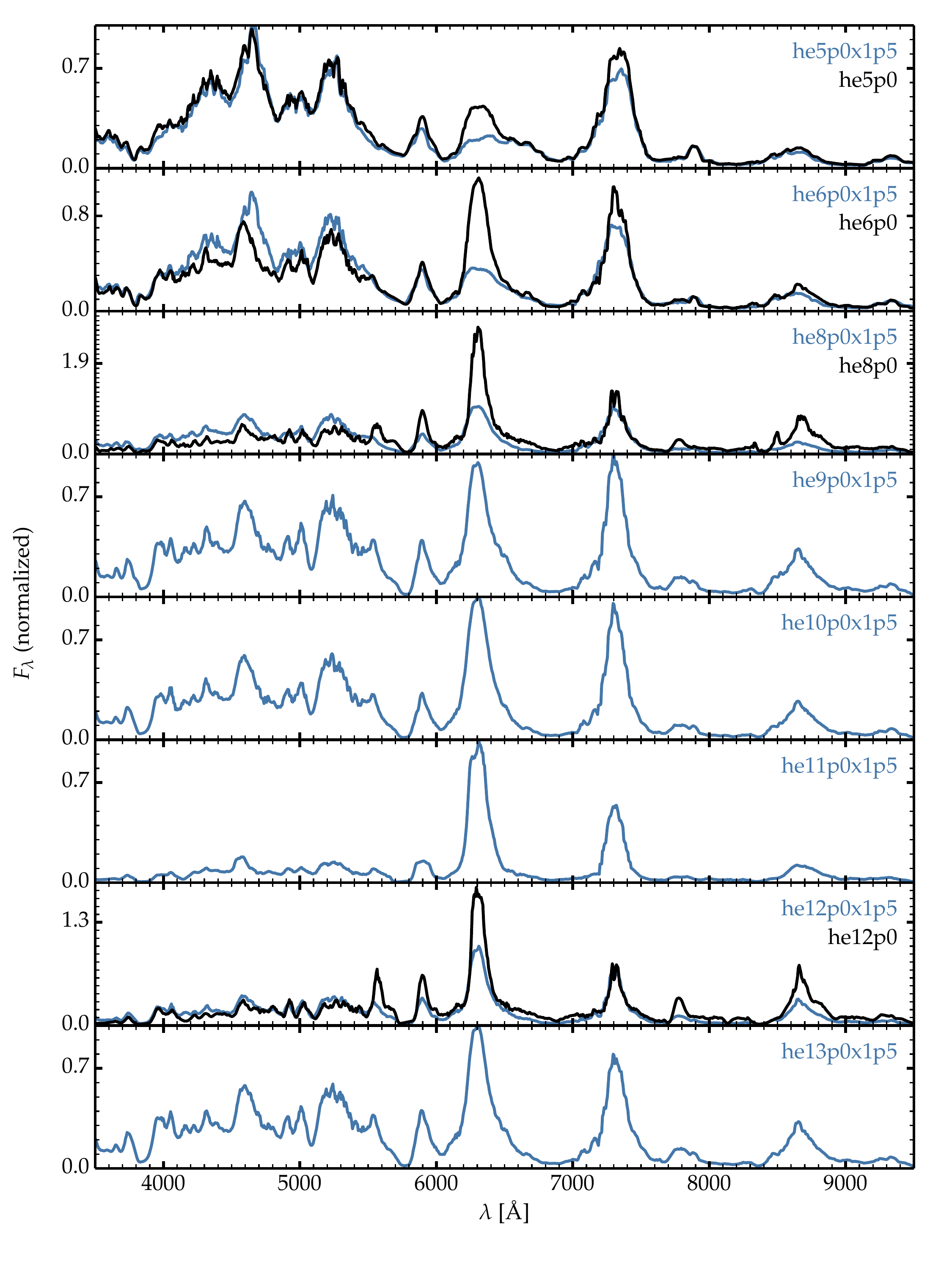}
\vspace{-0.5cm}
\caption{Optical spectra for explosions based on He-star progenitors evolved with a 50\% enhanced mass loss rate (blue), together with their counterparts with a nominal mass loss rate (black). In this stack, the spectra of models with the same He-star initial mass are normalized by the same factor (its value is such that the maximum flux of the ``x1p5'' model is unity).
\label{fig_montage_hex1p5_x1p0}
}
\end{figure*}

Because of their greater expansion rate, SNe in the x1p5 series suffer greater $\gamma$-ray leakage relative to the models with the nominal mass loss rate. For example, model he8p0x1p5 absorbs only 26\,\% of the total decay power emitted (compared to 55\,\% in model he8p0), which is representative of the offset for the whole sequence. This leads to a lower bolometric luminosity at the 10\,\% level (this small offset arises because of the larger \nifs\ mass of models in the x1p5 series), with moderate differences in absolute magnitudes (see Table~\ref{tab_mag}). The relative fraction of the total decay power absorbed that goes to the O-rich shell is comparable in both sets. This is because the O-rich shell mass relative to the total ejecta mass is within about 10\,\% in both sets.

We find that the ejecta ionization is systematically higher in the x1p5 model series, whatever the shell considered, i.e. O-rich, Si-rich, or Fe-rich. In the O-rich shell, the mean O ionization level is 0.93 (0.83) in model he5p0x1p5 (he5p0), 0.89 (0.57) in model he6p0x1p5 (he6p0), 0.69 (0.30) in model he8p0x1p5 (he8p0), and 0.50 (0.23) in model he12p0x1p5 (he12p0). Since the \oidoub\ doublet is a strong coolant for the O-rich material only if O is neutral or almost neutral, we can anticipate that the O-rich material will cool in a large part through Fe\two\ emission.

Figure~\ref{fig_montage_hex1p5_x1p0} shows a comparison of optical spectra between He-star explosions based on progenitors evolved with the nominal wind mass loss and those with a 50\,\% enhancement. For all initial masses for which there is a model in each set (masses 5, 6, 8, and 12\,\msun), the models from the x1p5 set show a weaker \oidoub\ and a stronger flux shortward of 5500\,\AA\ that is associated with the Fe\two\ emission (the same is seen when comparing model he10p0x2p0 and he10p0x1p5). As discussed earlier, this is associated with the greater ionization in the O-rich zones, which tend to cool through Fe\two\ emission rather than \oidoub\ (or lines from Na\one\ or Mg\one). This is also intricately related to the changes in C\two\ and Mg\two\ cooling within the O-rich zones, as also observed in the clumped models (Sect.~\ref{sect_fvol} and Tables~\ref{tab_cool_he3p3}\,$-$\,\ref{tab_cool_he8fvol0p2}). \caiidoub, which tends to form in the Fe/He and Si/S shells, is similar in both sets of models. So, the ratio of \oidoub\ to \caiidoub\ is lower in the x1p5 set for the same initial mass. The main limiting factor for the \oidoub\ line strength is not the slightly lower O abundance, but the greater ionization of the O-rich material (as was obtained in some models of the x1p0 set). The model with the strongest \oidoub\ in the x1p5 set is model he11p0x1p5, while model he13p0x1p5 is analogous to he10p0x1p5 or he9p0x1p5.

\begin{figure*}
\centering
\includegraphics[width=0.75\hsize]{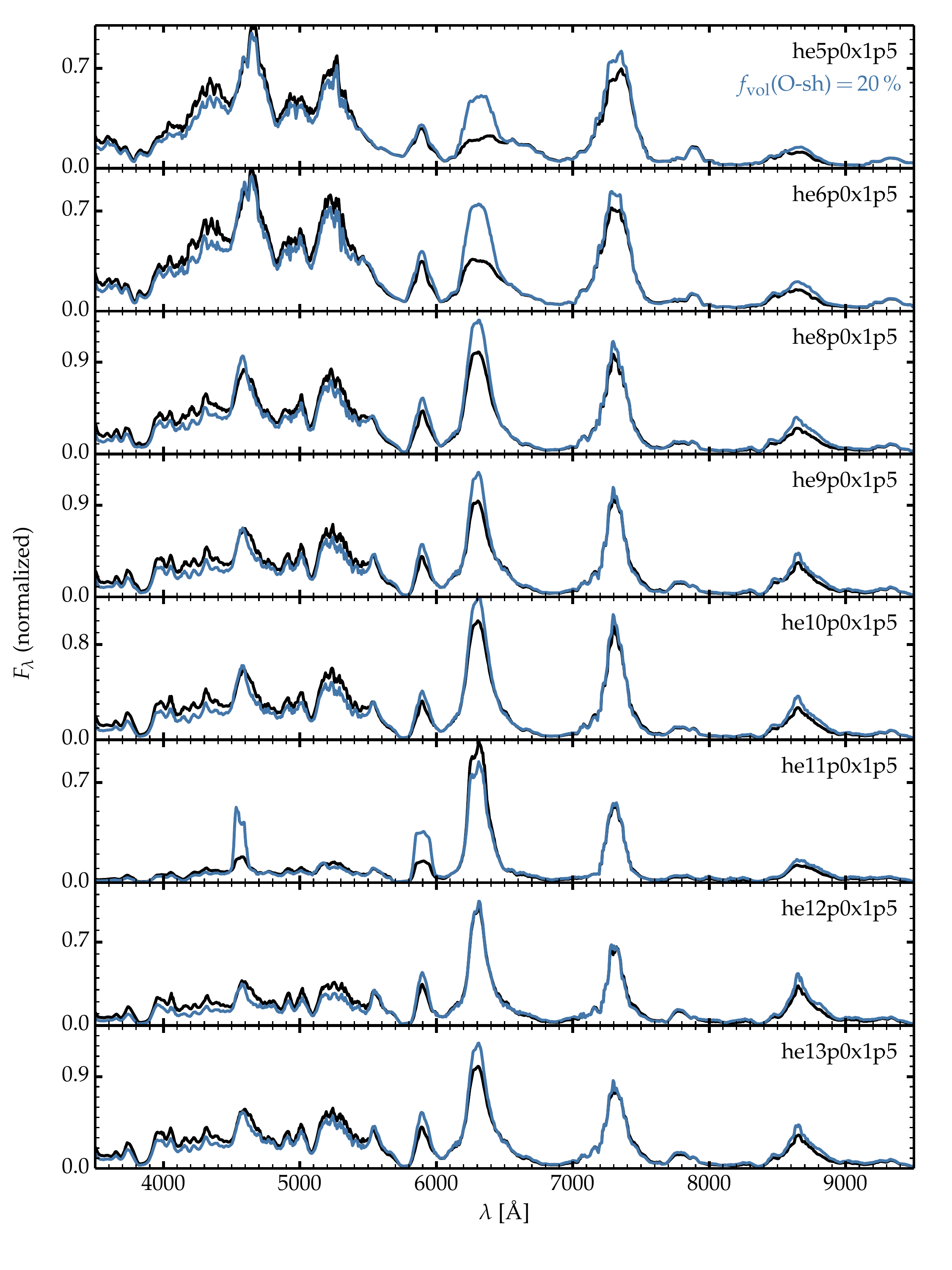}
\vspace{-0.5cm}
\caption{Same as Fig.~\ref{fig_montage_opt}, but now showing models evolved with a 50\% enhanced mass loss rate, together with their counterparts in which the O-rich material is given a volume filling factor of 20\%. In this stack, the spectra of models with the same He-star initial mass are normalized by the same factor (its value is such that the maximum flux of the unclumped model is unity).
\label{fig_montage_x1p5_O_fvol}
}
\end{figure*}

As discussed in Sect.~\ref{sect_ekin}, the ejecta kinetic energy and mass (or the density profile in velocity space) are critical ingredients for determining the radiative properties of SNe Ibc. Figure~\ref{fig_montage_x1p5_O_fvol} illustrates this point further by showing the impact of clumping within the O-rich shell for the x1p5 series. The impact is analogous to that obtained for the x1p0 series (see Fig.~\ref{fig_montage_O_fvol0p2} and Sect.~\ref{sect_O_fvol}). In model he11p0x1p5, which has the lowest ejecta ionization of the whole x1p5 set, clumping of the O-rich material enhances the recombination of Na and Mg and boosts the \nad\ and \mgi\ at the expense of \oidoub.

\section{The positron contribution to the SN luminosity}
\label{sect_positron}

   Many efforts have been devoted to identifying the birth of positron escape from SN ejecta. Late-time observations of thermonuclear explosions suggest that positrons escape from SN Ia ejecta after a few hundred days \citep{milne_positron_99,milne_positron_01,milne_snia_04}. This result is, however, uncertain because of the difficulty of inferring the bolometric luminosity of SNe at such late times and because of the uncertain decaying isotope at the origin of the SN luminosity, such as the relative importance of the \iso{57}Ni decay chain relative to that of \nifs\ \citep{graur_late_snia_16,graur_late_snia_20}.

    In Type Ibc SNe, similar observations are generally lacking. The challenge to identify a change of slope inherent to positron trapping\footnote{The bolometric luminosity should drop faster than 1\,mag per 100\,d as $\gamma$-rays increasingly leak from the ejecta, until a time is reached when the luminosity is powered primarily by positrons. At such times, the SN luminosity should follow a drop of 1\,mag per 100\,d, and be of the order of 3\,\% of the expected power for full $\gamma$-ray and positron trapping. In this context, an upturn in the bolometric light curve at late times is expected \citep{milne_positron_99}.} is compromised by the potential power contribution from ejecta interaction with a pre-SN wind. The systematically lower \nifs\ yields also imply a low SN luminosity which can then be weakly contrasted with the light contamination from the young star cluster hosting the SN (see, for example, \citealt{kuncarayakti_env_18}). While these shortcomings affect all such photometric methods, an alternate approach would be to search for spectral lines that are powered in part from positrons.

   The nonlocal energy deposition of positrons should start much earlier than their escape from the ejecta. In other words, even for full positron trapping within the ejecta (i.e., no escape to infinity), a significant fraction of positrons could leak from the \nifs-rich regions where they are first emitted. This is particularly true in core-collapse SNe because of their ``shrapnel'' distribution of \nifs. While the positron mean free path may be much smaller than the SN radius and prevent an escape to infinity, it may be large enough to allow positrons to step into a nearby clump rich in O or He. For the material that absorbs this positron, the gain in power would be small relative to the contribution from $\gamma$-rays. But the loss for the \nifs-rich material could be significant.

\begin{figure}
\centering
\includegraphics[width=\hsize]{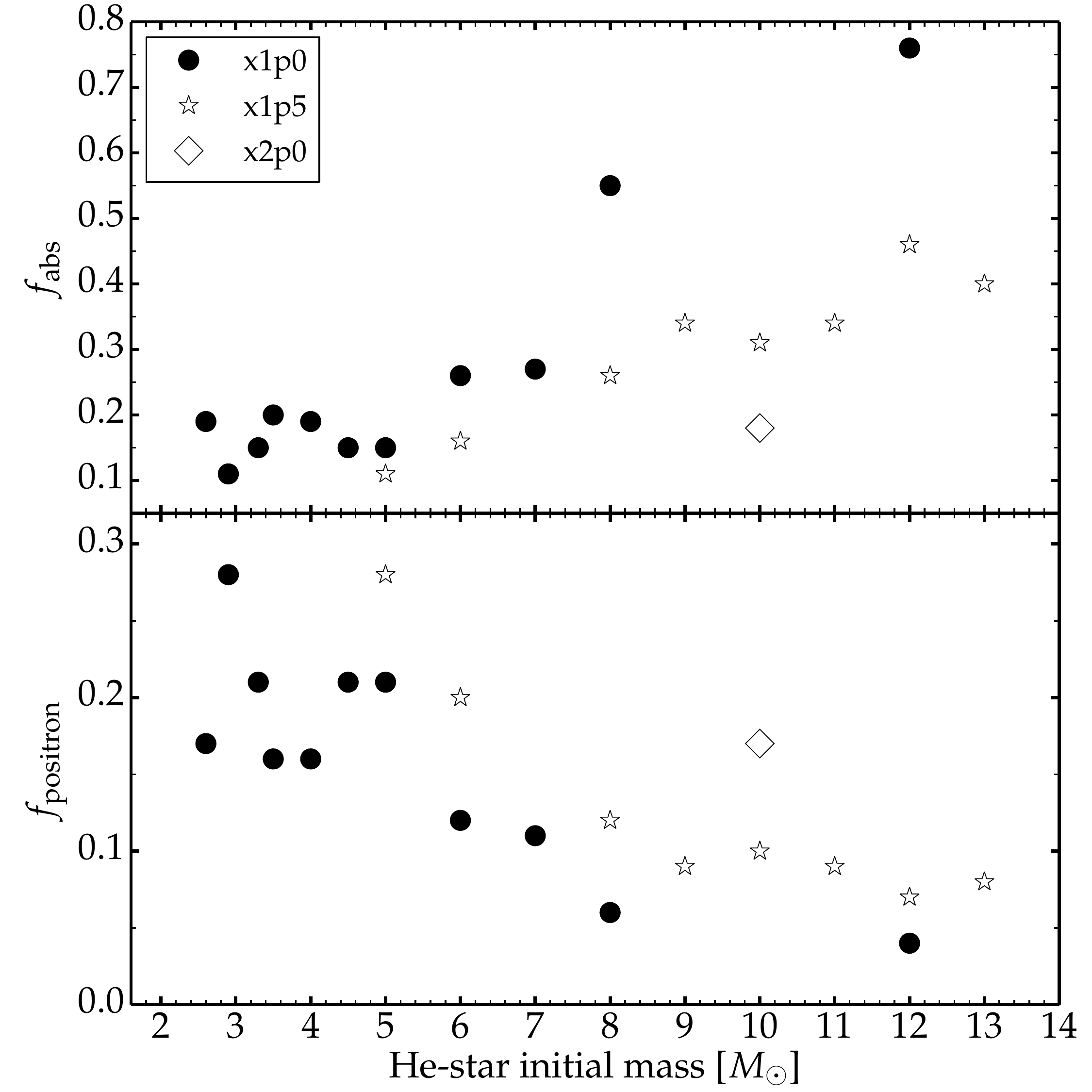}
\caption{Variation of the decay power absorbed in the ejecta (upper panel) and the fraction of that power that arises from positrons (assumed here to be absorbed locally; lower panel) for the models in the x1p0, x1p5, and x2p0 series.
\label{fig_ekin_eabs}
}
\end{figure}

  Figure~\ref{fig_ekin_eabs} shows the fraction of the decay power absorbed in the ejecta, as well as the fraction of that power that is due to positrons (absorbed locally) in the x1p0, x1p5, and x2p0 model series, and at 200\,d after explosion. In the absence of any $\gamma$-ray escape (i.e., full trapping of both $\gamma$-rays and positrons), positrons represent 3\,\% of the total decay power absorbed (or emitted in this case; \citealt{junde_a56_11}). With $\gamma$-ray escape, this fraction grows and reaches a maximum of 100\,\% (i.e., the ejecta is entirely powered by positrons) if all $\gamma$-rays escape. Figure~\ref{fig_ekin_eabs} indicates that lower mass ejecta models (lighter than he5p0 or he5p0x1p5) trap only about 20\,\% of the total decay power emitted, and consequently the fraction of that power due to positrons is large and in the range 15 to 30\,\%. Higher mass models trap more efficiently the $\gamma$-rays, so that the contribution from positrons is lower and of the order of 10\,\%. Model he12p0 absorbs 76\,\% of the total decay power,  and positrons contribute only 4\,\% to the bolometric luminosity.

We can go further and test the influence of positrons on the spectral properties. With the assumption of local deposition, positrons can only influence the regions initially rich in \nifs,  primarily the Fe/He shell, and the Si/S shell at a smaller level. To assess the impact that nonlocal energy deposition of positrons would have, we adopt the extreme postulate in which positrons are subject to the same opacity as $\gamma$-rays (local deposition of positron power is equivalent to assuming an infinite opacity).  In practice, positrons are subject to a variety of annihilation processes, likely dominated by interaction with bound electrons (see, for example, the discussion in \citealt{milne_positron_99}). For transport in a spherically-symmetric ejecta, a representative material opacity to positrons of about 10\,cm$^2$\,g$^{-1}$ is used. Our choice of a lower effective opacity applies heuristically to a complex, shrapnel like distribution of \nifs, with a short path length between \nifs-rich material and for example O-rich material.

\begin{figure*}
\centering
\includegraphics[width=0.75\hsize]{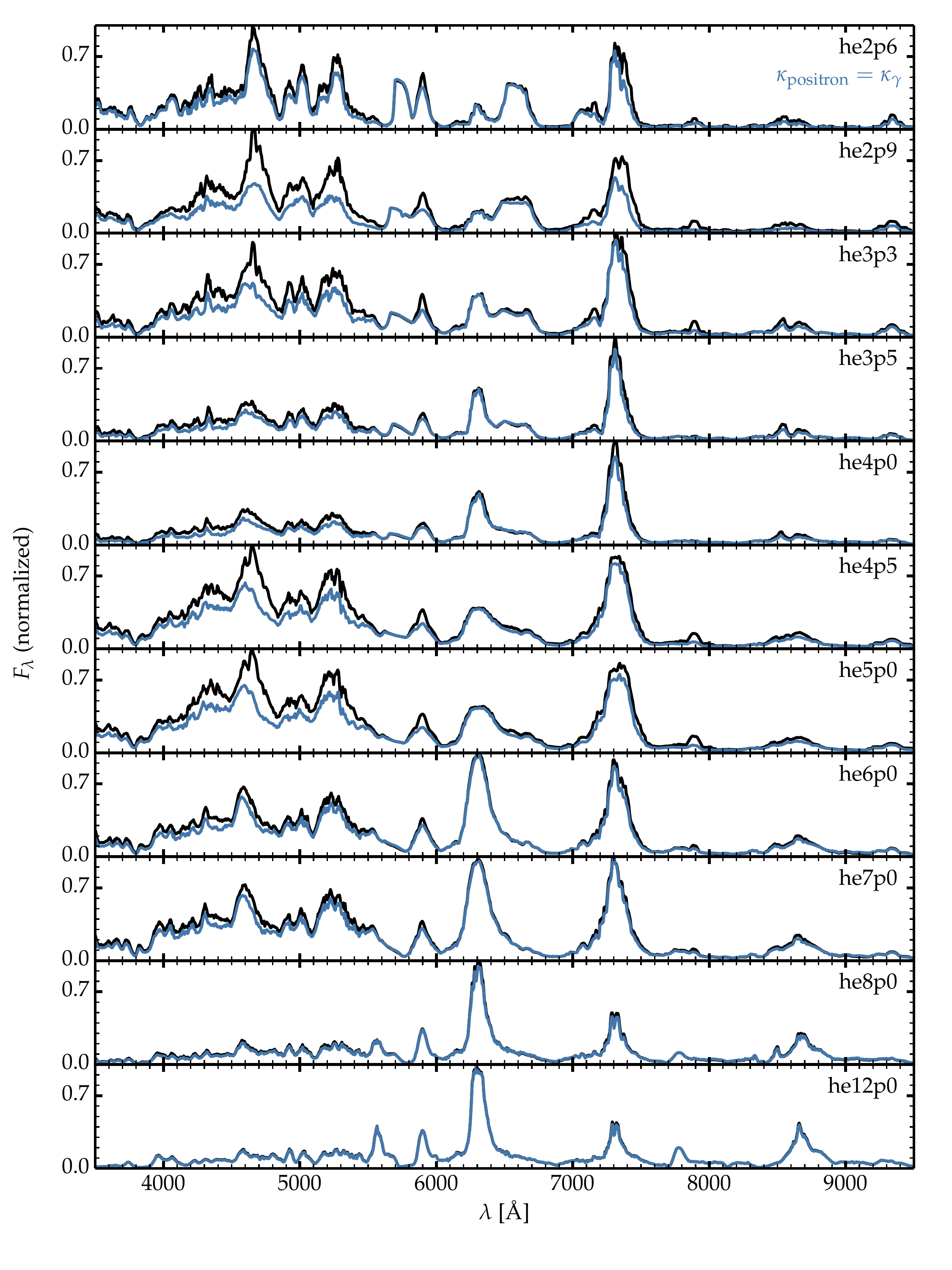}
\vspace{-0.5cm}
\caption{Same as Fig.~\ref{fig_montage_opt}, but now showing models with (black) and without (blue) local deposition of positron power.  For the latter, we adopt the same opacity for positrons as for $\gamma$-rays. In this stacking, the spectra of models with the same He-star initial mass are normalized by the same factor (its value is such that the maximum flux of the model with local positron trapping is unity). The flux offset resulting from these two assumptions is stronger in models with the greater $\gamma$-ray escape fraction (i.e., those models with the highest expansion rate, which are typically those with lower initial masses).
\label{fig_montage_positron}
}
\end{figure*}

Figure~\ref{fig_montage_positron} shows that the strongest impact on spectral properties occurs in lower mass ejecta where the positron contribution to the luminosity is significant, although one can see that in all models the main effect is to reduce the Fe\two\ emission shortward of 5500\,\AA. This reduction is sizable in models he2p9, he3p3, he4p5, and he5p0. In addition, lines that form exclusively in the Fe/He shell, such as [Ni\two]\,$\lambda$\,7378, [Ni\three]\,$\lambda$\,7889, and [Co\three]\,$\lambda$\,5889, are visibly quenched.

Identifying the nonlocal energy deposition of positrons from observed spectra is challenging. The global offset in Fe\two\ emission in the blue part of the optical range can also arise from a higher ionization of the O-rich material. Furthermore, the material from Fe/He may be more spread in space, producing a broad and weak emission from [Ni\two]\,$\lambda$\,7378 or [Ni\three]\,$\lambda$\,7889 that would be hard to discern. Line overlap is also an issue, challenging the determination of the relative contributions of [Co\three]\,$\lambda$\,5889 and \nad\ to the feature at 5900\,\AA. Since \nifs\ is expected to be localized in isolated blobs or clumps in some neutrino-driven explosions \citep{gabler_3dsn_21}, the acquisition of multiepoch high resolution spectra could help diagnose the birth of positron escape in SNe Ibc.

\section{Evolution until late times}
\label{sect_evol}

As the ejecta age, the escape fraction of $\gamma$-rays increases, the relative contribution from positrons increases, and the ejecta density and electron-scattering optical depth drop. All have the potential to alter
the observed spectrum. Therefore, in this section,  we extend our results from 200 to 428\,d for models he5p0, he6p0, and he8p0.  Starting from the results of the steady-state calculation at 200\,d, we compute the time-dependent radiative transfer at consecutive epochs, each step in time being set to 10\,\% of the current SN age. We use the same grid in velocity space at all epochs to avoid additional interpolations of time-dependent terms in the moment equations and preserve the same resolution of the various shells. We also assume a local energy deposition for positrons.

\begin{figure}
\centering
\includegraphics[width=\hsize]{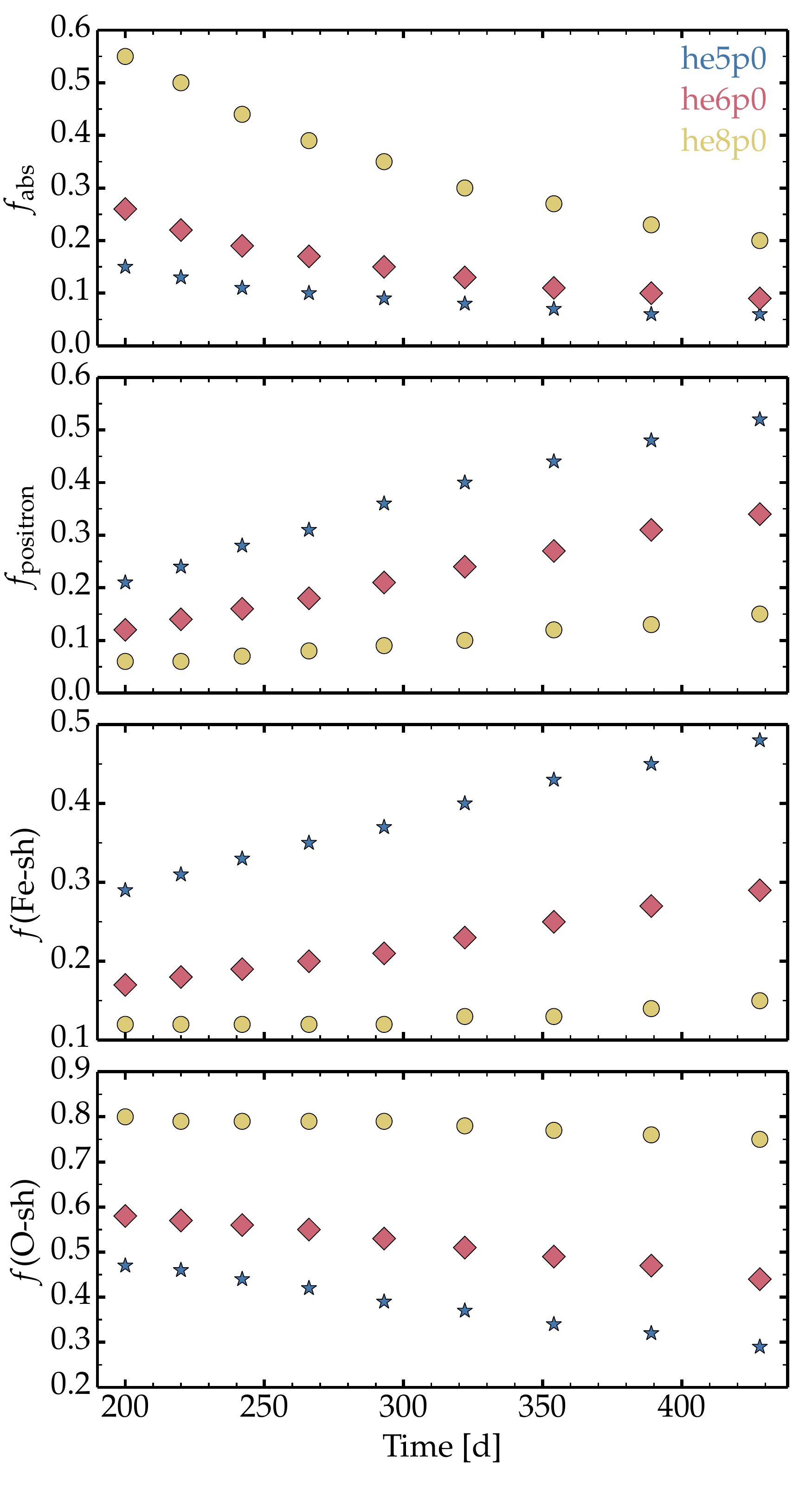}
\caption{Evolution of the decay power absorbed in the ejecta, the fraction of that power that arises from positrons (assumed here to be absorbed locally), and the fraction of the total decay power absorbed by the Fe-rich and O-rich shells in models he5p0 (blue star), he6p0 (red diamond), and he8p0 (yellow filled circle).
\label{fig_edep_evol}
}
\end{figure}

Figure~\ref{fig_edep_evol} illustrates the evolution from 200 to 428\,d of the fraction of decay power absorbed in the ejecta, $f_{\rm abs}$, the fraction of that power that arises from positrons, $f_{\rm positron}$, and the fraction of the total decay power absorbed by the Fe-rich and O-rich shells. As expected,  $\gamma$-ray increased leakage is seen in all models -- this is simply due to the optical depth which scales as $t^{-2}$. However, we assume positrons are locally absorbed, and
this leads to a ``flattening ''  in the variation of  $f_{\rm abs}$ with time. If positrons remain trapped,  $f_{\rm abs}$ will approach 0.03. Unfortunately the flattening may be hard to diagnose, since it applies to the power radiated over the entire electromagnetic spectrum (excluding $\gamma$-rays and X-rays), whereas observations at late times are typically limited to the optical.

The fraction of the decay power absorbed that is collected by the Fe-rich shell increases with time, weakly in model he8p0 because of the greater $\gamma$-ray trapping, but strongly in model he5p0 where the positron contribution is larger and grows significantly with time. In contrast, the fraction collected by the O-rich shell drops significantly in models he5p0 and he6p0, while it remains nearly constant in model he8p0 owing to the more efficient trapping of $\gamma$-rays.

\begin{figure*}
\centering
\includegraphics[width=\hsize]{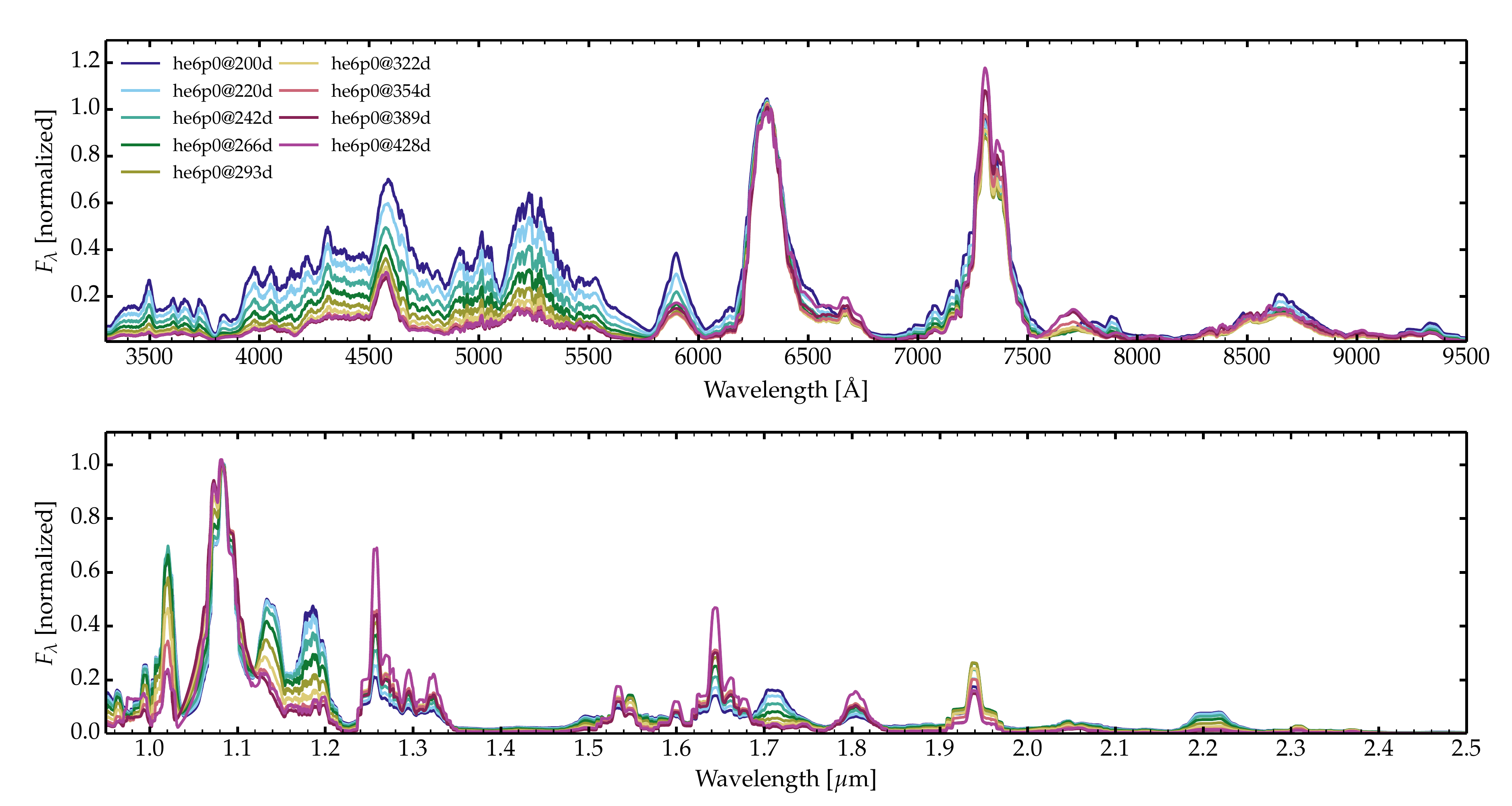}
\vspace{-0.2cm}
\caption{Sequence of  time-dependent radiative transfer calculations for ejecta model he6p0 from 200 to 428\,d after explosion. To cancel the drop in luminosity, due to both the exponential decrease in decay power emitted and the enhanced $\gamma$-ray escape with time, we normalize all spectra at 6330\,\AA\ in the optical (this wavelength corresponds roughly to the peak of \oidoub) and at 1.083\,$\mu$m in the near infrared. Other ways to normalize are possible but this choice shows the near constancy of the flux ratio of the \oidoub\ and \caiidoub\ doublets and the progressive drop in Fe emission below 5500\,\AA.
\label{fig_he6_evol}
}
\end{figure*}

The spectral evolution from 200 to 428\,d after explosion is shown in Fig.~\ref{fig_he6_evol} for model he6p0. The evolution for models he5p0 and he8p0 is qualitatively similar since the same physical processes are at work. We see the considerable strengthening of a few features (in a relative sense since the bolometric luminosity continuously drops), while the emission shortward of 5500\,\AA\ and associated with Fe\two\ ebbs. The strong feature around 6300\,\AA\ is due exclusively to  \oidoub, with the shoulder on the red side arising from \niidoub\ (with a narrow and weak feature associated with [Ni\two]\,$\lambda$\,6667). The other strong feature around 7300\,\AA\ is due primarily to \caiidoub\ but also to [Ni\two]\,$\lambda$\,7378. The flux ratio of the \oidoub\ doublet to the \caiidoub\ doublet is roughly constant over time. \nad\ is the only contributor to the 5900\,\AA\ feature. \mgi\  is clearly seen but overlaps with a forest of lines due to Fe\one\ and Fe\two\ (which cause both absorption and emission). The features around 7700\,\AA\ are due in equal parts to \kidoub\ and to the O\one\ triplet at 7774\,\AA. The emission around 8500\,\AA\ is due to both Fe\one\ emission and the Ca\two\ near-infrared triplet, and further to the red there are weaker features due to [C\one]\,$\lambda$\,9850 and [Co\two]\,1.019\,$\mu$m. In the near-infrared, the strongest feature is He\one\,1.083\,$\mu$m, with the presence of isolated and relatively strong lines associated with [Ni\two]\,1.939\,$\mu$m, [Ni\two]\,6.634\,$\mu$m and [Ne\two]\,12.81\,$\mu$m.

Ignoring the global reduction in the model luminosity, this spectral evolution reflects primarily the drop in ionization in all ejecta shells except the outermost He-rich zones when present. This recombination is weak for Fe (which is primarily Fe$^+$ in the Si/S and O-rich shells), but significant for O, Mg and Ca. At 428\,d, O is nearly neutral in the O-rich shell and collisional excitation of O\one\ becomes the main coolant. Mg becomes only partially ionized, which explains the strengthening of \mgi. Ca is once ionized in the Si/S shell.

\section{He\one\ and [N\two] lines}
\label{sect_hei}

   He\one\ and [N\two] lines represent important signatures since they may be used to constrain the SN type and the composition  of the progenitor star. The standard procedure is to determine the SN type at discovery, or at least as early as possible during the photospheric phase. It is nonetheless of interest to check whether one can still distinguish a Type Ib from a Type Ic at nebular times.

   Helium may be present but not sufficiently abundant or sufficiently excited by high-energy electrons to produce He\one\ lines. In the optical, the overlap of He\one\,5875\,\AA\ with \nad\  complicates the direct identification of that important He\one\ line. In that case, one may look for He\one\ lines in the near infrared. An alternative is to try to identify N lines since N is the most abundant metal species in the He/N shell. The presence of that shell in the progenitor implies a SN of Type Ib \citep{dessart_snibc_20}. As discussed in the preceding sections, the \niidoub\ doublet forms through collisional processes but He\one\ lines, including those at 10830\,\AA\ and 20581\,\AA, are nonthermally excited. Therefore, the sensitivity of these lines to the N and He abundances is very different.

\begin{figure*}
\centering
\includegraphics[width=0.33\hsize]{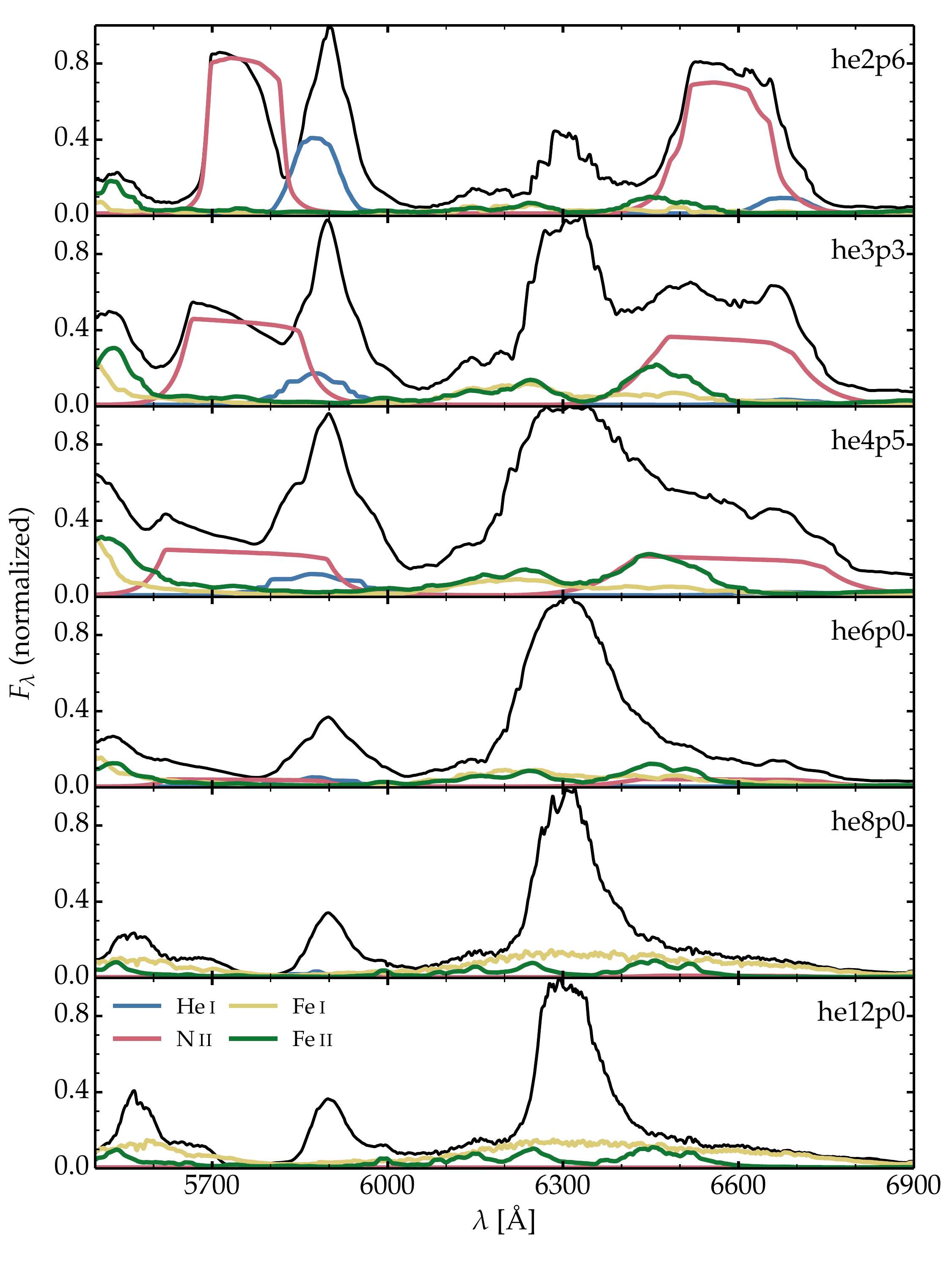}
\includegraphics[width=0.33\hsize]{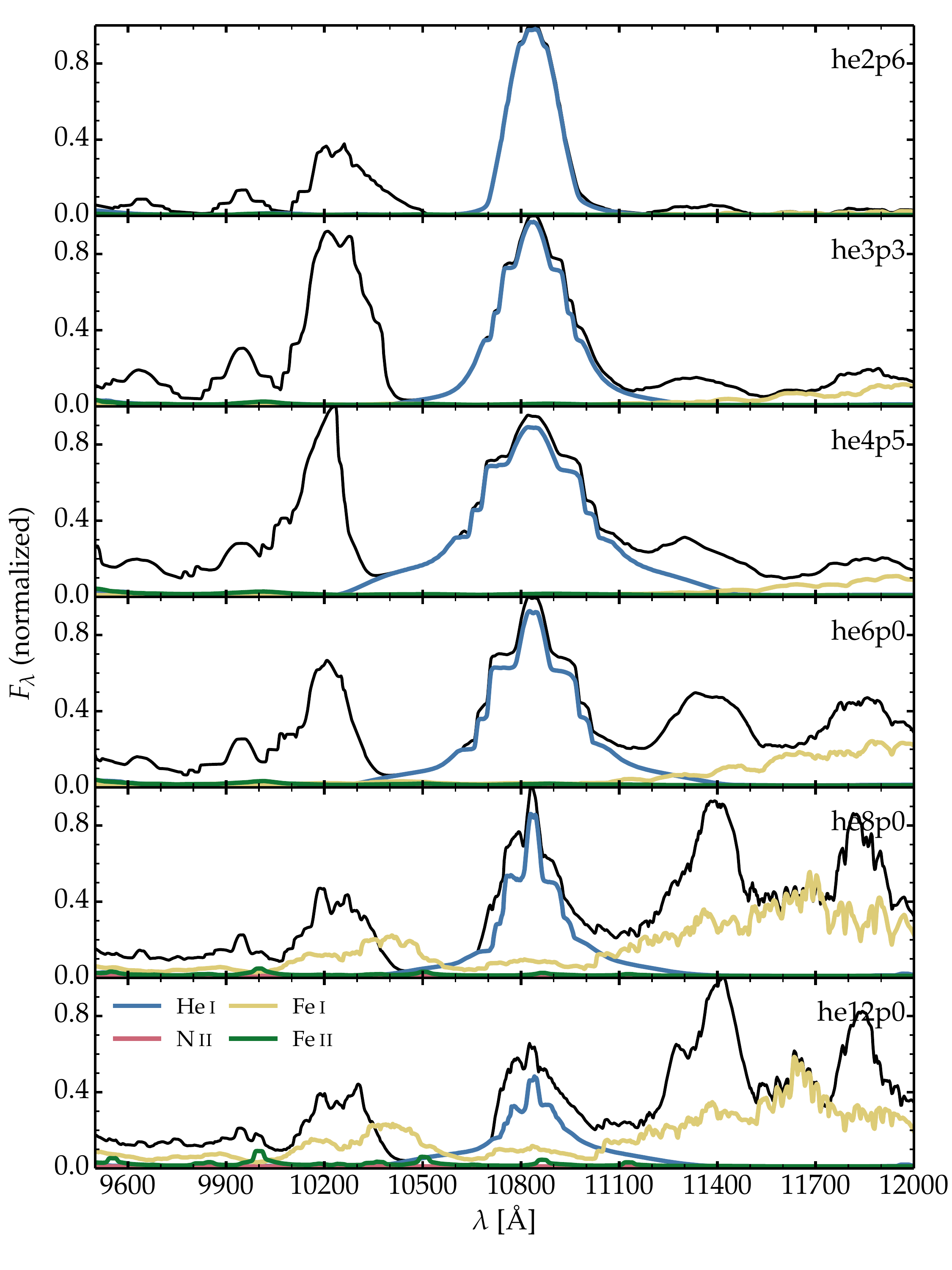}
\includegraphics[width=0.33\hsize]{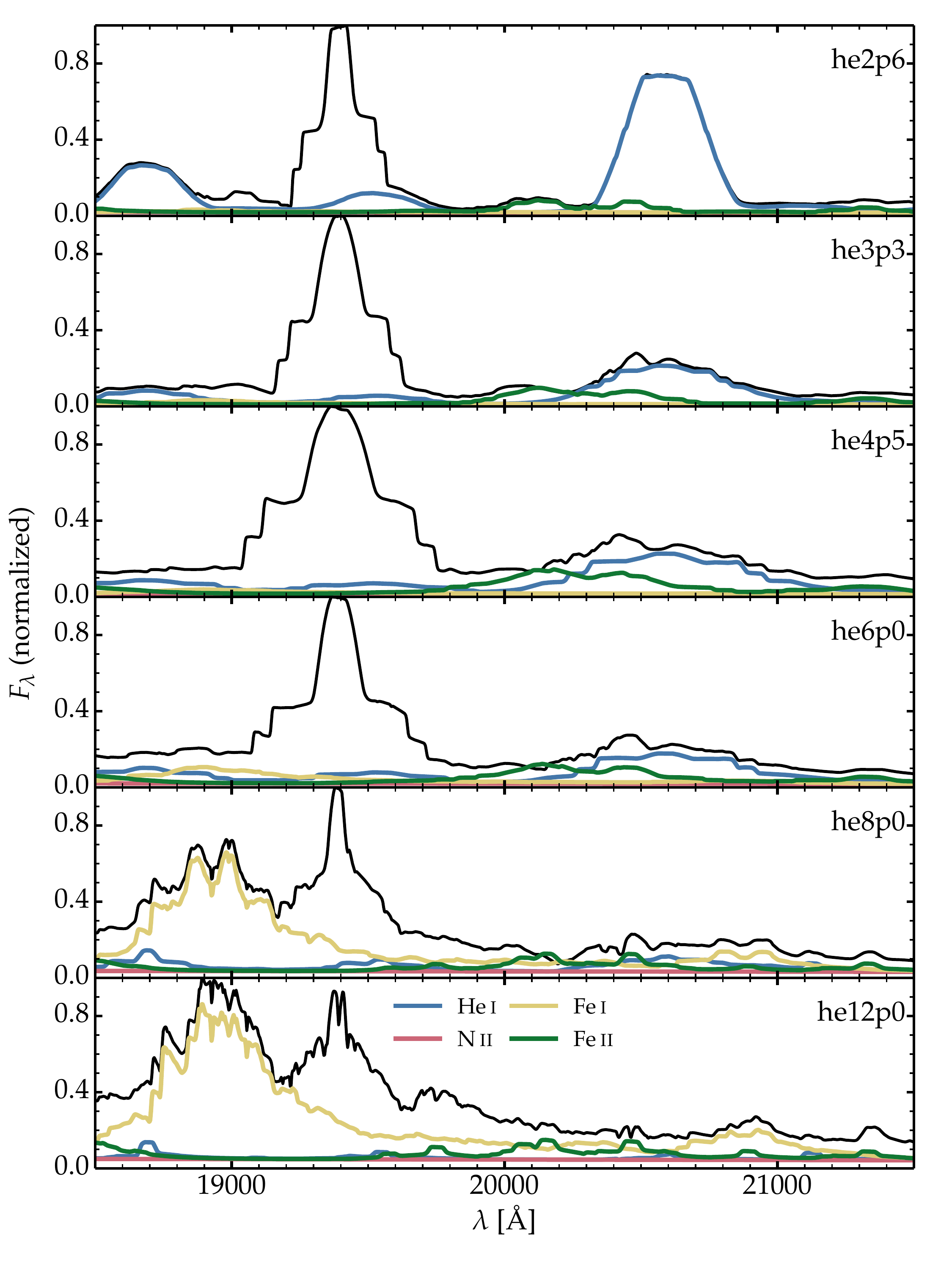}
\vspace{-0.2cm}
\caption{Montage of spectra for models he2p6, he3p3, he4p5, he6p0, he8p0, and he12p0  at 200\,d showing the wavelength region around 6300\,\AA\ (left), around 10800\,\AA\ (middle), and around 20000\,\AA\ (right). In all plots, we overplot the emission associated with bound-bound transitions from He\one, N\two, Fe\one, and Fe\two. The same species set and color coding is used even when some species or ions do not contribute flux to the selected spectral region.
\label{fig_he1_n2}
}
\end{figure*}

Figure~\ref{fig_he1_n2} shows three spectral regions hosting strong transitions from He\one, N\two, Fe\one, and Fe\two\ lines for a restricted set of models from the x1p0 series (the selection avoids duplications). In the optical, the He\one\,5875\,\AA\, line is present in the lower mass models up to he4p5, but it is typically weak and badly blended with \nad. Instead,  [N\two]\,$\lambda$\,5754.6 and the \niidoub\ doublet  can be used a a diagnostic for the presence of a  He/N shell -- they are strong and less blended up to model he4p5. In models he6p0, he8p0, and he12p0 these He\one\ and N\two\ lines are very weak. Although the \oidoub\ doublet in models he8p0 and he12p0 still exhibits a red shoulder, it is primarily due to Fe\one.

The He\one\,10830\,\AA\ line is present in all selected models (middle panel of Fig.~\ref{fig_he1_n2}), even in  he12p0 that lacks a He/N shell. Hence, He\one\,10830\,\AA\ can form either in the He/N shell (in Type Ib SN models that have such a shell) or in the He/C/O shell (in Type Ic models, which are characterized by the lack of a He/N shell). The He\one\,20581\,\AA\ line is present in all models up to he6p0, but vanishingly weak in models he8p0 and he12p0. Hence, this latter line is a good near-infrared discriminant between Type Ib and Type Ic SNe, as  found in photospheric-phase models (see, for example, Sect.~5 in \citealt{D15_SNIbc_I}).

\section{Comparison to observations}
\label{sect_obs}

The goal of this study was to generate observables in the nebular phase for ab~initio He-star explosion models. In a previous study based on the red-supergiant star explosion simulations of \citet{sukhbold_ccsn_16}, we demonstrated that a similar approach yielded a satisfactory agreement to the observations of low-luminosity and standard-luminosity Type II SNe \citep{D21_sn2p_neb}. The task is however more difficult for SNe Ibc because much uncertainty remains concerning their progenitors and their evolution (for example, relative to wind mass loss). Binary star evolution seems a prerequisite for the production of moderate mass ejecta, either He-rich  or He-poor  \citep{podsiadlowski_92,nomoto_bin_95,wlw_95_snibc,yoon_ibc_10,langer_araa_mdot_12}, and can  explain the dichotomy between SNe Ib and Ic \citep{yoon_wr_17,dessart_snibc_20}. The contribution of single and more massive Wolf-Rayet stars to observed SNe Ibc may be marginal. One such example may be SN\,2009jf \citep{valenti_09jf_11} which can be explained by single-star evolution, although there is still the possibility that it arose from an interacting binary.  Few conclusive identifications of their progenitors (or progenitor systems) have been made, and only robustly for SNe IIb (e.g., SN\,1993J -- \citealt{aldering_93j_94}, \citealt{vandyk_93j_02}, \citealt{maund_93j_04}; SN\,2011dh -- \citealt{vandyk_etal_13_11dh}, \citealt{folatelli_11dh_14}).

The progenitors used in the present study result from He-stars that evolve in isolation (although binary mass exchange is thought to cause stripping of the H-envelope to produce the He star). There is no consideration for binary mass exchange in the course of their evolution until core-collapse. Thus Case BB mass transfer\footnote{Case BB mass transfer applies to a star in a binary system that undergoes a first phase of mass transfer after the ignition of a hydrogen burning shell source and a further mass transfer after the ignition of a helium burning shell source; see discussion in \citet{langer_araa_mdot_12}.}, which can produce ultra-stripped Type Ic progenitors \citep{nomoto_94I_94,dewi_bin_02,dewi_pols_03,tauris_ulstr_13}, is not considered. Our grid of models covers He stars with masses extending from 2.6 to 12.0\,\msun, with the corresponding ejecta masses ranging from 0.8 to 5.3\,\msun.

Low mass models, with He-rich ejecta, have not yet been unambiguously associated with SNe Ib. Instead, only our He-star models above about 4\,\msun\ initially seem to reflect the observed nebular-phase properties of SNe Ib and Ic (see next sections). The nondetection of faint SNe Ib powered by a low \nifs\ mass (at the level of 0.001\,\msun\ inferred in low-energy SNe II; \citealt{pasto_low_lum_04}; \citealt{spiro_low_lum_14}; \citealt{lisakov_ll2p_18}) may be an observational bias. The SN would be too faint and fast evolving to be identified in the midst of massive luminous stars. Another possibility is that such SNe Ib undergo nuclear flashes that lead to the He-shell ejection before exploding (see for example \citealt{woosley_he_19}). Such a configuration would produce H-deficient interacting SNe. There is indeed a sizable population of such Type Ibn SNe (e.g., SN\,2006jc; \citealt{pasto_06jc_07}) in which the interaction power dominates at all times, hence compatible with a low \nifs\ mass and a dominance of He in the composition.

For the comparison to observations, we have selected a few SNe IIb, Ib, and Ic with high-quality optical spectra around 200\,d after the inferred time of explosion. Many objects for which good data exist around the time of bolometric maximum have no data at late times, or the available data are of a quality unsuitable for a fruitful model comparison. Our selection includes the Type IIb SNe 1993J and 2011dh even though our model grid is limited to He-stars initially. This is acceptable since the presence of H in the outer ejecta may only lead to a broad boxy H$\alpha$ at such late times. We include the Type Ib SN 2007C, and the Type Ic SNe 2004aw, 2007gr, and 2013ge. A description of their characteristics and the associated references are given in Table~\ref{tab_obs}. The goal here is to provide a basic compatibility check with a restricted sample of observations.

We have performed many simulations in this study, using the x1p0, x1p5, and x2p0 series, and variants of these with different forms of clumping or scaled velocity. As our simulations retain some level of arbitrariness, and since we only compare to the data at a single epoch we do not aim to find the  model that yields the best match to a given observation.  Instead, we wish to find the parameter space that yields a satisfactory agreement for each SN. The \nifs\ mass for our set of models  clusters around 0.05\,$-$\,0.1\,\msun\ and is thus in the ballpark of the values inferred for most of the selected SNe. One exception is SN\,2004aw, which is overenergetic and likely asymmetric \citep{taubenberger_04aw_06}. Consequently, the amount of \nifs\ at low velocity and powering the light curve at late times may be much smaller than that estimated from phases near bolometric maximum \citep{dessart_98bw_17}.

As discussed in the preceding sections, the Fe emission between 4100 and 5500\,\AA\ arises  largely from the Fe present in the O-rich shell with an abundance below solar. Only a fraction of that emission arises from the Fe, produced from the decay of \nifs\ and \cofs\ nuclei, in the Fe/He shell. In other words, the strength of the Fe emission is poorly correlated with the \nifs\ mass produced during the explosion. Any mismatch of this Fe emission between models and observations is therefore unlikely to arise from an offset in \nifs\ mass but is more likely to arise from an inadequate ionization in the O-rich shell (possibly arising from an inadequate description of the density of this emitting material because of clumping, etc.). It might also indicate a different metallicity for the progenitor.

\begin{table*}
\caption{
Characteristics of the Type IIb, Ib, and Ic SNe sample used for comparison to our models at 200\,d.
\label{tab_obs}
}
\begin{center}
\begin{tabular}{
l@{\hspace{4mm}}
|r@{\hspace{4mm}}c@{\hspace{4mm}}c@{\hspace{4mm}}
c@{\hspace{4mm}}c@{\hspace{4mm}}c@{\hspace{4mm}}
c@{\hspace{4mm}}c@{\hspace{4mm}}c@{\hspace{4mm}}
c@{\hspace{4mm}}c@{\hspace{4mm}}c@{\hspace{4mm}}
}
\hline
        SN       &   Type  &  $D$      &  $\mu$   & $t_{\rm expl}$   &    $E(B-V)$    &    $z$   & \nifs\     & References  \\
\hline
                     &      &  [Mpc]  &     [mag]  &     MJD [d]           &     [mag]        &      &   [\msun] &   \\
\hline
SN\,1993J   &  IIb   & 3.63   & 27.80 &  49074.0 & 0.08  & -0.00015   & 0.09 & J15, L94, M00, R94, R96, W94\\
SN\,2011dh &      IIb &   7.8 &  29.46 & 55713.0 &   0.07 &    0.00164 & 0.075 & E14, J15 \\
\hline
SN\,2007C  &   Ib  & 25.0 & 31.99 &   54096.0  & 0.47 & 0.00561 & 0.06\,$-$\,0.16 &    D11, P16, M20 \\
\hline
SN\,2004aw &  Ic &  68.2 &  34.17 & 53073.0 &   0.37 &    0.0160 & 0.15\,$-$\,0.30  & D20, M17, T06  \\
SN\,2007gr &   Ic &  9.3   &  29.84 & 54325.5 &   0.09 &    0.0017 &  0.076                   &  H07 \\
SN\,2013ge  & Ic &   23.7 &  31.87  & 56604.0  & 0.07 &  0.00436 &  0.109                 & D16, P16 \\
\hline
\end{tabular}
\end{center}
{\bf Notes:} D11: \citet{drout_11_ibc}; D16: \citet{drout_13ge_16}; D20: \citet{dessart_snibc_20}; E14: \citet{ergon_14_11dh}; H07: \citet{hunter_07gr}; J15: \citet{jerkstrand_15_iib}; L94: \citet{lewis_93j_94}; M00: \citet{matheson_93j_00a}; M17: \citet{mazzali_04aw_17}; M20: \citet{meza_anderson_ni56_20}; P16: \citet{prentice_ibc_16}; R94: \citet{richmond_93j_94}; R96: \citet{richmond_93j_96}; T06: \citet{taubenberger_04aw_06}; W94: \citet{woosley_94_93j}.
\end{table*}

\subsection{SN\,1993J}
\label{sect_93J}

\begin{figure}
\centering
\includegraphics[width=\hsize]{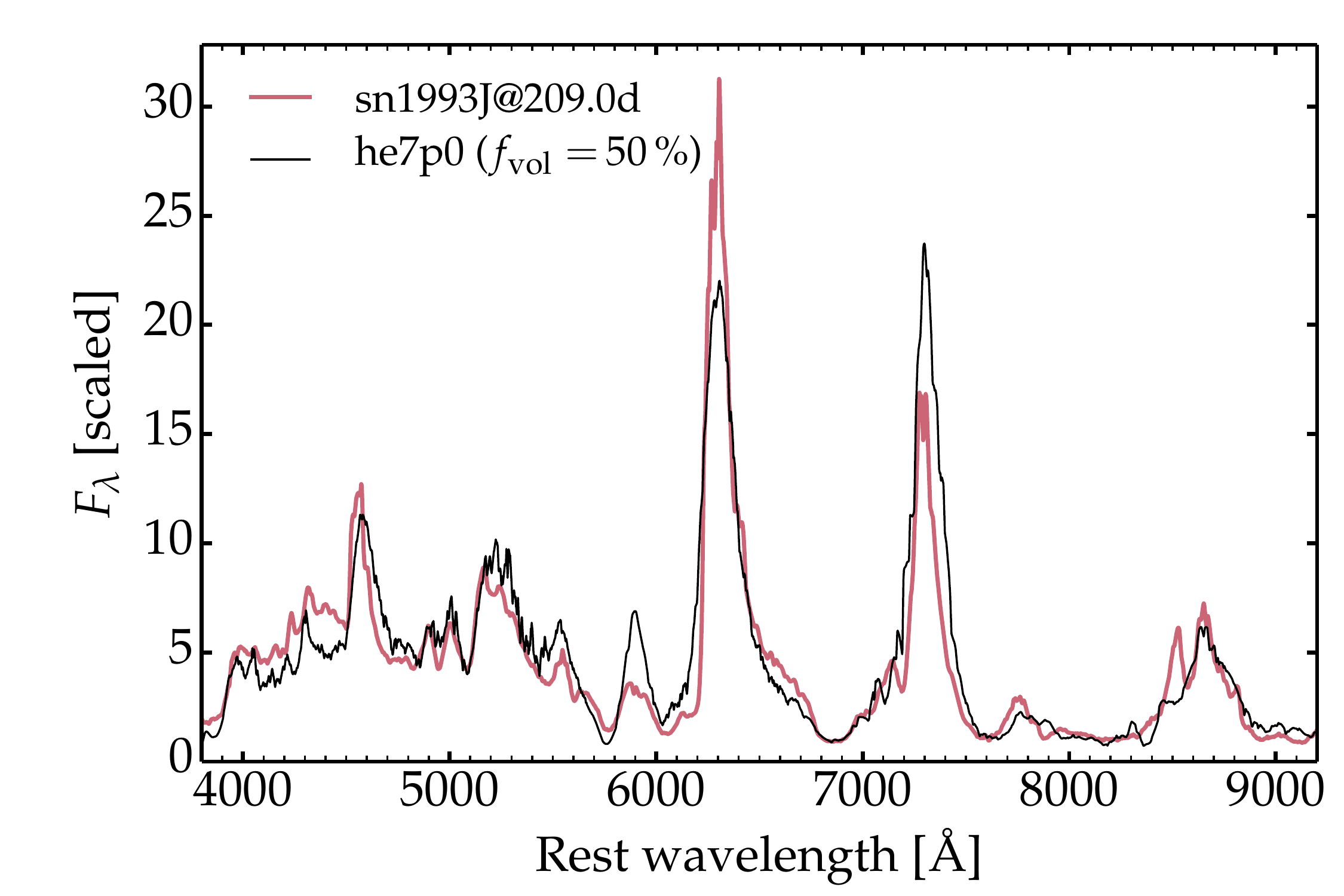}
\vspace{-0.2cm}
\caption{Comparison of the observations of SN\,1993J at 209\,d with the He-star explosion model he7p0 at 200\,d with a uniform volume filling factor of 50\,\%. Model and observation are normalized at 6900\,\AA. No H$\alpha$ contribution is included in the model, so that the red shoulder of the \oidoub\ doublet is due primarily to \niidoub.
\label{fig_93J}
}
\end{figure}

Figure~\ref{fig_93J} presents a comparison of SN\,1993J at 209\,d after explosion with the model he7p0 at 200\,d that assumes a 50\% volume filling factor. The model reproduces satisfactorily the Fe emission in the blue part of the optical, the strength of \oidoub, and the Ca\two\ near-infrared triplet. The \caiidoub\ doublet is overestimated, which may in part be due to an overestimate of the contribution from [Ni\,{\sc ii}]\,$\lambda$\,7378. The model also overestimates \nad, which could arise from a slight underestimate of the ionization in the O-rich shell. The red shoulder of the \oidoub\ line is due to \niidoub\ in the model, while in observations some contribution from H$\alpha$, omitted in the model, could be present.

\citet{jerkstrand_15_iib} presented a comparison between the 283\,d observation of SN\,1993J and a Type II SN model from the explosion of a single-star of 17\,\msun\ on the main sequence but excluding most of the H-rich layers.  Their model overestimates \oidoub\ but agrees otherwise with the observations at a level that is comparable to our fit. An important distinction is that their model corresponds to a trimmed Type II SN model, scaled to match the observed spectral line widths of SN\,1993J, while our model is not tuned (with the exception for clumping, which is also used by \citealt{jerkstrand_15_iib}).

Our he7p0 model, which most likely results from binary evolution, is H deficient, and corresponds to a star of 25.7\,\msun\ on the ZAMS \citep{woosley_he_19}. It produces the same O yield (i.e., 1.3\,\msun) as the single star model of 17\,\msun\ on the ZAMS used by Jerkstrand. The pre-SN mass of model he7p0 was 5.04\,\msun\ and the
    explosion energy $1.38 \times 10^{51}$\,erg. The 17\,\msun\ model of \citet{WH07} used by
    Jerkstrand had a helium core mass of 5.02\,\msun\ and an explosion
    energy of $1.2 \times 10^{51}$\,erg. One expects the explosion
    outcome, including the nucleosynthesis, to be similar if the
    hydrogen envelope is almost completely removed from the 17\,\msun\
    model before the shock passes through.

Inferences from pre-explosion imaging indicate that the progenitor of SN\,1993J was a G8I-K5I supergiant, and most likely double \citep{aldering_93j_94}. The companion star was likely detected by HST post-explosion imaging, and is likely to be an early B supergiant \citep{maund_93j_04}.   Lacking a suitable binary model, it might be more suitable to invoke a 17\,\msun\ progenitor, which is also more in line with previous inferences from light curve modeling (see, for example, \citealt{blinnikov_94_93j}; \citealt{woosley_94_93j}), and the binary model (15 and 14\,\msun) proposed by \citet{maund_93j_04}. Lower mass progenitors (e.g., a ZAMS mass of 13\,\msun) are inconsistent with the nebular-phase properties. In our grid of models, the \oidoub\ line flux is stronger than  \caiidoub\ (i.e., as observed in SN\,1993J) only for He-star masses above about 8\,\msun -- less massive models exhibit weak \oidoub\ and strong \caiidoub\ emission.

\subsection{SN\,2011dh}

\begin{figure*}
\centering
\includegraphics[width=0.495\hsize]{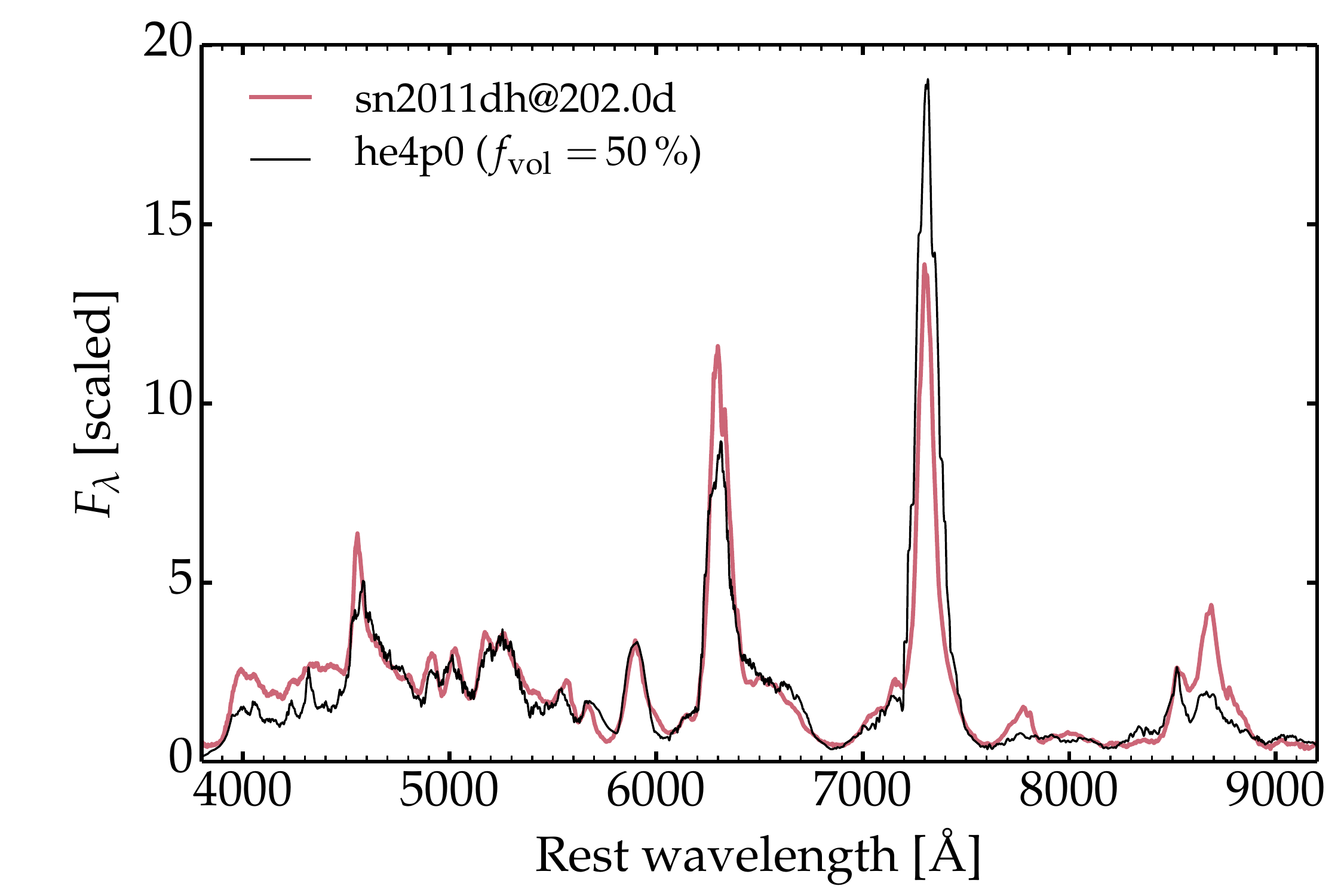}
\includegraphics[width=0.495\hsize]{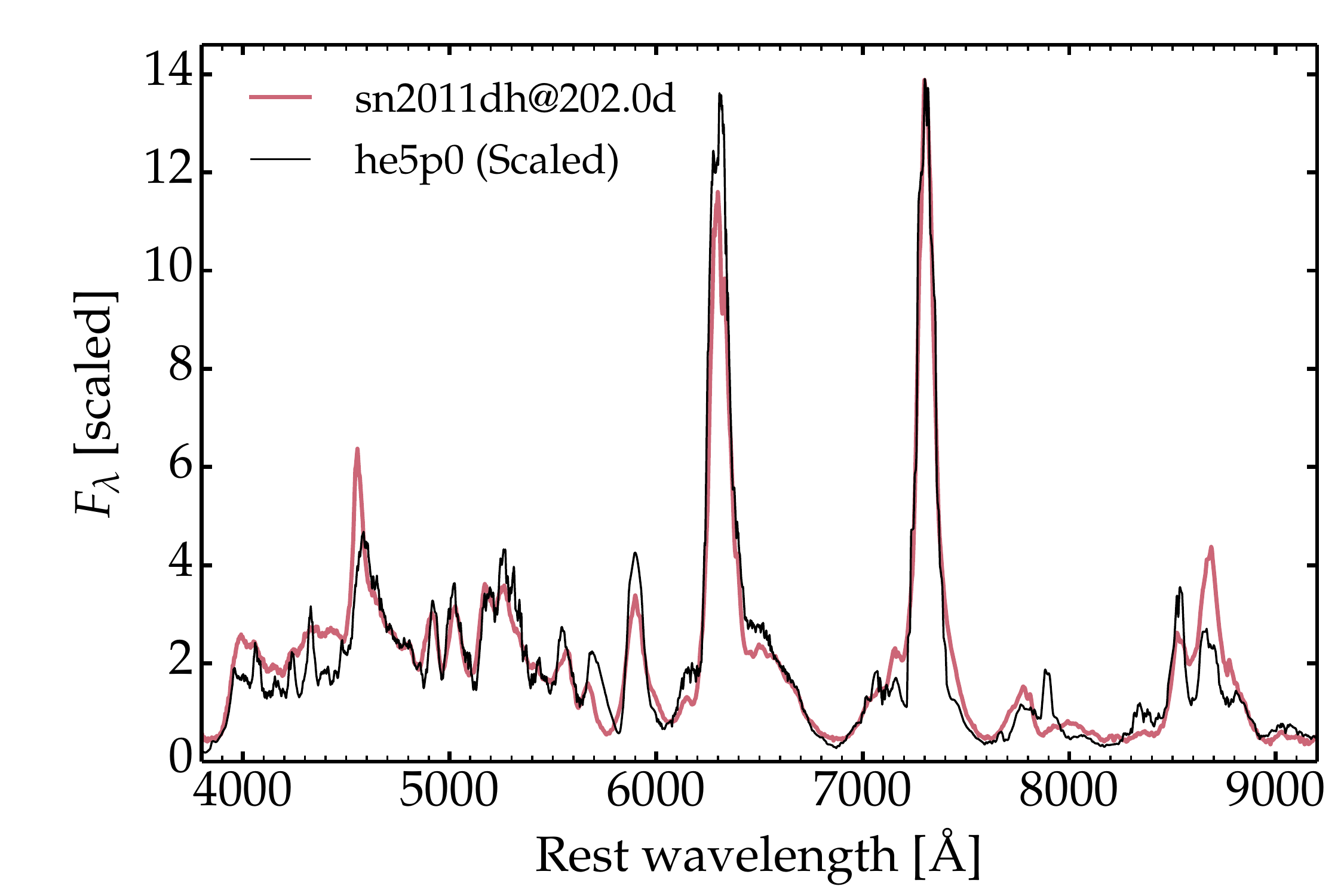}
\vspace{-0.2cm}
\caption{Left: Comparison of the observations of SN\,2011dh at 202\,d with the He-star explosion model he4p0 at 200\,d with a uniform volume filling factor of 50\,\%. Right: Same as left, but now using the he5p0 model but scaled down in $E_{\rm kin}$ and \nifs\ mass, and excluding bound-bound transitions from Ni\two. Model and observation are normalized at 7000\,\AA.
\label{fig_11dh}
}
\end{figure*}

The left panel of Fig.~\ref{fig_11dh} shows a comparison between the observations of Type IIb SN\,2011dh at 202\,d and the model he4p0 at 200\,d with a uniform volume filling factor of 50\%. This model reproduces relatively well the Fe\two\ emission in the blue part of the spectrum, the \nad\ doublet line, the strength and shape of \oidoub, and the strong bump at 6500\,\AA\ which we associate with \niidoub. The \caiidoub\ line is however overestimated and too broad, and the O\one\,7774\,\AA\ triplet and the Ca\two\ near-infrared triplet are underestimated.

The overestimate of the \caiidoub\ doublet strength and width may be due to a combination of an overestimate in the expansion rate and the contribution from [Ni\,{\sc ii}]\,$\lambda$\,7378. To further explore this object, and to test the sensitivity of our results to ejecta properties, we recalculated the model he5p0 but scaled down the ejecta kinetic energy by a factor of 2.0 and the \nifs\ mass by a factor of 2.62. In practice, we turn this excess \nifs\ into \nife\ to preserve the normalization of mass fractions to unity. The reduction in \nifs\ mass together with the greater $\gamma$-ray trapping efficiency keeps the luminosity the same as the unscaled model. However, the abundance of stable Ni is boosted, so that lines of Ni become much stronger in the model. Hence, in the right panel of Fig.~\ref{fig_11dh}, we show the scaled he5p0 model with Ni\two\ bound-bound transitions switched off. The resulting model yields a satisfactory match to the observations of SN\,2011dh at 202\,d. The \caiidoub\ doublet is now well-fitted, in both strength and width. The O\one\,7774\,\AA\ triplet line is also predicted, arising from the greater ejecta density. A major complexity in modeling SNe Ibc is that a different ejecta kinetic energy leads to different densities, which affects numerous properties (density squared processes, optical depth to $\gamma$-rays and low-energy photons, etc.).  Despite these shortcomings, a low- to moderate-mass He-star model seems suitable to explain the properties of SNe\,2011dh, in agreement with previous inferences (see, for example,  \citealt{bersten_etal_12_11dh}, \citealt{ergon_11dh_15}, \citealt{jerkstrand_15_iib}).

\subsection{SN\,2007C}
\label{sect_07C}

\begin{figure}
\centering
\includegraphics[width=\hsize]{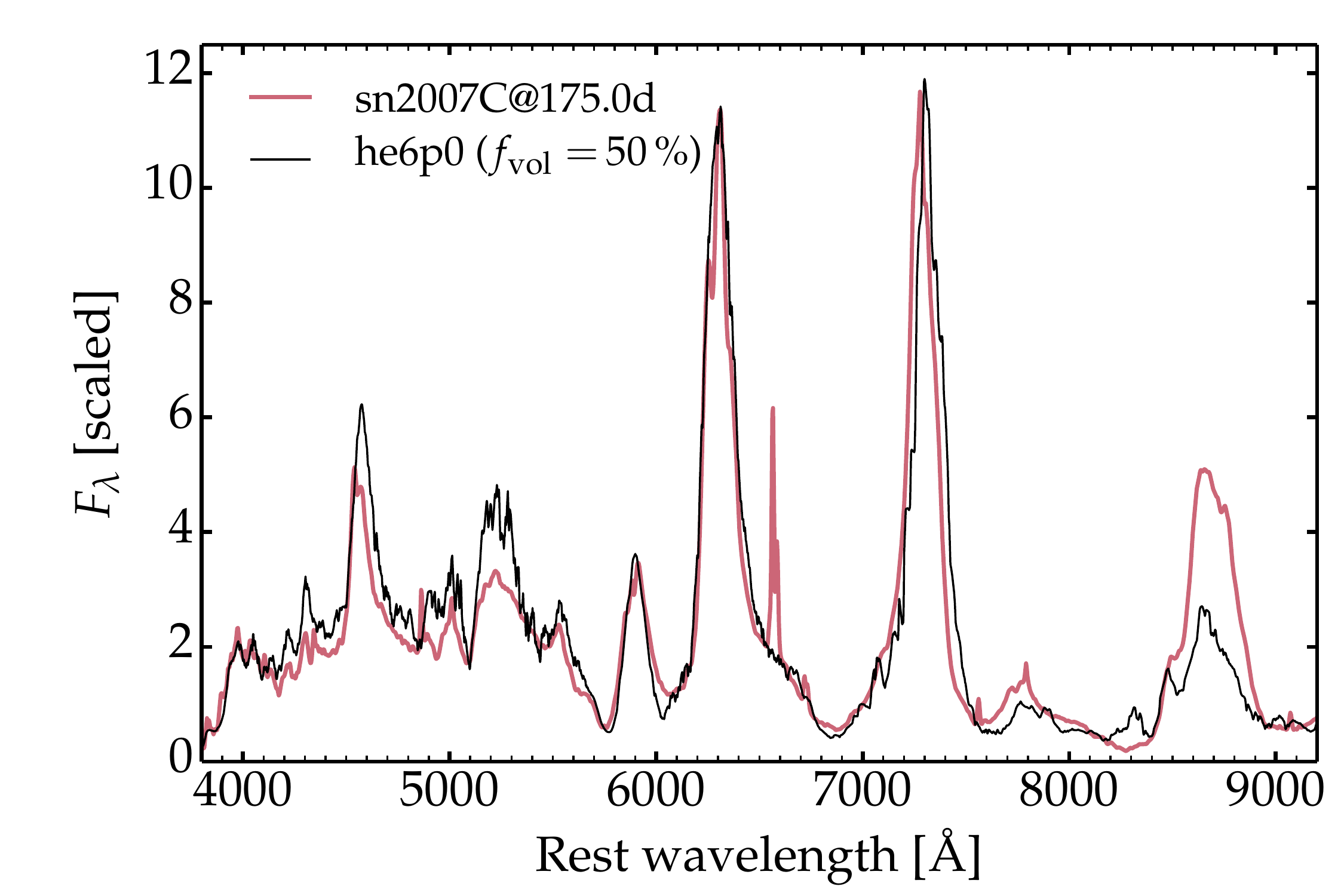}
\vspace{-0.2cm}
\caption{Comparison between the observations of SN\,2007C at 175\,d and the model he6p0 at 200\,d with a uniform volume filling factor of 50\,\%. Model and observation are normalized at 7000\,\AA.
\label{fig_07C}
}
\end{figure}

Figure~\ref{fig_07C} shows a comparison between the observations of the Type Ib SN\,2007C at 175\,d after explosion and model he6p0 at 200\,d with a uniform volume filling factor of 50\,\%. Most lines are reproduced satisfactorily, in particular the near equal strength between \oidoub\ and \caiidoub, the sizable Fe emission in the blue part of the optical, the \mgi\ line and \nad. Not well matched are the O\one\,7774\,\AA\ triplet and the Ca\two\ near-infrared triplet. Furthermore, the observations exhibit a blueshift of the peak emission for \mgi\ and \caiidoub\ (by about 700 to 1000\,\kms, respectively), while the model predicts a central peak at the rest wavelength. Such wavelength shifts are notorious in SNe Ibc \citep{taubenberger_ibc_09,milisavljevic_etal_10}, and may require the presence of asymmetry or dust attenuation. Both effects are ignored in the present work.

\subsection{SN\,2004aw}
\label{sect_04aw}

\begin{figure}
\centering
\includegraphics[width=\hsize]{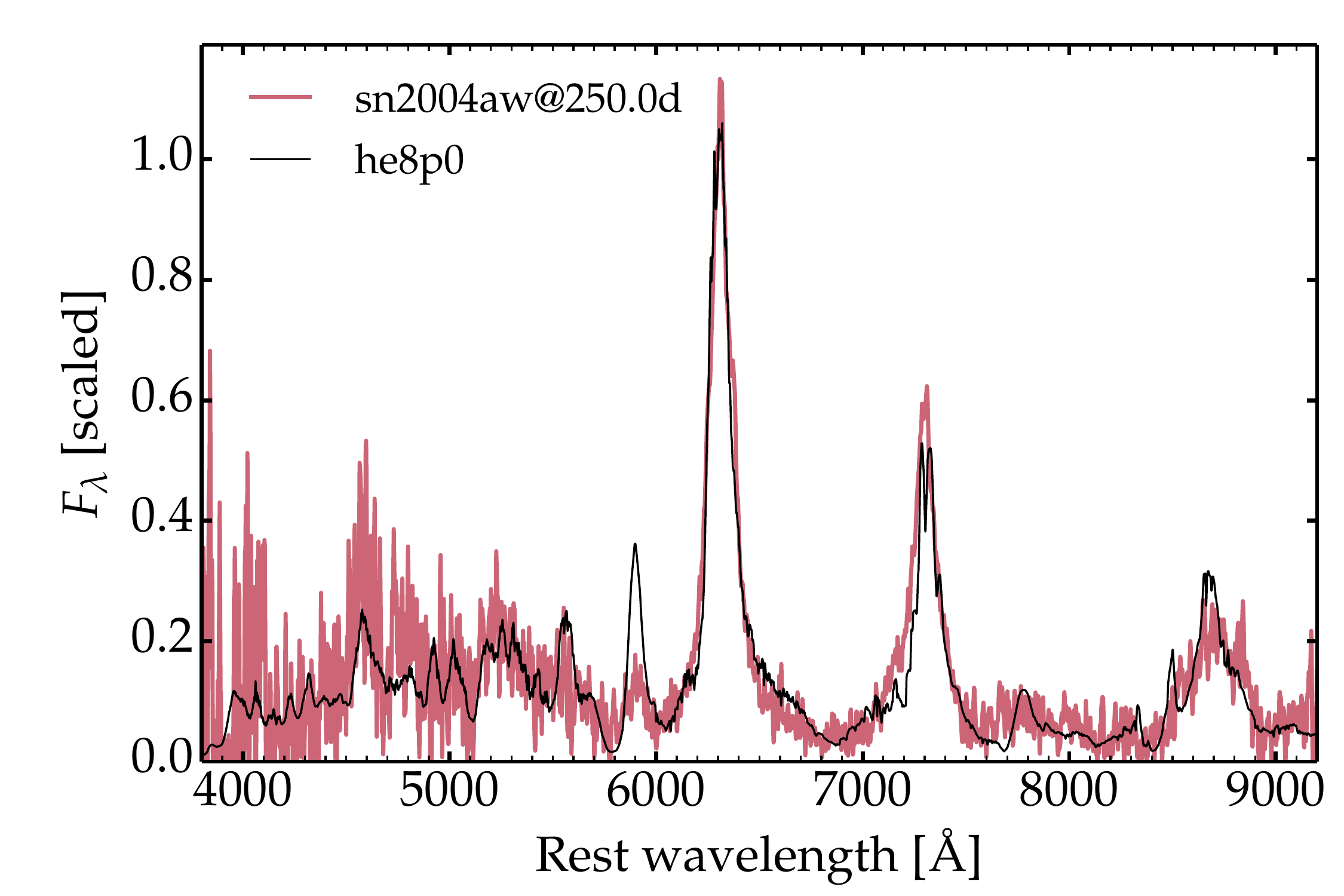}
\vspace{-0.2cm}
\caption{Comparison between the Type Ic SN\,2004aw at 250\,d with the model he8p0 at 200\,d. Model and observation are normalized at 6300\,\AA.
\label{fig_04aw}
}
\end{figure}

Figure~\ref{fig_04aw} presents a comparison between the Type Ic SN\,2004aw at 250\,d with the model he8p0 at 200\,d. The data is of poor quality but this is a good example where a model without any tuning (no clumping, etc.) yields a satisfactory match to an observation (despite the time offset of 50\,d). The model reproduces \mgi, \oidoub, \caiidoub, and the Ca\two\ near-infrared triplet, but it overpredicts the [O\one]\,5577.3\,\AA\ and \nad. SN\,2004aw  is overenergetic and likely asymmetric \citep{taubenberger_04aw_06}, so this comparison to model he8p0, which is both underenergetic and spherical, may be questionable. However, it demonstrates that energetic SNe Ic are probably systematically asymmetric, so that the material emitting at late times is both slow and dense (as in an underenergetic explosion) and has little to do with the high velocity material that contributed at the time of bolometric maximum \citep{dessart_98bw_17}. Similar conclusions were reached by \citet{mazzali_04aw_17}.

\subsection{SN\,2007gr and SN\,2013ge}
\label{sect_13ge}

\begin{figure*}
\centering
\includegraphics[width=0.495\hsize]{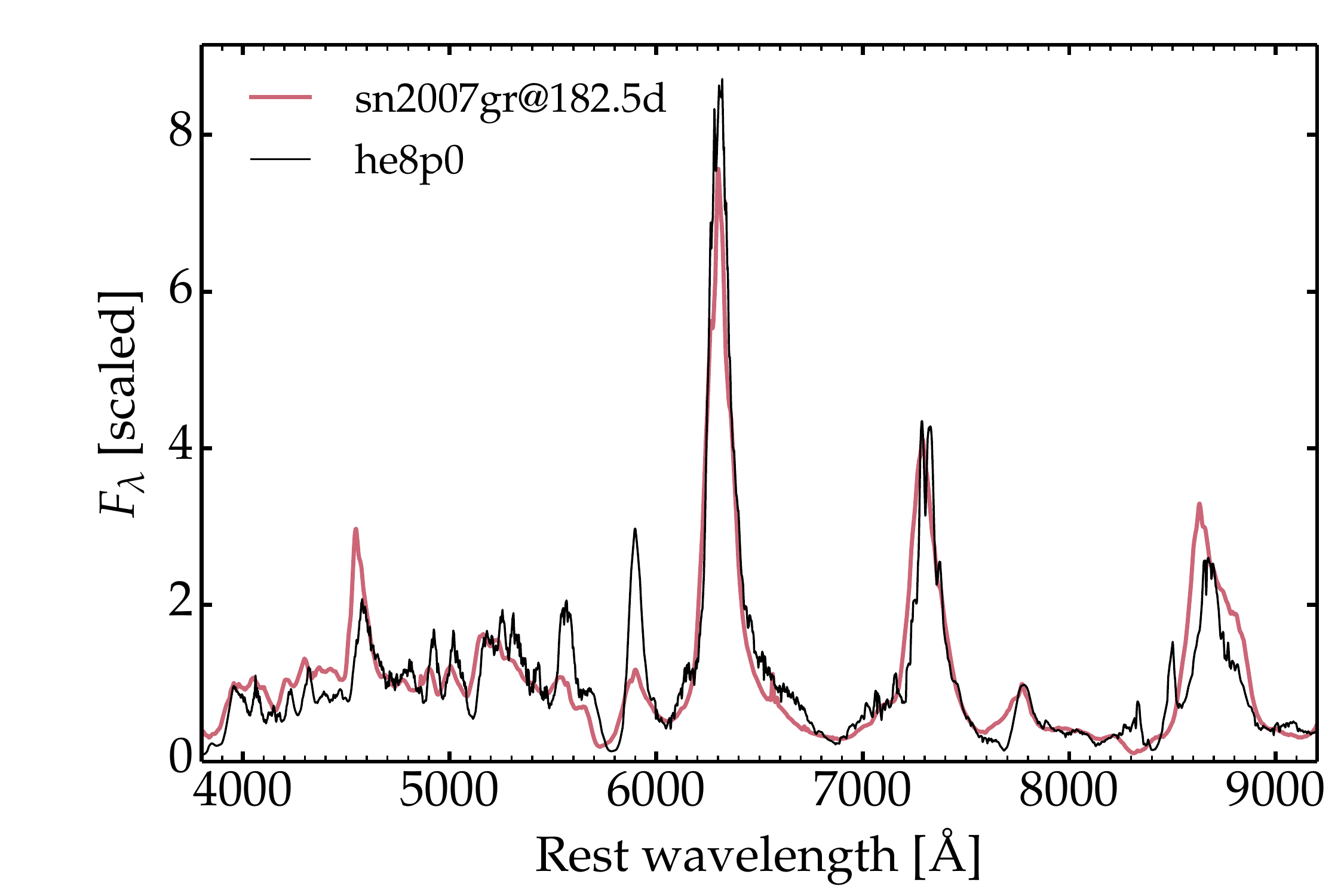}
\includegraphics[width=0.495\hsize]{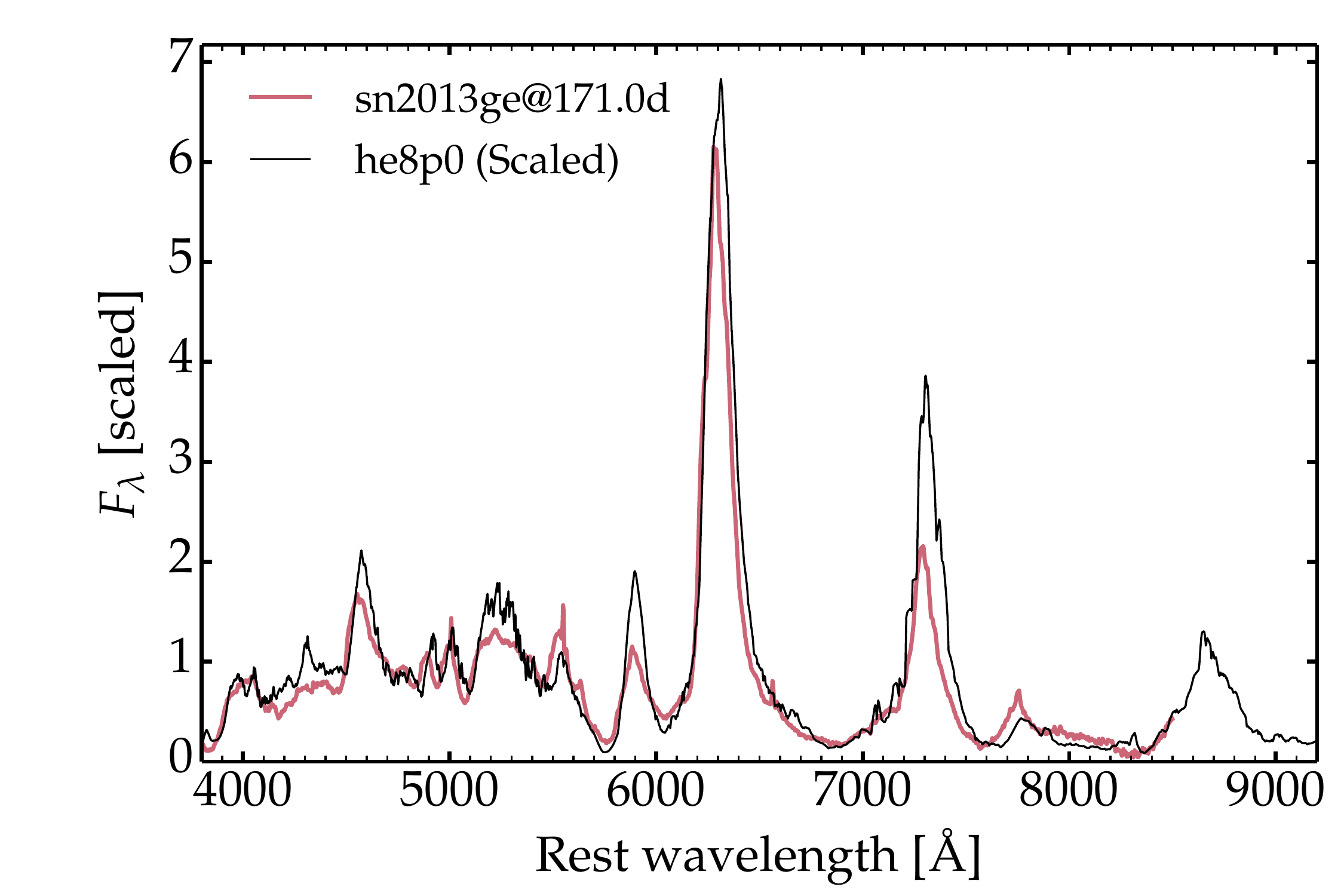}
\caption{Left: Comparison between the Type Ic SN\,2007gr at 182.5\,d with the model he8p0 at 200\,d. Right: Comparison between SN\,2013ge at 171\,d and the scaled he8p0 model at 200\,d. Model and observation are normalized at 5400\,\AA.
\label{fig_07gr_13ge}
}
\end{figure*}

The left panel of Fig.~\ref{fig_07gr_13ge} presents a comparison between the Type Ic SN\,2007gr at 182.5\,d with the model he8p0 at 200\,d. This is the same model used for SN\,2004aw. The right panel of Fig.~\ref{fig_07gr_13ge} presents a comparison between the Type Ic SN\,2013ge at 182.5\,d with the model he8p0 at 200\,d but now scaled up in $E_{\rm kin}$ by a factor of two. The optical spectra of SNe\,2007gr and 2013ge are very similar, and two models (he8p0, and the scaled he8p0) yield a similar result. Both models underestimate the width of most lines, in particular the presence of emission or absorption at the largest velocities (e.g., for \nad\ in SN\,2007gr), while the offset to the blue of emission peaks (e.g., \mgi) is not predicted by our models. A similar blueshift is observed in SN\,2007C (see Sect.\ref{sect_07C}).

Thus all three SNe Ic we have discussed require a higher mass model in order to explain the stronger \oidoub\ relative to \caiidoub. Conversely,  model he7p0 and lower mass models are suitable for explaining objects in which  \caiidoub\ is  at least as strong as \oidoub. This trend confirms the accepted notion than SNe Ic arise from more massive progenitors than SNe Ib and IIb
\cite[e.g.,][]{anderson_env_ccsn_15,aramyan_dist_16,maund_snibc_18}, although this result from nebular phase spectral modeling is subject to significant uncertainties associated with the adopted ejecta structure, the clumping, and the formation of molecules.

\section{Conclusions}
\label{sect_conc}

We have presented nebular-phase nonLTE steady-state and
time-dependent radiative transfer calculations with \cmfgen\ based on
the He-star explosion models of \citet{ertl_ibc_20}. These ejecta
result from ab~initio 1D neutrino-driven explosions calibrated
according to a prescribed ``explosion engine''. They utilize the
He-star models of \citet{woosley_he_19} that were first evolved to the
onset of core collapse. We consider several sets of progenitors
evolved with the nominal mass loss rate (x1p0 series), and variants in
which the mass loss rate is enhanced by 50\,\% (x1p5 series) and by
100\,\% (only one model done for a 10\,\msun\ progenitor He-star). To
handle macroscopic mixing without introducing microscopic mixing, we
implement the shuffled-shell technique of \citet{DH20_shuffle},
limiting the shuffling to regions interior to the outermost dominant
shell (i.e., the He/N shell in Type Ib SN models and the He/C or
He/C/O shell in Type Ic SN models). Limiting the shuffling to within
the CO core yields similar results.

Relative to a standard-energy Type II SN, the main differences in
nebular-phase spectra of a standard-energy SN Ibc arise from the
higher ejecta expansion rate and the lack of a massive H-rich ejecta
layer that reprocesses the radiation emitted from the metal-rich
regions. The decay power absorbed by the ejecta is very sensitive to
the ejecta dynamics and structure, and not just to the \nifs\ mass
produced in the explosion. Because of their higher expansion rate and
lower density, SN Ibc ejecta are typically more ionized than SN II
ejecta. Consequently, the SN Ibc radiation is more sensitive to the
ejecta density structure, the expansion rate, clumping, and probably
molecules (which may form, and at earlier times than in SNe II \citep{rho_20oi_21};
see, however, the case of SN\,2012au for a counterexample \citep{pandey_12au_21}).
In this paper, we have investigated a few of
these effects but much remains to be explored. An important caveat is
that our approach is 1D -- departures from spherical symmetry, which
will modify the density and the species' distribution, can strongly
affect the radiation from SNe Ibc.

Our selection of models covers initial He-star masses spanning the
range from 2.6 to 13\,\msun\ that, depending on the adopted pre-SN
mass loss recipe, give rise to ejecta masses between 0.79 and
5.32\,\msun. They exhibit a wide variety of nebular phase properties
at 200\,d after explosion. Progenitors with a residual He-rich shell
(i.e., the He/N shell), that would be unambiguously classified as a
Type Ib, yield optical spectra with strong emission below
5500\,\AA\ due to Fe\two, a relatively strong and broad
\niidoub\ arising from the outer ejecta, and \caiidoub\ stronger than
\oidoub. The relatively low O yield limits the maximum possible
\oidoub\ line flux, however, in our models another limiting factor is
the overionization of O in the O-rich shell for all the lower mass
ejecta. Clumping of the O-rich material can boost the \oidoub\ line
strength, but it remains weaker than \caiidoub\ in all He-star models
with an initial mass below $\sim$\,6\,\msun. Explosions of
  pre-SN stars with masses 3.15\,\msun\ to 4.45\,\msun, which
  correspond for the fiducial mass loss rate to He stars with initial
  masses 4.0 to 6.0 \msun\ are broadly compatible with the
nebular-phase spectra of some SNe IIb and Ib like 2011dh and
2007C. Helium stars with these masses correspond to ZAMS masses in the
range 18\,$-$\,23\,\msun.

We find no observational counterparts at nebular times for our
lighter models (i.e., initial He-star masses below 4\,\msun;
  pre-SN masses below 3.15 \msun), which may be an
observational bias (such objects are not detected and we have a too
small sample of nebular-phase spectra for SNe Ib). From a theoretical
point of view, there is no reason to expect a lack of SNe
Ib with a \nifs\ mass of order 0.001\msun\ when many such counterparts
are found in the population of low-luminosity SNe II
\citep{pasto_low_lum_04}.  An alternative is that such low-mass He
stars undergo nuclear flashes in the final phases of
  evolution prior to core collapse (see discussion in
\citealt{woosley_he_19}) and contribute to the population of SNe Ibn
like SN\,2006jc \citep{pasto_06jc_07}. Each scenario may occur in
nature.

As the initial He-star mass is increased above 6\,\msun\ (pre-SN mass of 4.45\,\msun),
the O yields and $M_{\rm ej}$ increase, but the $E_{\rm kin}$ levels off or even decreases. In those
models, the \oidoub\ is of comparable strength or stronger than
\caiidoub, the Fe\two\ emission below 5500\,\AA\ is weaker, and the
optical spectrum closely resembles observed SNe Ic like SNe\,2007gr
and 2013ge.

Variations in expansion rate affect the width of lines, which are
primarily Doppler broadened, but the main impact of variations in
$E_{\rm kin}$ is to modulate the ejecta density and its ionization, in
particular in the O-rich zones. When O has a low ionization,
\oidoub\ is the main coolant for that region in our present
simulations. If the O ionization drops to zero (i.e., O is neutral),
this tends to boost the strength of \mgi\ and \nad. If the O
ionization is raised (with a mixture of neutral O and O$^+$), the
\oidoub\ weakens and the Fe\two\ emission below
5500\,\AA\ strengthens. Reducing the $E_{\rm kin}$, introducing
clumping in the O-rich zones, or adopting a larger SN age leads
  to a relative strengthening of \oidoub. The spectral consequences
of modulations in $E_{\rm kin}$ are difficult to anticipate since they
alter both the ejecta density (which affects density-squared
processes, critical densities, optical depth, etc.) and the total
decay power absorbed (since it affects the $\gamma$-ray mean free
path).

Although there is considerable scatter in the radiation properties of
our nominal mass loss SN models at 200\,d, three classes emerge. For
initial He masses less than 6\,\msun, the \oidoub\ is
systematically weaker than the \caiidoub, while for masses
$\gtrsim$\,7\,\msun, emission in \oidoub\ dominates that in \caiidoub.
At intermediate masses (6 to 7\,\msun) the \oidoub\ and
\caiidoub\ emission are comparable.  Unfortunately, the \oidoub\ is
highly sensitive to the O ionization in the O-rich shell, and
consequently the connection between the \oidoub\ line strength and O
yield is complicated. Variations in pre-explosion mass loss, ejecta
clumping, and the energetic properties of the SN, can significantly
influence the \oidoub\ line strength.

 We surmise that the density structure of our 1D explosion models is
 likely responsible for some of the disagreements with
 observations. Spherical explosion models cannot predict clumping, and
 they yield a density structure that differs significantly from the
 predictions of 3D explosions. Angular averages are unlikely to be
 suitable alternatives because much of the radiative transfer depends
 on nonlinear processes. While this matters little for the
 nebular phase properties of SNe II (i.e., because O is nearly neutral
 in the O-rich shell in most SNe II; \citealt{D21_sn2p_neb}), this
 shortcoming is a major concern in SNe Ibc.

We generally assumed that positrons are absorbed where they
are emitted. Relaxing this assumption by assigning the same
mean-free path to positrons and $\gamma$ rays (which is an extreme postulate)
leads to a reduction of the emission from the Fe/He shell, causing a
reduction of the Fe\two\ emission in the range 4100 to 5500\,\AA. Some
specific lines forming exclusively in the Fe/He shell are also
affected, such as [Fe\three]\,$\lambda$\,4658\,\AA,
[Co\three]\,$\lambda$\,5889, [Ni\two]\,$\lambda$\,7378, or
[Ni\three]\,$\lambda$\,7889. Because the material from the Fe/He shell
may be distributed throughout a large ejecta volume, the associated
weak emission from these lines may be spread in velocity space (and
thus in wavelength) and therefore hard to identify in observations.

An important limitation to our simulations is the neglect of
molecular cooling, especially in view of the often partial ionization
of O in the O-rich shell. Indeed, the formation of CO in the O/C shell
and of SiO in the O/Si shell is expected to alter the ionization and
temperature of the corresponding material. While it would take away
some of the available power for the O-rich material, it could also
significantly affect some line emissions through a change in
ionization and temperature. Observations of many well observed SNe Ibc suggest that
molecule formation may take place soon after bolometric maximum,
sometimes as soon as the ejecta turns optically thin. Molecule
formation has been reported, for example, for SN\,2007gr
\citep{hunter_07gr}, SN\,2011dh \citep{ergon_11dh_15}, SN\,2013ge
\citep{drout_13ge_16}, and SN\,2020oi \citep{rho_20oi_21}. The
importance of molecular cooling is well known \citep{liljegren_co_20}.
CO molecule formation is, however, not systematic \citep{pandey_12au_21}.

Our He-star models, likely arising from binary mass exchange, are
evolved with wind mass loss (which directly peels off the outer
layers, reduces the total mass, and shrinks the size of the burning
core). Their metal composition at stellar death, just before core
collapse, is comparable to the cores of single stars of
much lower mass on the ZAMS. For example, model he6p0 corresponds to a
ZAMS mass of 23.3\,\msun\ but has roughly the same O yield as
a single 15.2\,\msun\ star from \citet{sukhbold_ccsn_16} that dies
with a massive H-rich envelope and produces a SN II-P. One should thus
be cautious when comparing He-star models (directly affected by wind
mass loss) with the ``He-core'' of single progenitor stars leading to
SNe II (which are hardly affected, and if so only indirectly, by
red-supergiant mass loss). The assumption of prompt removal of the
H-rich envelope at helium ignition in our He-star progenitor
models is certainly too simplistic to explain all manner of
  supernovae in stripped stars. It is in tension, for example, with
the Type IIb SN progenitors that are notorious for having
residual H at the time of their explosion (which testifies that
wind mass loss never affected directly the H-deficient layers).
  Presumably in these systems, the binary properties were such that
  the mass exchange happened very late and/or was
  inefficient. Nevertheless, assuming full loss of the envelope at the
  time the star first becomes a supergiant seems a reasonable
  approximation for forming a common SN Ibc. The approximation is
  partly justified because the predictions it makes for remnant masses,
  explosion energies, ejecta masses, and light curves are in
  good accord with observations \citep{ertl_ibc_20,woosley_ibc_21}.
  Considerable uncertainty is also subsumed into the poorly known
  mass loss rate.

He-star models have a more modest metal yield at the time of core
  collapse relative to single star models with the same ZAMS mass. We
  find, however, that the nebular phase properties of the SN Ibc they
  produce often suggest similar O yields as SNe II. Hence, the
  inference of a similar O yield from a nebular-phase spectrum in a SN
  II and a SN Ibc would indicate a larger ZAMS mass for the
  latter. This offset may explain the association of SNe Ibc with
higher mass progenitors
\citep{anderson_env_ccsn_15,kuncarayakti_env_18,maund_snibc_18}.

One class of nebular observations that our models disagree with is
those showing strong \mgi\ emission. An example of such a supernova is
the Type Ic SN 2002ap \citep[e.g.,][]{taubenberger_ibc_09}. In SN
2002ap, \mgi\ can be readily identified in the spectrum at day 130
(post explosion), and becomes more prominent with time as the Fe
forest disappears \citep{mazzali_02ap_02}. Its peak intensity is
second only to \oidoub\ at this and later times. The early prominence
of \mgi\ in SN 2002ap may be due to its high $E_{\rm kin}/M_{\rm ej}$,
which fosters a lower density faster expanding ejecta -- SN 2002ap was
classified by \cite{kinugasa_02ap_02} as a hypernova. This idea is
consistent with the evolution of \mgi\ in our time-dependent
models in which \mgi\ is much more distinct/prominent at 482\,d than
at 200\,d. It is also consistent with the spectral evolution seen for
SN 2011dh - at 328\,d its central intensity rivals that of
\oidoub\ and \caiidoub\ \citep{shivvers_11dh_13} -- this can be
contrasted with its strength at 202 days as shown in
Fig.~\ref{fig_11dh}. Since the models used here would have great
  difficulty developing greater kinetic energies in a neutrino-powered
  paradigm, this rapid expansion may be yet another indicator of a
  nonstandard central engine for these sorts of events.

The wide diversity of spectral properties from our grid of spherical
He-star explosion models contrasts with the relative uniformity of
observed properties of SNe Ibc at nebular times. This could
  actually be an indicator of as yet unexplored diversity in
  nature. Much of the diversity in models arises because of
changes in ionization that result from changes in the density
structure, even for constant nucleosynthetic yields.
Numerous models, in particular those with lower
initial He-star masses, show a larger ionization and stronger flux in
the blue part of the optical range than observed. The subset of these
ejecta that are also characterized by a low ejecta mass, low kinetic
energy, and low \nifs\ mass, are probably missed by current surveys so
the disagreement with the existing dataset may be an observational
bias.

We find that most observations (here limited to few of the best
observed SNe Ibc) can be explained by using only a few of our He-star
explosion models (between models he4p0 and he8p0), though often with an
adjustment to the density structure (e.g., by introducing
clumping or scaling the kinetic energy). There is
much degeneracy relative to yields when the ionization conditions are
similar: this is seen for models he8p0 and he12p0, which have nearly
identical spectra although model he12p0 has nearly twice as much O
than model he8p0. This degeneracy is illustrated in observations by
the similarity of nebular phase spectra from GRB/SNe Ic and
super-luminous SNe Ic, which are thought to arise from different
explosion mechanisms and probably different progenitors
\citep{jerkstrand_slsnic_17,d19_slsn_ic}, or from the Type IIb
SN\,1993J and the Type Ic SN\,2013ge (see Sects.~\ref{sect_93J} and
\ref{sect_13ge}). Some process must be at work in nature to produce
the degeneracy of nebular-phase spectra of SNe Ibc. We can speculate
that the observed degeneracy in SN Ibc nebular phase spectra arises
from the fact that at late times we probe the inner dense clumped
O-rich (and likely asymmetric) ejecta, while the more complex
composition and density distribution affect the outer ejecta and can
only be probed during the photospheric phase. Understanding this
degeneracy requires further work.

Based on their observed luminosity, typical SNe Ibc are
  sometimes inferred to produce more \nifs\ than SNe II-Plateau
  \citep{drout_11_ibc,lyman_ibc_16,prentice_ibc_16,anderson_nifs_19,meza_anderson_ni56_20},
  though this is not true of the models. While this offset seems real,
  it is exacerbated by the absence of SNe Ibc with a low \nifs\ mass,
  a potential overestimate of the reddening, and the potential
  bridging between SNe Ibc and SN Ic broad-lined and GRB/SNe (in which
  a different mechanism of explosion is at work, likely tied to fast
  progenitor-core rotation). Indeed, ab~initio explosion models of SNe
  Ibc suggest a maximum \nifs\ mass production of about
  0.15\,\msun\ and a maximum explosion energy of about $2 \times
  10^{51}$\,erg \citep{ertl_ibc_20}. Although much needed, inferring
  the explosion energy of an observed SN Ibc is difficult because
  there is a degeneracy in light curve and spectra with $E_{\rm kin} /
  M_{\rm ej}$ \citep[see Sect.~6 of][]{dessart_snibc_20}. Furthermore,
  determining the original \nifs\ mass synthesized in SNe Ibc from
  observations at late times is difficult because of the
  incomplete $\gamma$-ray trapping (also influenced by asymmetry, and
  perhaps by clumping and inhomogeneities) and because a fraction of
  the Fe emission at late times arises from the primordial Fe rather
  than from the decayed \nifs\ and \cofs\ (for example, the Fe in the
  O-rich shell, as well as in the He-rich shell in SNe Ib, is
  primordial).  So, it is hard to infer the original \nifs\ mass from
  either the photometry or the spectra, although it would be useful to
  know how far off such inferences would be. This deserves a careful
  study of its own.

Future work will be devoted to improving the physical realism of our
simulations, and to performing multiepoch studies. We first need to
implement molecular cooling in the ejecta, particularly for dominant
molecules like CO and SiO.  Performing multiepoch simulations is also
crucial, since a model consistent at one epoch may not by consistent
at another epoch, and this will lead to insights into deficiencies
into both the models and ejecta structure.

Another effort will be to base the calculations on a realistic density
structure for the ejecta, as obtained in 3D explosion simulations
(see, for example, the results for the Type IIb SN that led to Cas A;
\citealt{wongwathanarat_casa_17}). Such models will also allow for a
more physical treatment of mixing, in particular for \nifs. They will
also enable a treatment of asphericity and in particular the
assessment of the impact of large scale asymmetry on the nebular phase
properties. There is indeed evidence from the observed skewness of
emission line profiles that even standard-energy SN Ibc ejecta are
asymmetric
\citep{taubenberger_ibc_09,milisavljevic_etal_10}. Alternate power
sources and asymmetry will also allow for a description of exceptional
late time properties, for example with the hybrid line profile
morphology observed in SN\,2016gkg \citep{hanin_16gkg_20}. Our study
suggests that metal yields are just one component influencing the
nebular-phase properties of SNe Ibc, and that a proper
characterization of these events requires a global and
physically-consistent modeling of their complex ejecta structure.

\begin{acknowledgements}

This work was supported by the ``Programme National de Physique Stellaire'' of CNRS/INSU co-funded by CEA and CNES. DJH thanks NASA for partial support through the astrophysical theory grant 80NSSC20K0524. HTJ acknowledges funding by the Deutsche Forschungsgemeinschaft (DFG, German Research Foundation) through Sonderforschungsbereich (Collaborative Research Centre) SFB-1258 \say{Neutrinos and Dark Matter in Astro- and Particle Physics (NDM)} and under Germany's Excellence Strategy through Cluster of Excellence ORIGINS (EXC-2094)---390783311. This work was granted access to the HPC resources of  CINES under the allocation  2019 -- A0070410554 and 2020 -- A0090410554 made by GENCI, France. This research has made use of NASA's Astrophysics Data System Bibliographic Services.

\end{acknowledgements}

% \bibliographystyle{aa}
% \bibliography{/Users/ldessart/Bibliography/new_sn_library_luc}

\appendix

\section{Additional information on initial ejecta properties used for the radiative transfer calculations.}

\begin{table*}
    \caption{Additional ejecta model properties from \citet{ertl_ibc_20} to complement those given in Table~\ref{tab_prog}.
\label{tab_prog_bis}
}
\begin{center}
\begin{tabular}{l@{\hspace{1mm}}c@{\hspace{2mm}}c@{\hspace{2mm}}c@{\hspace{2mm}}c@{\hspace{2mm}}c@{\hspace{2mm}}c@{\hspace{2mm}}
c@{\hspace{2mm}}c@{\hspace{2mm}}c@{\hspace{2mm}}c@{\hspace{2mm}}c@{\hspace{2mm}}c@{\hspace{2mm}}
}
\hline
Model    &       \ekin\  &        \mej\  &  \mhesh\  &     \mosh\  &    \msish\  &  \mnish\  &     \mni\  &  \avxheinnish\ & \avxcainnish\
&  \avxnistableinnish\ &    \avxcainosh\ &  \avxcainhesh\ \\
        & [10$^{51}$\,erg] & [\msun]  & [\msun]  &[\msun]  &[\msun]  &[\msun]  &[\msun]  & & & & & \\
\hline
he2p6   &   1.34(-1)  &   7.91(-1) &  7.29(-1)  &   3.97(-2)  &   1.77(-3)  &   2.03(-2)  &   1.22(-2)  &   2.03(-1)  &   5.79(-3)  &   7.32(-2)  &   5.54(-5)  &   7.01(-5)  \\
he2p9   &   3.72(-1)  &   9.31(-1) &  8.01(-1)  &   8.29(-2)  &   9.63(-3)  &   3.75(-2)  &   2.32(-2)  &   2.23(-1)  &   8.32(-3)  &   3.78(-2)  &   2.56(-5)  &   6.85(-5)  \\
he3p3   &   5.48(-1)  &   1.20(0)  &  8.64(-1)  &   2.50(-1)  &   2.37(-2)  &   6.34(-2)  &   4.00(-2)  &   1.84(-1)  &   7.58(-3)  &   5.54(-2)  &   3.02(-5)  &   6.95(-5)  \\
he3p5   &   4.10(-1)  &   1.27(0)  &  8.80(-1)  &   3.21(-1)  &   1.95(-2)  &   4.51(-2)  &   2.92(-2)  &   2.07(-1)  &   5.66(-3)  &   5.19(-2)  &   3.40(-5)  &   7.03(-5)  \\
he4p0   &   6.33(-1)  &   1.62(0)  &  9.26(-1)  &   5.77(-1)  &   4.53(-2)  &   7.06(-2)  &   4.45(-2)  &   1.94(-1)  &   8.49(-3)  &   4.73(-2)  &   3.71(-5)  &   7.00(-5)  \\
he4p5   &   1.17(0)   &   1.89(0)  &  9.49(-1)  &   7.55(-1)  &   6.00(-2)  &   1.28(-1)  &   8.59(-2)  &   1.84(-1)  &   6.66(-3)  &   4.62(-2)  &   3.33(-5)  &   6.83(-5)  \\
he5p0   &   1.51(0)   &   2.21(0)  &  9.95(-1)  &   1.02(0)   &   5.06(-2)  &   1.47(-1)  &   9.77(-2)  &   1.99(-1)  &   4.80(-3)  &   5.04(-2)  &   4.00(-5)  &   6.99(-5)  \\
he6p0   &   1.10(0)   &   2.82(0)  &  1.06(0)   &   1.61(0)   &   4.69(-2)  &   1.06(-1)  &   7.04(-2)  &   2.14(-1)  &   5.14(-3)  &   4.56(-2)  &   4.11(-5)  &   7.02(-5)  \\
he7p0   &   1.38(0)   &   3.33(0)  &  6.66(-1)  &   2.43(0)   &   7.55(-2)  &   1.47(-1)  &   1.02(-1)  &   1.81(-1)  &   7.51(-3)  &   2.69(-2)  &   5.62(-5)  &   7.06(-5)  \\
he8p0   &   7.10(-1)  &   3.95(0)  &  6.11(-1)  &   3.21(0)   &   3.78(-2)  &   8.65(-2)  &   5.46(-2)  &   2.10(-1)  &   9.01(-3)  &   3.83(-2)  &   4.07(-5)  &   7.07(-5)  \\
he12p0  &   8.10(-1)  &   5.32(0)  &  1.28(-6)  &   5.15(0)   &   6.19(-2)  &   1.15(-1)  &   7.90(-2)  &   1.77(-1)  &   1.03(-2)  &   2.02(-2)  &   6.25(-5)  &   6.88(-5)  \\
\hline
\end{tabular}
\end{center}
{\bf Notes:} The columns give the ejecta kinetic energy, the ejecta mass, the mass of the He-rich shell, the O-rich shell, the Si-rich shell, and the \nifs-rich shell, and the \nifs\ mass. The subsequent columns give the average mass fraction of several isotopes in different shells (due to space constraints, O stands for O-rich shell, Ni for the \nifs-rich shell etc).
\end{table*}

\begin{table*}
\caption{Explosion properties for the models evolved with the nominal mass loss rate.
  \label{tab_ertliron}
  }
\begin{tabular}{lcccccccc}
\hline
$M_{\rm He,i}$   &  $E_{\rm kin}$  &  \nifs   &      ``Tr''   &  $\alpha$  &  \nifs + ``Tr''/2  &  \nifs + ``Tr'' & $0.75 \times$ (\nifs +``Tr''+$\alpha$) & This work \\
\vspace{0.01cm}
[\msun]  &  [foe] &   [\msun]  & [\msun]  & [\msun]  & [\msun]  & [\msun]  & [\msun]  & [\msun] \\
\hline
  2.6  &  0.15   &   0.0033  &    0.0047   &   0.0160  &    0.0057  &    0.0080   &   0.0180  &  0.0122  \\  
  2.9  &  0.38   &   0.0082  &    0.0098   &   0.0291  &    0.0131  &    0.0180   &   0.0353  &  0.0232  \\  
  3.3  &  0.59   &   0.0182  &    0.0167   &   0.0380  &    0.0266  &    0.0349   &   0.0547  &  0.0400  \\  
  3.5  &  0.42   &   0.0167  &    0.0091   &   0.0279  &    0.0212  &    0.0258   &   0.0403  &  0.0292  \\  
  4.0  &  0.65   &   0.0285  &    0.0166   &   0.0365  &    0.0368  &    0.0451   &   0.0612  &  0.0445  \\  
  4.5  &  1.28   &   0.0337  &    0.0368   &   0.0609  &    0.0521  &    0.0705   &   0.0986  &  0.0859  \\  
  5.0  &  1.50   &   0.0366  &    0.0436   &   0.0693  &    0.0584  &    0.0802   &   0.1121  &  0.0977  \\  
  6.0  &  1.07   &   0.0273  &    0.0307   &   0.0538  &    0.0427  &    0.0580   &   0.0839  &  0.0704  \\  
  7.0  &  1.37   &   0.0497  &    0.0383   &   0.0637  &    0.0688  &    0.0880   &   0.1138  &  0.102  \\  
  8.0  &  0.70   &   0.0244  &    0.0192   &   0.0378  &    0.0340  &    0.0436   &   0.0611  &  0.0546  \\  
 12.0  &  0.81   &   0.0559  &    0.0152   &   0.0344  &    0.0635  &    0.0711   &   0.0791  &  0.0790  \\  
\hline
\end{tabular}
\tablefoot{The table gives the results of the explosion calculations with \photb\ and presented in \citet{ertl_ibc_20}.  In our work, the \nifs\ masses
  were taken from the \kepler\ approximation to the \photb\ neutrino simulation.
  \nifs\ +``Tr''/2 is a realistic lower bound, \nifs\ +``Tr'' is a reasonable upper bound, and $0.75 \times$ (\nifs\ +``Tr''+$\alpha$)
  is a generous upper bound. Our values (rightmost column) lie between the lower and generous upper bounds. Only the 4.0\,\msun\ He-star model
  is below the reasonable upper bound. \citet{ertl_ibc_20} and \citet{woosley_ibc_21} used the
     generous upper bounds for their light curve analysis, so our values are
     consistent with theirs.
}
\end{table*}

\begin{figure*}
\centering
\includegraphics[width=0.495\hsize]{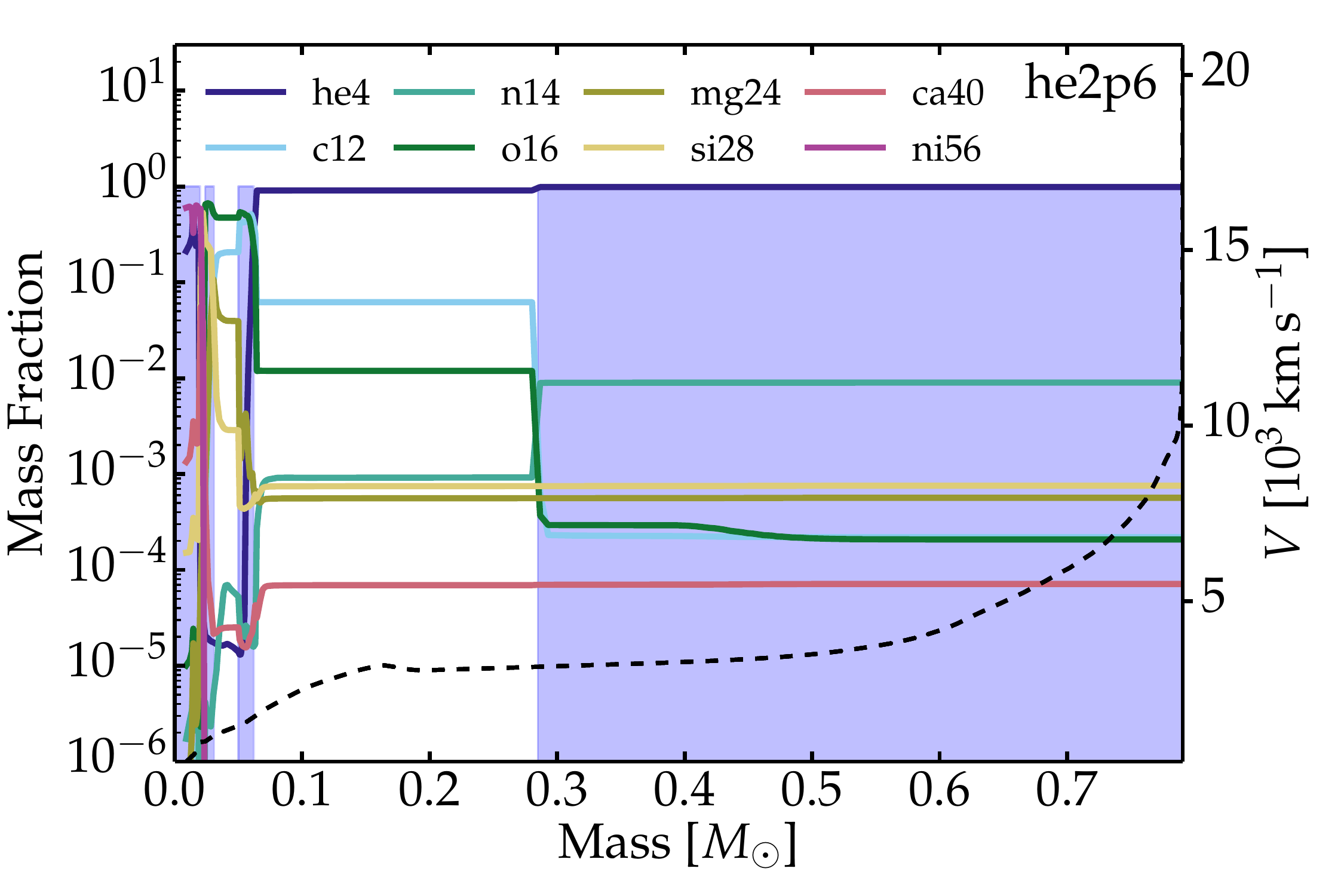}
\includegraphics[width=0.495\hsize]{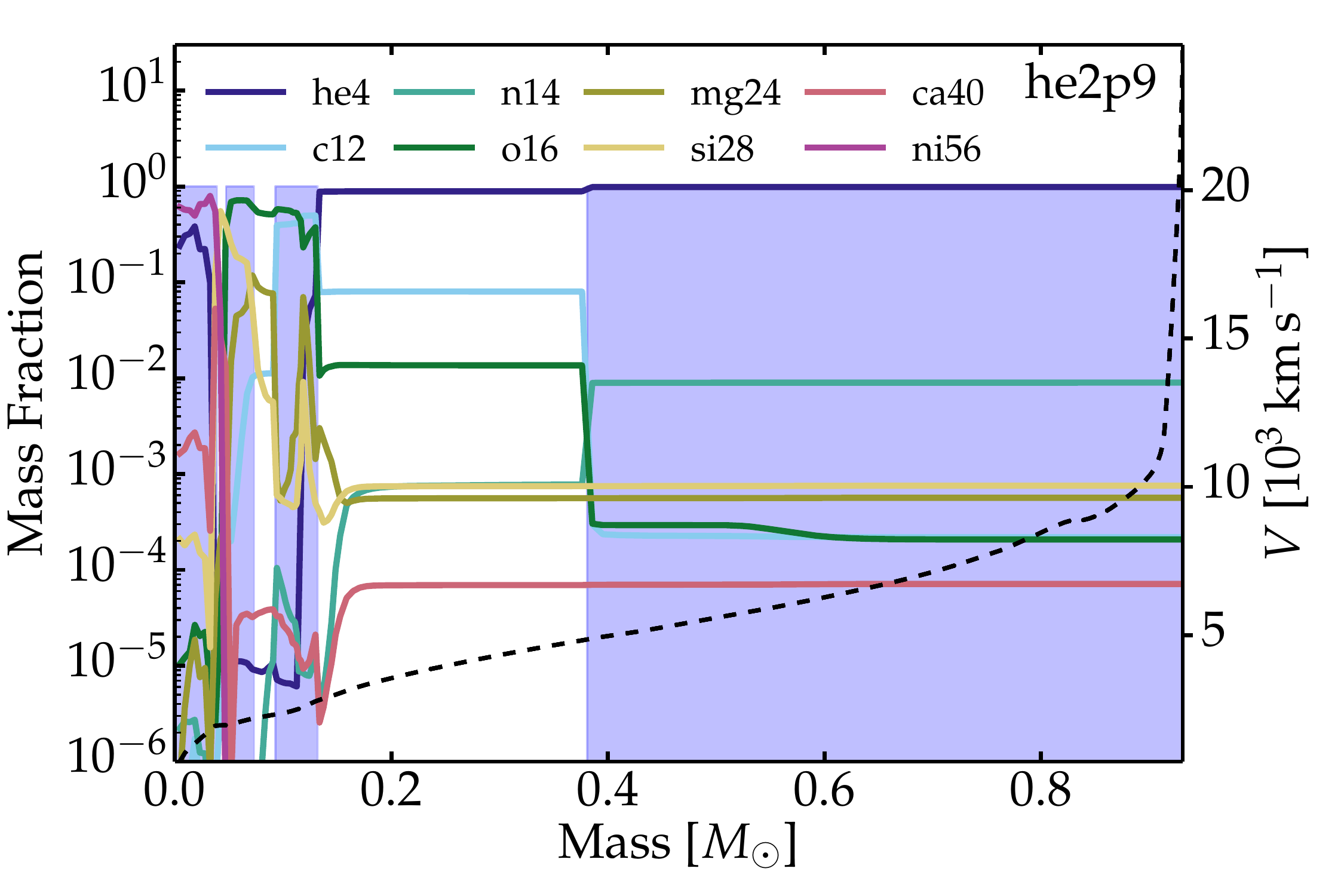}
\includegraphics[width=0.495\hsize]{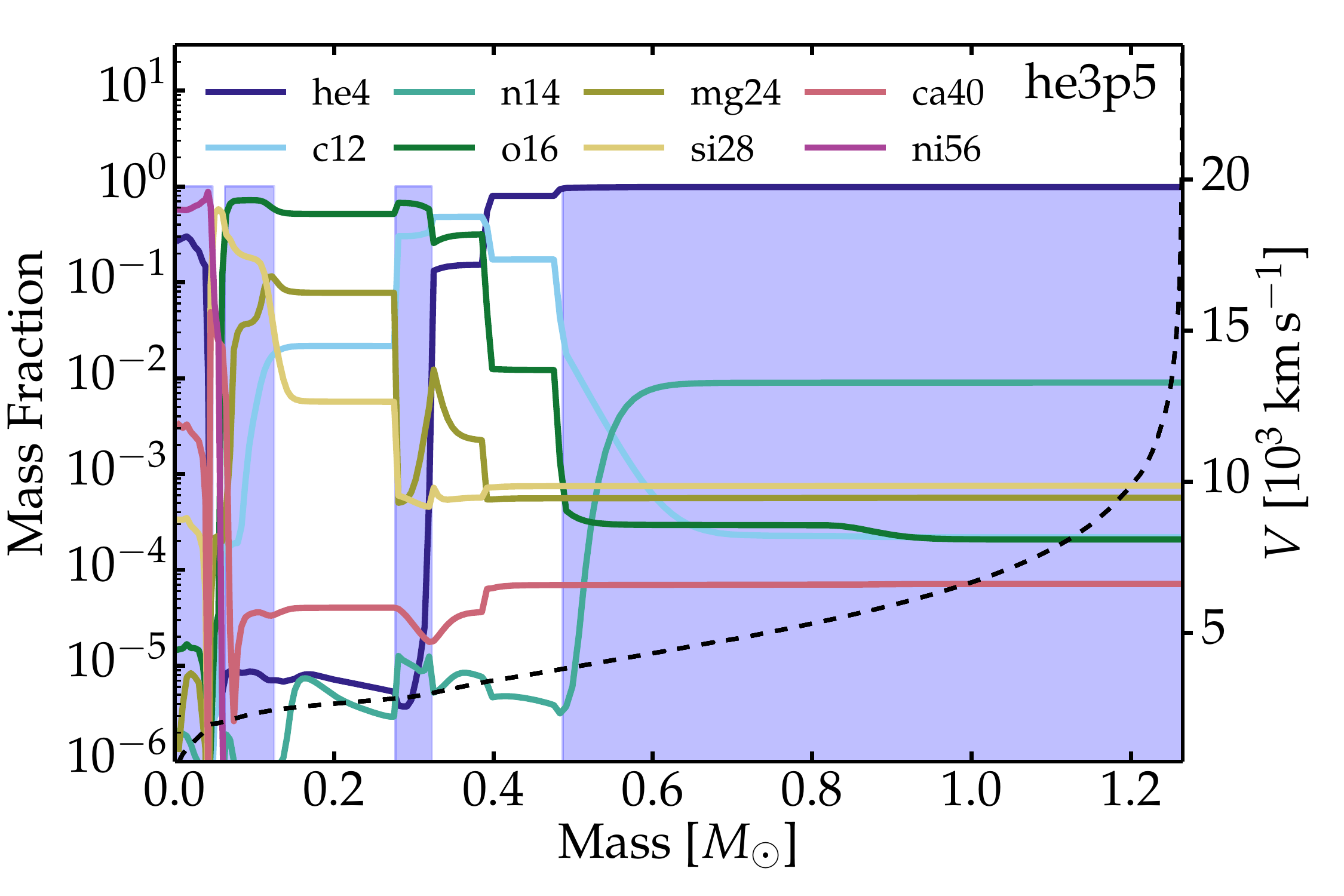}
\includegraphics[width=0.495\hsize]{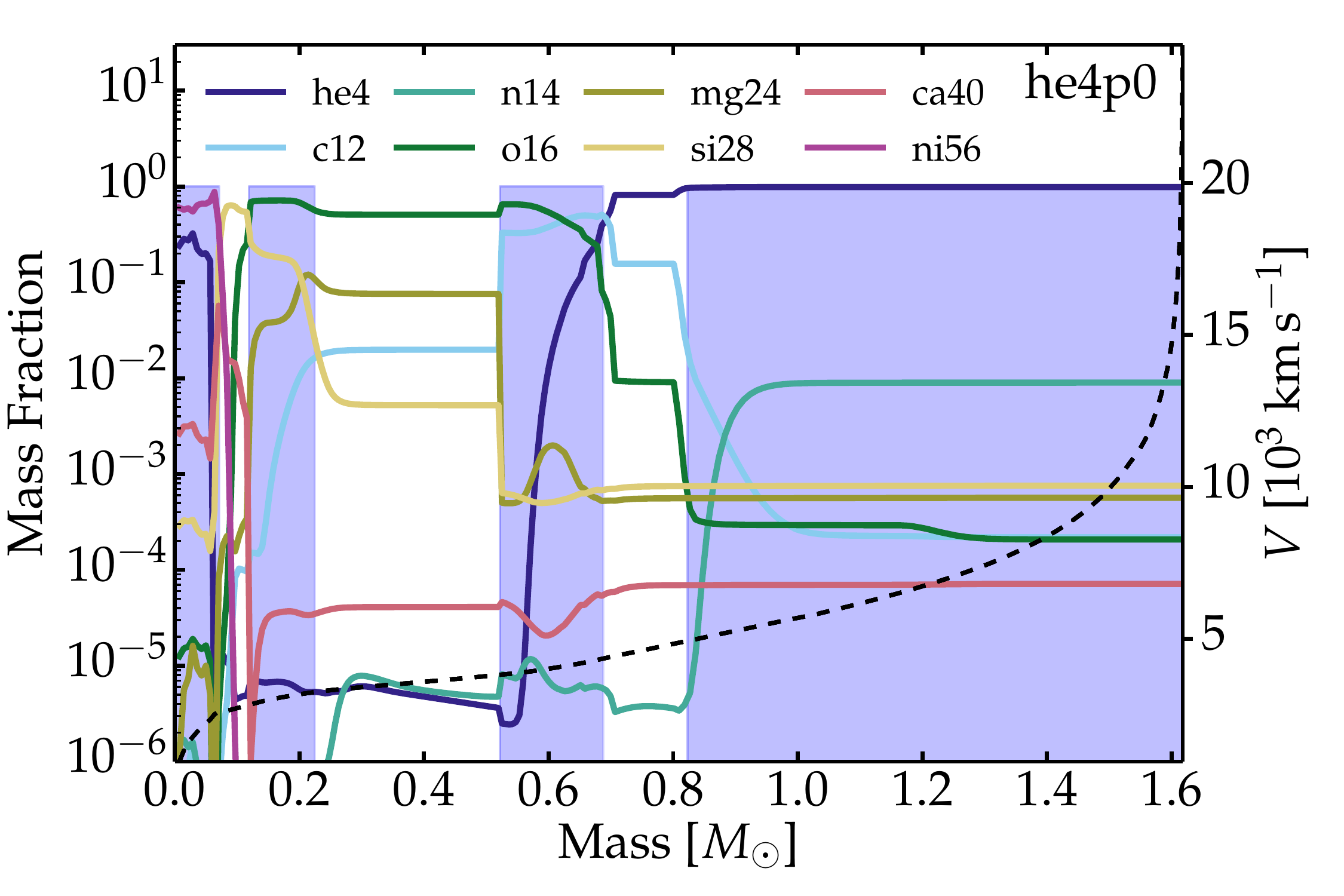}
\includegraphics[width=0.495\hsize]{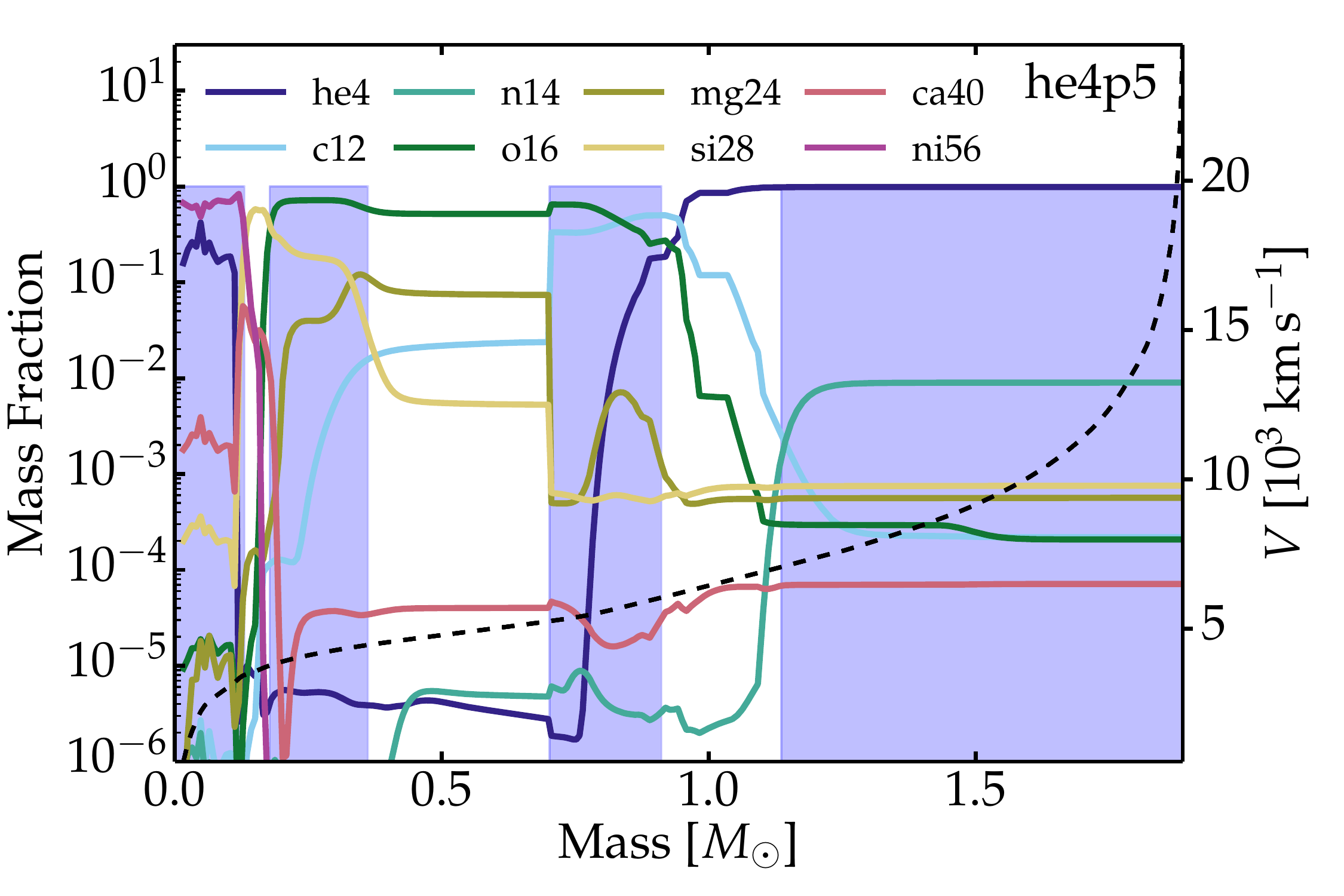}
\includegraphics[width=0.495\hsize]{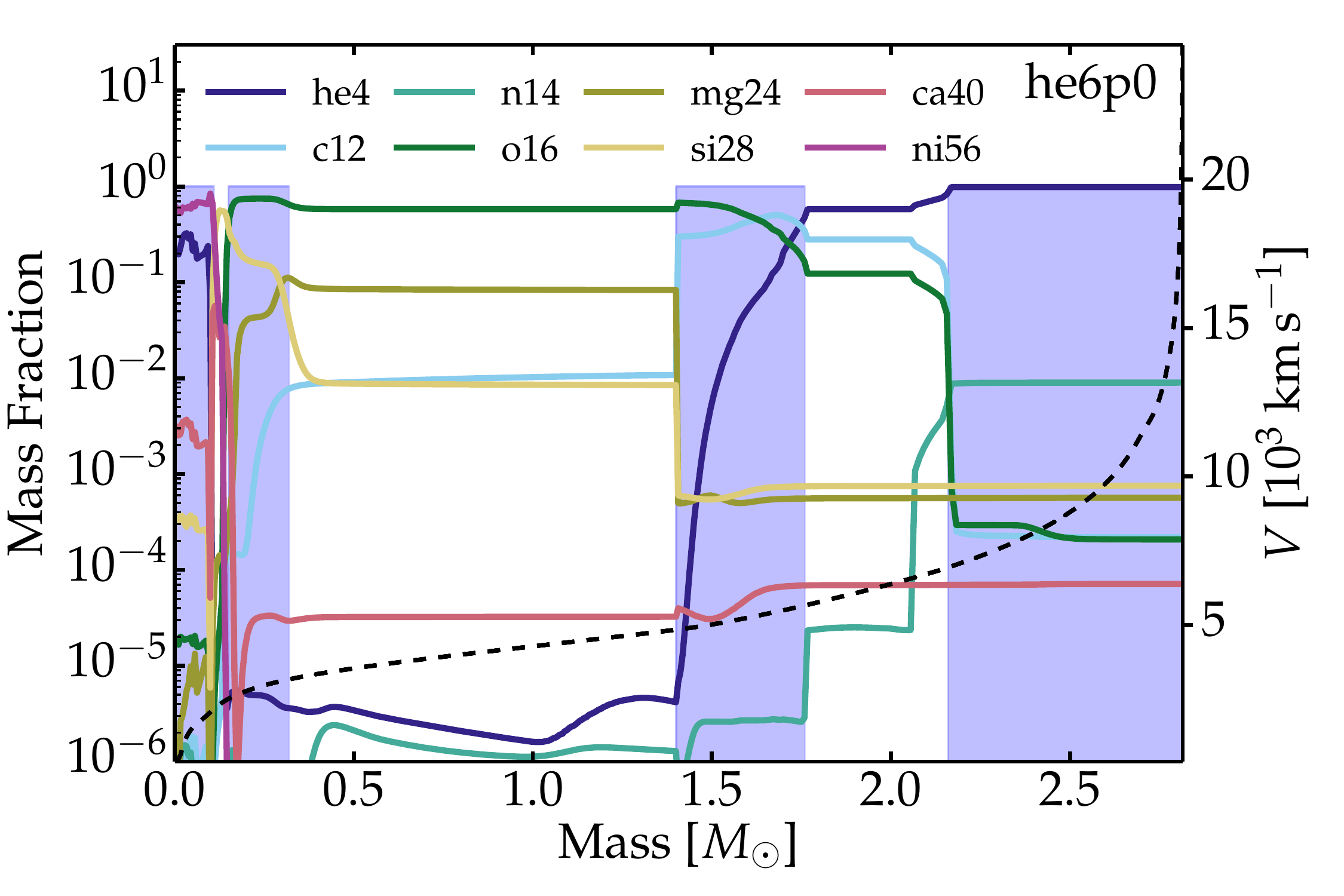}
\vspace{-0.2cm}
\caption{Chemical composition versus Lagrangian mass in the he2p6 he2p9, he3p5, he4p0, he4p5, and he6p0 from \citet{ertl_ibc_20} at 200\,d (the original mass fraction is shown for \nifs) -- ejecta properties are summarized in Table~\ref{tab_prog}. Models he3p3, he5p0, and he8p0 are shown in Fig.~\ref{fig_prog_comp}. The shaded areas denote dominant shells, of a given composition, that are used in the shuffling procedure.
\label{fig_prog_comp_all}
}
\end{figure*}

\begin{figure*}
\centering
\includegraphics[width=0.495\hsize]{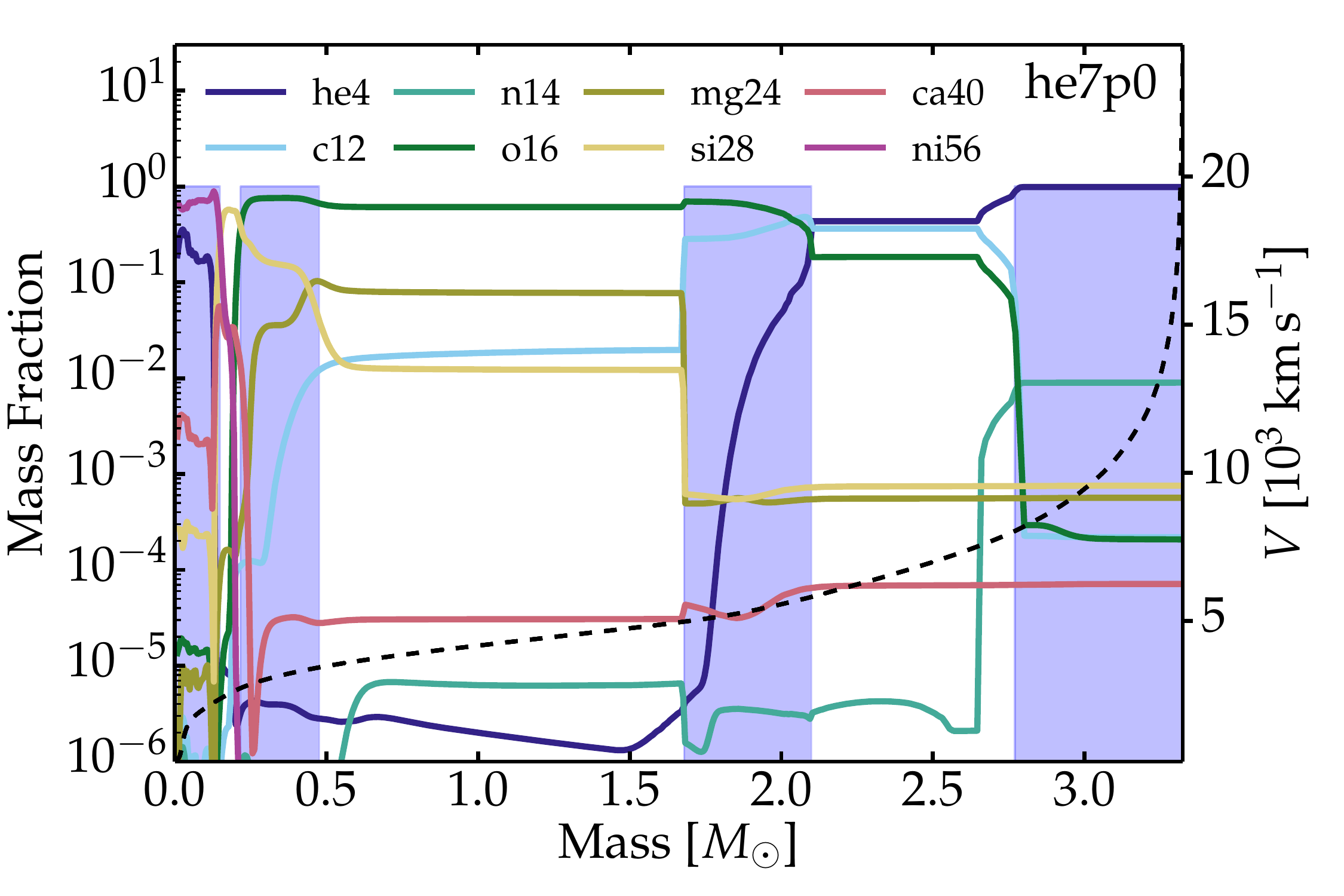}
\includegraphics[width=0.495\hsize]{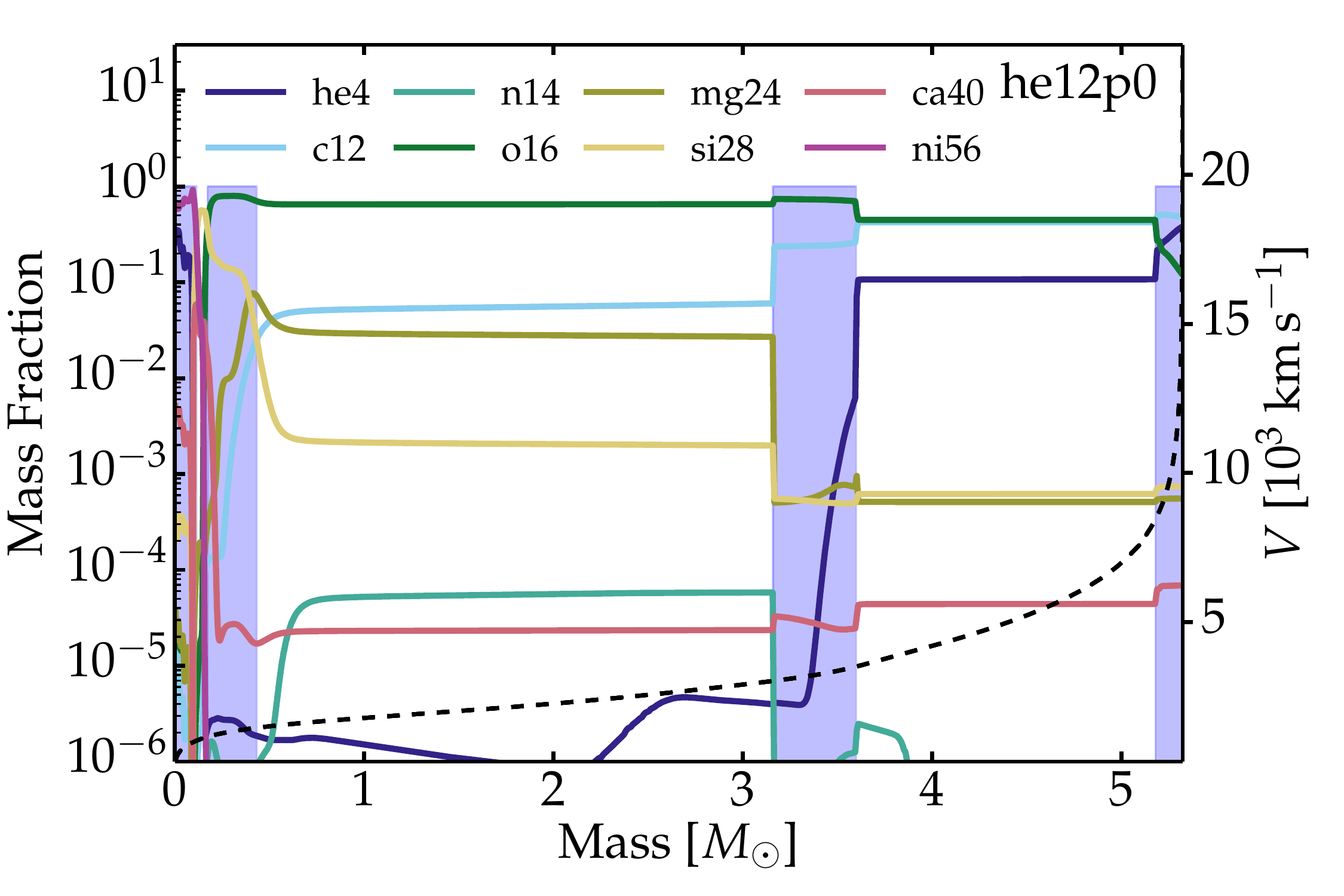}
\vspace{-0.2cm}
\caption{Same as Fig.~\ref{fig_prog_comp_all}, but now for models he7p0 and he12p0.
\label{fig_prog_comp_all_bis}
}
\end{figure*}

\clearpage

\begin{table*}
    \caption{Properties of the main-subshells of model he2p6 as selected from the \citet{ertl_ibc_20} unmixed ejecta. Each of these sub-shells is split in three equal mass parts and shuffled in mass space with the Lagrangian mass $M_{\rm sh}$, which is chosen to be at the interface between the He/C and the He/N shells, or between the O/C and He/C shells when the He/N shell is absent (as in the most massive progenitors). The lower part of the table indicates the average mass fraction of each species within each of these sub-shells. With the exception of \nifs\ and \cofs, all isotopes in the nuclear network are considered stable and included in the total mass fraction of the corresponding element. So, \cofn\ (\nife) include all Co (Ni) isotopes except \cofs\ (\nifs).
\label{tab_he2p60}
}
\begin{center}
\begin{tabular}{l@{\hspace{4mm}}c@{\hspace{4mm}}c@{\hspace{4mm}}c@{\hspace{4mm}}c@{\hspace{4mm}}
c@{\hspace{4mm}}c@{\hspace{4mm}}c@{\hspace{4mm}}
}
\hline
    Model  &      Fe/He &      Si/S  &      O/Si  &    O/Ne/Mg &        O/C &       He/C &       He/N \\
\hline
           &    [\msun] &    [\msun] &    [\msun] &    [\msun] &    [\msun] &    [\msun] &    [\msun] \\
\hline
    he2p6 &      0.018 &      0.006 &      0.006 &      0.018 &      0.012 &      0.222 &       0.483 \\
 \hline
    \hline
   Element &      \multicolumn{7}{c}{Average element mass fraction in each shell} \\
\hline
        He &      0.280 &   1.91e-03 &   1.85e-05 &   1.63e-05 &      0.029 &      0.903 &       0.985 \\
         C &   1.09e-06 &   4.92e-04 &      0.049 &      0.201 &      0.424 &      0.065 &    4.28e-04 \\
         N &   3.38e-06 &   9.20e-06 &   3.63e-05 &   9.56e-05 &   6.12e-05 &   9.97e-04 &    8.96e-03 \\
         O &   1.98e-05 &      0.272 &      0.615 &      0.475 &      0.473 &      0.016 &    2.75e-04 \\
\hline
\end{tabular}
\end{center}
\end{table*}

\begin{table*}
    \caption{Same as Table~\ref{tab_he2p60}, but now for model he2p9.
\label{tab_he2p90}
}
\begin{center}
\begin{tabular}{l@{\hspace{4mm}}c@{\hspace{4mm}}c@{\hspace{4mm}}c@{\hspace{4mm}}c@{\hspace{4mm}}
c@{\hspace{4mm}}c@{\hspace{4mm}}c@{\hspace{4mm}}
}
\hline
    Model  &      Fe/He &      Si/S  &      O/Si  &    O/Ne/Mg &        O/C &       He/C &       He/N \\
\hline
           &    [\msun] &    [\msun] &    [\msun] &    [\msun] &    [\msun] &    [\msun] &    [\msun] \\
\hline
    he2p9 &      0.039 &      0.006 &      0.027 &      0.018 &      0.039 &      0.249 &       0.513 \\
\hline
\hline
   Element &      \multicolumn{7}{c}{Average element mass fraction in each shell} \\
\hline
        He &      0.229 &   1.03e-05 &   1.05e-05 &   9.28e-06 &      0.037 &      0.883 &       0.985 \\
         C &   1.00e-06 &   6.10e-05 &   3.54e-03 &      0.019 &      0.397 &      0.082 &    5.22e-04 \\
         N &   1.75e-06 &   1.90e-06 &   8.87e-06 &   3.58e-05 &   5.68e-05 &   7.20e-04 &    8.95e-03 \\
         O &   1.79e-05 &      0.098 &      0.657 &      0.535 &      0.445 &      0.018 &    2.95e-04 \\
\hline
\end{tabular}
\end{center}
\end{table*}

\begin{table*}
    \caption{Same as Table~\ref{tab_he2p60}, but now for model he3p3.
\label{tab_he3p30}
}
\begin{center}
\begin{tabular}{l@{\hspace{4mm}}c@{\hspace{4mm}}c@{\hspace{4mm}}c@{\hspace{4mm}}c@{\hspace{4mm}}
c@{\hspace{4mm}}c@{\hspace{4mm}}c@{\hspace{4mm}}
}
\hline
    Model  &      Fe/He &      Si/S  &      O/Si  &    O/Ne/Mg &        O/C &       He/C &       He/N \\
\hline
           &    [\msun] &    [\msun] &    [\msun] &    [\msun] &    [\msun] &    [\msun] &    [\msun] \\
\hline
    he3p3 &      0.063 &      0.018 &      0.066 &      0.111 &      0.039 &      0.135 &       0.733 \\
\hline
\hline
   Element &      \multicolumn{7}{c}{Average element mass fraction in each shell} \\
\hline
        He &      0.229 &   1.22e-05 &   8.77e-06 &   7.14e-06 &   1.36e-03 &      0.547 &       0.984 \\
         C &   8.39e-07 &   3.51e-05 &   3.03e-03 &      0.014 &      0.295 &      0.307 &    1.21e-03 \\
         N &   1.51e-06 &   1.10e-06 &   7.90e-06 &   1.98e-05 &   3.58e-05 &   1.26e-05 &    8.04e-03 \\
         O &   1.67e-05 &      0.112 &      0.687 &      0.603 &      0.636 &      0.110 &    8.04e-04 \\
\hline
\end{tabular}
\end{center}
\end{table*}

\begin{table*}
    \caption{Same as Table~\ref{tab_he2p60}, but now for model he3p5.
\label{tab_he3p50}
}
\begin{center}
\begin{tabular}{l@{\hspace{4mm}}c@{\hspace{4mm}}c@{\hspace{4mm}}c@{\hspace{4mm}}c@{\hspace{4mm}}
c@{\hspace{4mm}}c@{\hspace{4mm}}c@{\hspace{4mm}}
}
\hline
    Model  &      Fe/He &      Si/S  &      O/Si  &    O/Ne/Mg &        O/C &       He/C &       He/N \\
\hline
           &    [\msun] &    [\msun] &    [\msun] &    [\msun] &    [\msun] &    [\msun] &    [\msun] \\
\hline
    he3p5 &      0.045 &      0.015 &      0.063 &      0.153 &      0.045 &      0.165 &       0.773  \\
\hline
\hline
   Element &      \multicolumn{7}{c}{Average element mass fraction in each shell} \\
\hline
        He &      0.221 &   1.26e-05 &   8.31e-06 &   6.89e-06 &   2.63e-03 &      0.532 &       0.984  \\
         C &   7.79e-07 &   4.22e-05 &   5.12e-03 &      0.023 &      0.306 &      0.298 &    9.49e-04  \\
         N &   1.30e-06 &   1.17e-06 &   8.05e-06 &   3.48e-05 &   3.57e-05 &   1.24e-05 &    8.09e-03  \\
         O &   1.53e-05 &      0.076 &      0.671 &      0.525 &      0.635 &      0.130 &    7.87e-04  \\
\hline
\end{tabular}
\end{center}
\end{table*}

\begin{table*}
    \caption{Same as Table~\ref{tab_he2p60}, but now for model he4p0.
\label{tab_he4p00}
}
\begin{center}
\begin{tabular}{l@{\hspace{4mm}}c@{\hspace{4mm}}c@{\hspace{4mm}}c@{\hspace{4mm}}c@{\hspace{4mm}}
c@{\hspace{4mm}}c@{\hspace{4mm}}c@{\hspace{4mm}}
}
\hline
    Model  &      Fe/He &      Si/S  &      O/Si  &    O/Ne/Mg &        O/C &       He/C &       He/N \\
\hline
           &    [\msun] &    [\msun] &    [\msun] &    [\msun] &    [\msun] &    [\msun] &    [\msun] \\
\hline
    he4p0 &      0.069 &      0.048 &      0.105 &      0.297 &      0.168 &      0.135 &       0.793 \\
\hline
\hline
   Element &      \multicolumn{7}{c}{Average element mass fraction in each shell} \\
\hline
        He &      0.208 &   7.90e-06 &   6.26e-06 &   4.98e-06 &      0.070 &      0.797 &       0.985 \\
         C &   8.62e-07 &   5.40e-05 &   5.17e-03 &      0.020 &      0.399 &      0.171 &    7.70e-04 \\
         N &   1.00e-06 &   1.18e-06 &   7.33e-06 &   3.25e-05 &   2.17e-05 &   7.63e-06 &    8.10e-03 \\
         O &   1.58e-05 &      0.092 &      0.660 &      0.513 &      0.482 &      0.013 &    8.23e-04 \\
\hline
\end{tabular}
\end{center}
\end{table*}

\begin{table*}
    \caption{Same as Table~\ref{tab_he2p60}, but now for model he4p5.
\label{tab_he4p50}
}
\begin{center}
\begin{tabular}{l@{\hspace{4mm}}c@{\hspace{4mm}}c@{\hspace{4mm}}c@{\hspace{4mm}}c@{\hspace{4mm}}
c@{\hspace{4mm}}c@{\hspace{4mm}}c@{\hspace{4mm}}
}
\hline
    Model  &      Fe/He &      Si/S  &      O/Si  &    O/Ne/Mg &        O/C &       He/C &       He/N \\
\hline
           &    [\msun] &    [\msun] &    [\msun] &    [\msun] &    [\msun] &    [\msun] &    [\msun] \\
\hline
    he4p5 &      0.129 &      0.048 &      0.183 &      0.339 &      0.210 &      0.225 &       0.673 \\
\hline
\hline
   Element &      \multicolumn{7}{c}{Average element mass fraction in each shell} \\
\hline
        He &      0.210 &   6.65e-06 &   4.96e-06 &   3.67e-06 &      0.040 &      0.773 &       0.985 \\
         C &   1.00e-06 &   4.03e-05 &   4.49e-03 &      0.022 &      0.393 &      0.157 &    3.58e-04 \\
         N &   9.70e-07 &   7.46e-07 &   6.39e-06 &   2.68e-05 &   1.88e-05 &   9.34e-05 &    8.52e-03 \\
         O &   1.65e-05 &      0.069 &      0.686 &      0.525 &      0.497 &      0.049 &    6.75e-04 \\
\hline
\end{tabular}
\end{center}
\end{table*}

\begin{table*}
    \caption{Same as Table~\ref{tab_he2p60}, but now for model he5p0.
\label{tab_he5p00}
}
\begin{center}
\begin{tabular}{l@{\hspace{4mm}}c@{\hspace{4mm}}c@{\hspace{4mm}}c@{\hspace{4mm}}c@{\hspace{4mm}}
c@{\hspace{4mm}}c@{\hspace{4mm}}c@{\hspace{4mm}}
}
\hline
    Model  &      Fe/He &      Si/S  &      O/Si  &    O/Ne/Mg &        O/C &       He/C &       He/N \\
\hline
           &    [\msun] &    [\msun] &    [\msun] &    [\msun] &    [\msun] &    [\msun] &    [\msun] \\
\hline
    he5p0 &      0.153 &      0.042 &      0.162 &      0.621 &      0.246 &      0.204 &       0.753 \\
\hline
\hline
   Element &      \multicolumn{7}{c}{Average element mass fraction in each shell} \\
\hline
        He &      0.219 &   5.11e-06 &   4.88e-06 &   2.63e-06 &      0.099 &      0.696 &       0.985 \\
         C &   1.06e-06 &   7.12e-05 &   4.46e-03 &      0.020 &      0.402 &      0.203 &    4.23e-04 \\
         N &   9.35e-07 &   1.19e-06 &   6.12e-06 &   1.96e-05 &   1.22e-05 &   2.50e-05 &    8.37e-03 \\
         O &   1.71e-05 &      0.108 &      0.714 &      0.554 &      0.466 &      0.082 &    8.04e-04 \\
\hline
\end{tabular}
\end{center}
\end{table*}

\begin{table*}
    \caption{Same as Table~\ref{tab_he2p60}, but now for model he6p0.
\label{tab_he6p00}
}
\begin{center}
\begin{tabular}{l@{\hspace{4mm}}c@{\hspace{4mm}}c@{\hspace{4mm}}c@{\hspace{4mm}}
c@{\hspace{4mm}}c@{\hspace{4mm}}c@{\hspace{4mm}}c@{\hspace{4mm}}
}
\hline
    Model  &      Fe/He &      Si/S  &      O/Si  &    O/Ne/Mg &        O/C &     He/C/O &       He/N \\
\hline
           &    [\msun] &    [\msun] &    [\msun] &    [\msun] &    [\msun] &    [\msun] &    [\msun] \\
\hline
    he6p0 &      0.108 &      0.039 &      0.171 &      1.080 &      0.360 &      0.402 &       0.633 \\
\hline
\hline
   Element &      \multicolumn{7}{c}{Average element mass fraction in each shell} \\
\hline
        He &      0.235 &   4.16e-04 &   4.74e-06 &   2.93e-06 &      0.089 &      0.611 &       0.985 \\
         C &   1.09e-06 &   4.40e-05 &   2.80e-03 &   9.96e-03 &      0.387 &      0.257 &    5.24e-04 \\
         N &   1.00e-06 &   7.72e-07 &   6.21e-06 &   1.47e-05 &   1.70e-05 &   6.97e-04 &    8.95e-03 \\
         O &   1.99e-05 &      0.055 &      0.715 &      0.586 &      0.497 &      0.113 &    3.75e-04 \\
\hline
\end{tabular}
\end{center}
\end{table*}

\begin{table*}
    \caption{Same as Table~\ref{tab_he2p60}, but now for model he7p0.
\label{tab_he7p00}
}
\begin{center}
\begin{tabular}{l@{\hspace{4mm}}c@{\hspace{4mm}}c@{\hspace{4mm}}c@{\hspace{4mm}}c@{\hspace{4mm}}
c@{\hspace{4mm}}c@{\hspace{4mm}}c@{\hspace{4mm}}
}
\hline
    Model  &      Fe/He &      Si/S  &      O/Si  &    O/Ne/Mg &        O/C &     He/C/O &       He/N \\
\hline
           &    [\msun] &    [\msun] &    [\msun] &    [\msun] &    [\msun] &    [\msun] &    [\msun] \\
\hline
    he7p0 &      0.147 &      0.069 &      0.258 &      1.206 &      0.420 &      0.669 &       0.543 \\
\hline
\hline
   Element &      \multicolumn{7}{c}{Average element mass fraction in each shell} \\
\hline
        He &      0.205 &   6.68e-06 &   3.74e-06 &   2.13e-06 &      0.034 &      0.475 &       0.984 \\
         C &   1.11e-06 &   3.67e-05 &   3.45e-03 &      0.019 &      0.346 &      0.336 &    1.09e-03 \\
         N &   6.36e-07 &   5.65e-07 &   6.39e-06 &   2.14e-05 &   1.40e-05 &   6.80e-04 &    8.96e-03 \\
         O &   1.54e-05 &      0.058 &      0.729 &      0.617 &      0.595 &      0.169 &    6.36e-04 \\
\hline
\end{tabular}
\end{center}
\end{table*}

\begin{table*}
    \caption{Same as Table~\ref{tab_he2p60}, but now for model he8p0.
\label{tab_he8p00}
}
\begin{center}
\begin{tabular}{l@{\hspace{4mm}}c@{\hspace{4mm}}c@{\hspace{4mm}}c@{\hspace{4mm}}c@{\hspace{4mm}}
c@{\hspace{4mm}}c@{\hspace{4mm}}c@{\hspace{4mm}}
}
\hline
    Model  &      Fe/He &      Si/S  &      O/Si  &    O/Ne/Mg &        O/C &     C/He/O &     He/C/N \\
\hline
           &    [\msun] &    [\msun] &    [\msun] s&    [\msun] &    [\msun] &    [\msun] &    [\msun] \\
\hline
    he8p0 &      0.084 &      0.045 &      0.150 &      1.950 &      0.366 &      0.804 &       0.543 \\
\hline
\hline
   Element &      \multicolumn{7}{c}{Average element mass fraction in each shell} \\
\hline
        He &      0.237 &   9.17e-06 &   4.45e-06 &   3.37e-06 &      0.021 &      0.339 &       0.972 \\
         C &   1.41e-06 &   7.15e-05 &   5.41e-03 &      0.018 &      0.311 &      0.415 &    9.56e-03 \\
         N &   6.01e-07 &   1.26e-06 &   8.20e-06 &   2.47e-05 &   1.47e-05 &   3.44e-04 &    8.79e-03 \\
         O &   2.11e-05 &      0.157 &      0.749 &      0.607 &      0.642 &      0.227 &    4.74e-03 \\
\hline
\end{tabular}
\end{center}
\end{table*}

\begin{table*}
    \caption{Same as Table~\ref{tab_he2p60}, but now for model he12p0.
\label{tab_he12p00}
}
\begin{center}
\begin{tabular}{l@{\hspace{4mm}}c@{\hspace{4mm}}c@{\hspace{4mm}}c@{\hspace{4mm}}c@{\hspace{4mm}}
c@{\hspace{4mm}}c@{\hspace{4mm}}c@{\hspace{4mm}}
}
\hline
    Model  &      Fe/He &      Si/S  &      O/Si  &    O/Ne/Mg &        O/C &    O/C/He &     C/He/O \\
\hline
           &    [\msun] &    [\msun] &    [\msun] &    [\msun] &    [\msun] &    [\msun] &    [\msun] \\
\hline
   he12p0 &      0.114 &      0.057 &      0.261 &      2.727 &      0.438 &      1.581 &       0.143 \\
\hline
\hline
   Element &      \multicolumn{7}{c}{Average element mass fraction in each shell} \\
\hline
        He &      0.185 &   7.76e-06 &   2.50e-06 &   2.08e-06 &   1.01e-03 &      0.107 &       0.296 \\
         C &   4.38e-06 &   3.81e-05 &   5.96e-03 &      0.055 &      0.241 &      0.422 &       0.496 \\
         N &   1.58e-07 &   3.68e-07 &   6.30e-06 &   7.24e-05 &   2.03e-05 &   5.59e-06 &    3.50e-07 \\
         O &   1.24e-05 &      0.056 &      0.760 &      0.654 &      0.728 &      0.448 &       0.189 \\
\hline
\end{tabular}
\end{center}
\end{table*}

\clearpage

\begin{table*}
    \caption{Same as Table~\ref{tab_he2p60}, but now for model he5p0x1p5.
\label{tab_he5p00x1p5}
}
\begin{center}
\begin{tabular}{l@{\hspace{4mm}}c@{\hspace{4mm}}c@{\hspace{4mm}}c@{\hspace{4mm}}c@{\hspace{4mm}}
c@{\hspace{4mm}}c@{\hspace{4mm}}c@{\hspace{4mm}}
}
\hline
    Model  &      Fe/He &      Si/S  &      O/Si  &    O/Ne/Mg &        O/C &       He/C &       He/N \\
\hline
           &    [\msun] &    [\msun] &    [\msun] &    [\msun] &    [\msun] &    [\msun] &    [\msun] \\
\hline
he5p0x1p5 &      0.168 &      0.051 &      0.162 &      0.321 &      0.186 &      0.231 &       0.703 \\
\hline
\hline
   Element &      \multicolumn{7}{c}{Average element mass fraction in each shell} \\
\hline
        He &      0.214 &   4.97e-06 &   4.77e-06 &   3.22e-06 &      0.019 &      0.466 &       0.985 \\
         C &   1.16e-06 &   4.27e-05 &   2.76e-03 &      0.019 &      0.376 &      0.351 &    4.52e-04 \\
         N &   9.38e-07 &   6.72e-07 &   5.49e-06 &   2.38e-05 &   1.80e-05 &   7.34e-05 &    8.48e-03 \\
         O &   1.75e-05 &      0.054 &      0.707 &      0.563 &      0.505 &      0.138 &    6.05e-04 \\
\hline
\end{tabular}
\end{center}
\end{table*}

\begin{table*}
    \caption{Same as Table~\ref{tab_he2p60}, but now for model he6p0x1p5.
\label{tab_he6p00x1p5}
}
\begin{center}
\begin{tabular}{l@{\hspace{4mm}}c@{\hspace{4mm}}c@{\hspace{4mm}}c@{\hspace{4mm}}c@{\hspace{4mm}}
c@{\hspace{4mm}}c@{\hspace{4mm}}c@{\hspace{4mm}}
}
\hline
    Model  &      Fe/He &      Si/S  &      O/Si  &    O/Ne/Mg &        O/C &       He/C &       He/N \\
\hline
           &    [\msun] &    [\msun] &    [\msun] &    [\msun] &    [\msun] &    [\msun] &    [\msun] \\
\hline
he6p0x1p5 &      0.186 &      0.057 &      0.171 &      0.636 &      0.198 &      0.450 &       0.603 \\
\hline
\hline
   Element &      \multicolumn{7}{c}{Average element mass fraction in each shell} \\
\hline
        He &      0.268 &   4.72e-06 &   4.38e-06 &   2.34e-06 &      0.020 &      0.399 &       0.986 \\
         C &   1.40e-06 &   4.36e-05 &   3.28e-03 &      0.024 &      0.356 &      0.417 &    2.57e-04 \\
         N &   1.14e-06 &   6.35e-07 &   5.41e-06 &   2.11e-05 &   1.50e-05 &   1.36e-04 &    8.95e-03 \\
         O &   1.73e-05 &      0.066 &      0.728 &      0.576 &      0.575 &      0.160 &    2.67e-04 \\
\hline
\end{tabular}
\end{center}
\end{table*}

\begin{table*}
    \caption{Same as Table~\ref{tab_he2p60}, but now for model he7p0x1p5.
\label{tab_he7p00x1p5}
}
\begin{center}
\begin{tabular}{l@{\hspace{4mm}}c@{\hspace{4mm}}c@{\hspace{4mm}}c@{\hspace{4mm}}
c@{\hspace{4mm}}c@{\hspace{4mm}}c@{\hspace{4mm}}c@{\hspace{4mm}}c@{\hspace{4mm}}
}
\hline
    Model  &      Fe/He &      Si/S  &      O/Si  &    O/Ne/Mg &        O/C &    C/He/O &    He/C &       He/N \\
\hline
           &    [\msun] &    [\msun] &    [\msun] &    [\msun] &    [\msun] &    [\msun] &    [\msun] &    [\msun] \\
\hline
he7p0x1p5 &      0.318 &      0.081 &      0.261 &      0.765 &      0.267 &      0.447 &      0.321 &       0.363 \\
\hline
\hline
   Element &      \multicolumn{8}{c}{Average element mass fraction in each shell} \\
\hline
        He &      0.292 &   3.67e-06 &   3.08e-06 &   1.63e-06 &      0.015 &      0.290 &      0.659 &       0.986 \\
         C &   2.45e-06 &   4.54e-05 &   3.70e-03 &      0.029 &      0.339 &      0.485 &      0.280 &    2.62e-04 \\
         N &   1.10e-06 &   4.62e-07 &   5.05e-06 &   1.94e-05 &   1.31e-05 &   6.28e-06 &   1.37e-04 &    8.98e-03 \\
         O &   2.45e-05 &      0.054 &      0.732 &      0.594 &      0.590 &      0.194 &      0.042 &    2.41e-04 \\
\hline
\end{tabular}
\end{center}
\end{table*}

\begin{table*}
    \caption{Same as Table~\ref{tab_he2p60}, but now for model he8p0x1p5.
\label{tab_he8p00x1p5}
}
\begin{center}
\begin{tabular}{l@{\hspace{4mm}}c@{\hspace{4mm}}c@{\hspace{4mm}}c@{\hspace{4mm}}c@{\hspace{4mm}}
c@{\hspace{4mm}}c@{\hspace{4mm}}c@{\hspace{4mm}}c@{\hspace{4mm}}
}
\hline
    Model  &      Fe/He &      Si/S  &      O/Si  &    O/Ne/Mg &        O/C &       C/He/O &   He/C  &   He/C/N \\
\hline
           &    [\msun]  &    [\msun] &    [\msun] &    [\msun] &    [\msun] &    [\msun] &    [\msun] &    [\msun] \\
\hline
he8p0x1p5 &      0.162 &      0.051 &      0.225 &      1.320 &      0.294 &      0.537 &      0.510 &       0.153 \\
\hline
\hline
   Element &      \multicolumn{8}{c}{Average element mass fraction in each shell} \\
\hline
        He &      0.295 &   6.27e-06 &   4.59e-06 &   2.50e-06 &      0.022 &      0.258 &      0.619 &       0.890 \\
         C &   1.84e-06 &   5.03e-05 &   3.80e-03 &      0.014 &      0.330 &      0.497 &      0.318 &       0.090 \\
         N &   1.26e-06 &   7.58e-07 &   6.57e-06 &   1.66e-05 &   1.69e-05 &   6.10e-06 &   1.84e-07 &    3.58e-03 \\
         O &   2.09e-05 &      0.055 &      0.730 &      0.601 &      0.622 &      0.225 &      0.045 &    2.94e-03 \\
\hline
\end{tabular}
\end{center}
\end{table*}

\begin{table*}
    \caption{Same as Table~\ref{tab_he2p60}, but now for model he9p0x1p5.
\label{tab_he9p00x1p5}
}
\begin{center}
\begin{tabular}{l@{\hspace{4mm}}c@{\hspace{4mm}}c@{\hspace{4mm}}c@{\hspace{4mm}}c@{\hspace{4mm}}
c@{\hspace{4mm}}c@{\hspace{4mm}}c@{\hspace{4mm}}
}
\hline
    Model  &      Fe/He &      Si/S  &      O/Si  &    O/Ne/Mg &        O/C &     C/O/He &      C/He/O \\
\hline
           &    [\msun] &    [\msun] &    [\msun] &    [\msun] &    [\msun] &    [\msun] &    [\msun] \\
\hline
he9p0x1p5 &      0.174 &      0.066 &      0.249 &      1.290 &      0.291 &      0.759 &       0.273 \\
\hline
\hline
   Element &      \multicolumn{7}{c}{Average element mass fraction in each shell} \\
\hline
        He &      0.211 &   5.80e-06 &   3.57e-06 &   2.05e-06 &   3.41e-03 &      0.140 &       0.407 \\
         C &   1.40e-06 &   3.60e-05 &   3.17e-03 &      0.020 &      0.281 &      0.479 &       0.472 \\
         N &   6.11e-07 &   5.06e-07 &   6.13e-06 &   2.75e-05 &   1.80e-05 &   7.74e-06 &    4.29e-07 \\
         O &   1.49e-05 &      0.059 &      0.744 &      0.624 &      0.689 &      0.360 &       0.102 \\
\hline
\end{tabular}
\end{center}
\end{table*}

\begin{table*}
    \caption{Same as Table~\ref{tab_he2p60}, but now for model he10p0x1p5.
\label{tab_he10p0x1p5}
}
\begin{center}
\begin{tabular}{l@{\hspace{4mm}}c@{\hspace{4mm}}c@{\hspace{4mm}}c@{\hspace{4mm}}c@{\hspace{4mm}}
c@{\hspace{4mm}}c@{\hspace{4mm}}c@{\hspace{4mm}}
}
\hline
    Model  &      Fe/He &      Si/S  &      O/Si  &    O/Ne/Mg &        O/C &     C/O/He &     C/O/He \\
\hline
           &    [\msun] &    [\msun] &    [\msun] &    [\msun] &    [\msun] &    [\msun] &    [\msun] \\
\hline
he10p0x1p5 &      0.144 &      0.063 &      0.264 &      1.224 &      0.300 &      0.927 &       0.253 \\
\hline
\hline
   Element &      \multicolumn{7}{c}{Average element mass fraction in each shell} \\
\hline
        He &      0.203 &   6.50e-06 &   3.62e-06 &   2.03e-06 &   1.82e-03 &      0.100 &       0.199 \\
         C &   1.28e-06 &   3.78e-05 &   3.17e-03 &      0.018 &      0.270 &      0.451 &       0.518 \\
         N &   5.72e-07 &   5.35e-07 &   6.35e-06 &   2.54e-05 &   1.85e-05 &   7.68e-06 &    8.46e-07 \\
         O &   1.56e-05 &      0.049 &      0.742 &      0.636 &      0.701 &      0.428 &       0.264 \\
\hline
\end{tabular}
\end{center}
\end{table*}

\begin{table*}
    \caption{Same as Table~\ref{tab_he2p60}, but now for model he11p0x1p5.
\label{tab_he11p0x1p5}
}
\begin{center}
\begin{tabular}{l@{\hspace{4mm}}c@{\hspace{4mm}}c@{\hspace{4mm}}c@{\hspace{4mm}}c@{\hspace{4mm}}
c@{\hspace{4mm}}c@{\hspace{4mm}}c@{\hspace{4mm}}
}
\hline
    Model  &      Fe/He &      Si/S  &      O/Si  &    O/Ne/Mg &        O/C &    C/O/He &     C/O/He \\
\hline
           &    [\msun] &    [\msun] &    [\msun] &    [\msun] &    [\msun] &    [\msun] &    [\msun] \\
\hline
he11p0x1p5 &      0.063 &      0.042 &      0.171 &      1.536 &      0.312 &      1.053 &       0.243 \\
\hline
\hline
   Element &      \multicolumn{7}{c}{Average element mass fraction in each shell} \\
\hline
        He &      0.244 &   1.04e-05 &   4.34e-06 &   3.32e-06 &   2.28e-03 &      0.123 &       0.192 \\
         C &   2.42e-06 &   4.24e-05 &   4.80e-03 &      0.031 &      0.267 &      0.466 &       0.513 \\
         N &   5.79e-07 &   6.89e-07 &   7.35e-06 &   5.21e-05 &   2.60e-05 &   6.57e-06 &    6.18e-07 \\
         O &   1.78e-05 &      0.056 &      0.749 &      0.620 &      0.705 &      0.392 &       0.276 \\
\hline
\end{tabular}
\end{center}
\end{table*}

\begin{table*}
    \caption{Same as Table~\ref{tab_he2p60}, but now for model he12p0x1p5.
\label{tab_he12p0x1p5}
}
\begin{center}
\begin{tabular}{l@{\hspace{4mm}}c@{\hspace{4mm}}c@{\hspace{4mm}}c@{\hspace{4mm}}c@{\hspace{4mm}}
c@{\hspace{4mm}}c@{\hspace{4mm}}c@{\hspace{4mm}}
}
\hline
    Model  &      Fe/He &      Si/S  &      O/Si  &    O/Ne/Mg &        O/C &     C/O/He &     C/O/He \\
\hline
           &    [\msun] &    [\msun] &    [\msun] &    [\msun] &    [\msun] &    [\msun] &    [\msun] \\
\hline
he12p0x1p5 &      0.126 &      0.054 &      0.252 &      1.587 &      0.336 &      1.128 &       0.213 \\
 \hline
    \hline
   Element &      \multicolumn{7}{c}{Average element mass fraction in each shell} \\
\hline
        He &      0.195 &   7.94e-06 &   3.57e-06 &   2.44e-06 &   2.64e-03 &      0.134 &       0.219 \\
         C &   1.26e-06 &   3.14e-05 &   5.60e-03 &      0.033 &      0.265 &      0.470 &       0.515 \\
         N &   4.26e-07 &   4.16e-07 &   6.78e-06 &   5.18e-05 &   2.42e-05 &   5.88e-06 &    5.10e-07 \\
         O &   1.58e-05 &      0.035 &      0.740 &      0.625 &      0.706 &      0.375 &       0.247 \\
\hline
\end{tabular}
\end{center}
\end{table*}

\begin{table*}
    \caption{Same as Table~\ref{tab_he2p60}, but now for model he13p0x1p5.
\label{tab_he13p0x1p5}
}
\begin{center}
\begin{tabular}{l@{\hspace{4mm}}c@{\hspace{4mm}}c@{\hspace{4mm}}c@{\hspace{4mm}}c@{\hspace{4mm}}
c@{\hspace{4mm}}c@{\hspace{4mm}}c@{\hspace{4mm}}
}
\hline
    Model  &      Fe/He &      Si/S  &      O/Si  &    O/Ne/Mg &        O/C &     C/O/He &      C/O/He \\
\hline
           &    [\msun] &    [\msun] &    [\msun] &    [\msun] &    [\msun] &    [\msun] &    [\msun] \\
\hline
he13p0x1p5&      0.189 &      0.090 &      0.285 &      1.695 &      0.339 &      1.179 &       0.173 \\
 \hline
    \hline
   Element &      \multicolumn{7}{c}{Average element mass fraction in each shell} \\
\hline
        He &      0.229 &   5.80e-06 &   3.42e-06 &   2.04e-06 &   2.23e-03 &      0.135 &       0.229 \\
         C &   1.55e-06 &   5.43e-05 &   6.94e-03 &      0.040 &      0.258 &      0.465 &       0.512 \\
         N &   6.33e-07 &   8.76e-07 &   6.81e-06 &   6.02e-05 &   2.13e-05 &   5.44e-06 &    4.89e-07 \\
         O &   1.75e-05 &      0.165 &      0.768 &      0.634 &      0.711 &      0.380 &       0.239 \\
\hline
\end{tabular}
\end{center}
\end{table*}

\begin{table*}
    \caption{Same as Table~\ref{tab_he2p60}, but now for model he10p0x2p0.
\label{tab_he10Ax2p0}
}
\begin{center}
\begin{tabular}{l@{\hspace{4mm}}c@{\hspace{4mm}}c@{\hspace{4mm}}c@{\hspace{4mm}}c@{\hspace{4mm}}
c@{\hspace{4mm}}c@{\hspace{4mm}}c@{\hspace{4mm}}
}
\hline
    Model  &      Fe/He &      Si/S  &      O/Si  &    O/Ne/Mg &        O/C &     C/O/He &     C/O/He \\
\hline
           &    [\msun] &    [\msun] &    [\msun] &    [\msun] &    [\msun] &    [\msun] &    [\msun] \\
\hline
 he10p0x2p0 &      0.159 &      0.072 &      0.186 &      0.705 &      0.162 &      0.534 &       0.363 \\
\hline
\hline
   Element &      \multicolumn{7}{c}{Average element mass fraction in each shell} \\
\hline
        He &      0.144 &   1.59e-05 &   5.76e-06 &   3.19e-06 &   6.03e-03 &      0.114 &       0.171 \\
         C &   4.13e-07 &   6.40e-05 &   6.29e-03 &      0.033 &      0.329 &      0.507 &       0.545 \\
         N &   1.95e-06 &   9.22e-06 &   1.91e-05 &   3.05e-05 &   1.65e-05 &   9.41e-06 &    2.37e-06 \\
         O &   1.16e-05 &      0.188 &      0.718 &      0.551 &      0.633 &      0.357 &       0.265 \\
\hline
\end{tabular}
\end{center}
\end{table*}

\clearpage

\begin{table*}
  \caption{Fractional decay powers absorbed in the He-rich shell, the O-rich shell, the Si-rich shell, the Fe-rich shell, and then the fractional decay power absorbed that arises from positrons, and finally the fraction of the total decay power emitted that is absorbed in the ejecta. The SN age is 200\,d. The models are he2p6 to he12p0 (nominal mass loss rate; x1p0 series), he5p0x1p5 to he13p0x1p5 (50\% enhancement in mass loss rate; x1p5 series), and model he10p0x2p0 (100\% enhancement in mass loss rate; x2p0 series).
\label{tab_edep}
}
\begin{center}
\begin{tabular}{lcccccc}
\hline
%          Model  f(He-sh)    f(O-sh)   f(Si-sh)   f(Fe-sh)    f(ekin)    f(eabs)
           Model &     $f$(He-sh)  &  $f$(O-sh) &  $f$(Si-sh) &  $f$(Fe-sh)  &  $f_{\rm positron}$  &  $f_{\rm abs}$ \\
\hline
         he2p6   &     0.71   &      0.12   &      0.14  &       0.03  &       0.17  &       0.19  \\
         he2p9   &     0.52   &      0.15   &      0.28  &       0.06  &       0.28  &       0.11  \\
         he3p3   &     0.30   &      0.32   &      0.20  &       0.19  &       0.21  &       0.15  \\
         he3p5   &     0.33   &      0.38   &      0.23  &       0.06  &       0.16  &       0.20  \\
         he4p0   &     0.25   &      0.47   &      0.20  &       0.08  &       0.16  &       0.19  \\
         he4p5   &     0.22   &      0.41   &      0.12  &       0.26  &       0.21  &       0.15  \\
         he5p0   &     0.16   &      0.47   &      0.08  &       0.29  &       0.21  &       0.15  \\
\hline
\end{tabular}
\end{center}
\end{table*}

\begin{table*}
    \caption{Fractional decay power absorbed, bolometric luminosity, UVOIR luminosity (accounts for the radiation falling between 1000\,\AA\ and 2.5\,$\mu$m), and optical and NIR absolute magnitudes at 200\,d for the explosion models he2p6 to he12p0 (standard mass loss rate; x1p0 series), he5p0x1p5 to he13p0x1p5 (50\% enhancement in mass loss rate; x1p5 series), and model he10p0x2p0 (100\% enhancement in mass loss rate; x2p0 series).
\label{tab_mag}
}
\begin{center}
\begin{tabular}{lccccccccccc}
\hline
Model & $f_{\rm abs}$ & $L_{\rm bol}$ &  $L_{UVOIR}$&  $U$ &       $B$&        $V$ &       $R$ &       $I$  &      $J$  &      $H$ &     $K$ \\
      &              &  \multicolumn{2}{c}{[10$^{40}$\,erg\,s$^{-1}$]} & [mag] & [mag]& [mag]& [mag]& [mag]& [mag]& [mag]& [mag] \\
\hline
   he2p6      &  0.19  &     0.43    &     0.41 &   -9.93 &  -10.18 &  -10.51 &  -10.62 &  -10.60 &   -9.73 &  -10.28 &  -10.84 \\
   he2p9      &  0.11  &     0.56    &     0.52 &  -10.07 &  -10.53 &  -10.83 &  -10.88 &  -10.97 &   -9.94 &  -10.75 &  -10.81 \\
   he3p3      &  0.15  &     1.38    &     1.31 &  -10.91 &  -11.53 &  -11.88 &  -11.95 &  -12.18 &  -11.12 &  -11.72 &  -11.59 \\
   he3p5      &  0.20  &     1.17    &     1.10 &  -10.61 &  -11.15 &  -11.62 &  -11.97 &  -12.16 &  -11.20 &  -11.59 &  -11.51 \\
   he4p0      &  0.19  &     1.64    &     1.53 &  -10.81 &  -11.44 &  -11.94 &  -12.43 &  -12.64 &  -11.62 &  -11.94 &  -11.85 \\
   he4p5      &  0.15  &     2.93    &     2.74 &  -11.81 &  -12.42 &  -12.71 &  -12.69 &  -12.96 &  -11.91 &  -12.44 &  -12.41 \\
\hline
\end{tabular}
\end{center}
\end{table*}

\begin{table*}
  \caption{Summary of the main coolants in the ejecta model he3p3. A representative location is chosen in each dominant shell. The coolants are widely different between shells in part because of the distinct compositions. For each shell, we list the top five coolants (as percentage of the total decay power absorbed at that location), and give the cumulative cooling power associated (usually $\sim$\,90\,\% of the total).
\label{tab_cool_he3p3}
}
\begin{center}
\begin{tabular}{|l|l|l|l|}
\hline
He/N &                          He/C/O &                         O/Ne/Mg  &                          Si/S                    \\
\hline
$V=5077$\,\kms &                $V=4411$\,\kms &               $V=4073$\,\kms  &                     $V=3661$\,\kms    \\
$T=10322$\,K  &                 $T=6208$\,K  &                 $T=4865$\,K   &                       $T=4344$\,K               \\
N\,{\sc ii} COL (34.8\,\%)  &   C\,{\sc ii} COL (63.6\,\%)  &  Mg\,{\sc ii} COL (59.7\,\%)   &       Fe\,{\sc ii} COL (21.9\,\%)   \\
He\,{\sc i} NT (22.8\,\%)  &    He\,{\sc i} NT (14.0\,\%)  &   O\,{\sc i} COL (16.1\,\%)   &         Ca\,{\sc ii} COL (18.6\,\%)   \\
Fe\,{\sc iii} COL (20.9\,\%)  & Mg\,{\sc ii} COL (6.6\,\%)  &  Ne\,{\sc ii} COL (8.3\,\%)   &        Co\,{\sc ii} COL (14.8\,\%)   \\
C\,{\sc ii} COL (5.9\,\%)  &    O\,{\sc ii} COL (5.5\,\%)  &   O\,{\sc ii} NT (7.9\,\%)   &          Ni\,{\sc ii} COL (14.4\,\%)   \\
O\,{\sc ii} COL (4.5\,\%)  &    C\,{\sc ii} NT (4.8\,\%)  &    O\,{\sc i} NT (3.1\,\%)   &           Fe\,{\sc iii} COL (9.5\,\%)   \\
\hline
\end{tabular}
\end{center}
\end{table*}

\begin{table*}
    \caption{Summary of main coolants in the dominant shells of ejecta model he5p0.
\label{tab_cool_he5p0}
}
\begin{center}
\begin{tabular}{|l|l|l|l|}
\hline
He/N &                          He/C/O &                       O/Ne/Mg &                      Si/S  \\
\hline
$V=8207$\,\kms &                $V=7620$\,\kms &               $V=7135$\,\kms &               $V=6140$\,\kms  \\
$T=11389$\,K  &                 $T=6762$\,K  &                 $T=5107$\,K  &                 $T=3986$\,K   \\
N\,{\sc ii} COL (31.3\,\%)  &   C\,{\sc ii} COL (62.1\,\%)  &  Mg\,{\sc ii} COL (49.1\,\%)  & Ca\,{\sc ii} COL (41.5\,\%)   \\
Fe\,{\sc iii} COL (20.7\,\%)  & He\,{\sc i} NT (13.7\,\%)  &   Ne\,{\sc ii} COL (24.3\,\%)  & Co\,{\sc ii} COL (12.2\,\%)   \\
He\,{\sc i} NT (18.2\,\%)  &    O\,{\sc ii} COL (10.8\,\%)  &  O\,{\sc i} COL (9.4\,\%)  &    Fe\,{\sc ii} COL (12.2\,\%)   \\
O\,{\sc ii} COL (9.6\,\%)  &    Ne\,{\sc ii} COL (3.4\,\%)  &  O\,{\sc ii} NT (7.8\,\%)  &    Ni\,{\sc ii} COL (11.5\,\%)   \\
C\,{\sc ii} COL (8.2\,\%)  &    C\,{\sc ii} NT (3.2\,\%)  &    C\,{\sc ii} NT (2.4\,\%)  &    Si\,{\sc ii} NT (5.1\,\%)   \\
\hline
\end{tabular}
\end{center}
\end{table*}

\begin{table*}
    \caption{Summary of the main coolants in the dominant shells of ejecta model he8p0.
\label{tab_cool_he8p0}
}
\begin{center}
\begin{tabular}{|l|l|l|l|}
\hline
He/N &                          O/C &                             O/Si &                          Fe/He   \\
\hline
$V=6146$\,\kms &                $V=4929$\,\kms &                  $V=3467$\,\kms &                $V=3390$\,\kms   \\
$T=8789$\,K  &                  $T=4551$\,K  &                    $T=4718$\,K  &                  $T=3826$\,K    \\
N\,{\sc ii} COL (35.5\,\%)  &   O\,{\sc i} COL (64.0\,\%)  &      Mg\,{\sc ii} COL (37.9\,\%)  &  Ca\,{\sc ii} COL (35.0\,\%)    \\
He\,{\sc i} NT (29.9\,\%)  &    C\,{\sc ii} NT (8.0\,\%)  &       O\,{\sc i} COL (24.9\,\%)  &    Fe\,{\sc ii} COL (21.4\,\%)    \\
Fe\,{\sc iii} COL (20.6\,\%)  & O\,{\sc i} NT (7.2\,\%)  &        O\,{\sc i} NT (10.1\,\%)  &     Co\,{\sc ii} COL (15.9\,\%)    \\
S\,{\sc ii} COL (3.4\,\%)  &    Mg\,{\sc ii} COL (5.4\,\%)  &     O\,{\sc ii} NT (7.1\,\%)  &     Ni\,{\sc ii} COL (13.3\,\%)    \\
He\,{\sc i} COL (2.6\,\%)  &    C\,{\sc ii} COL (4.9\,\%)  &      Ca\,{\sc ii} COL (5.1\,\%)  &   He\,{\sc i} NT (11.5\,\%)    \\
\hline
\end{tabular}
\end{center}
\end{table*}

\begin{table*}
    \caption{Summary of the main coolants in the dominant shells of ejecta model he3p3 in which we enforced a uniform volume filling factor of 20\,\%.
\label{tab_cool_he3p3fvol0p2}
}
\begin{center}
\begin{tabular}{|l|l|l|l|}
\hline
He/N &                         He/C/O &                       O/Ne/Mg &                            Si/S      \\
\hline
$V=5077$\,\kms &               $V=4411$\,\kms &               $V=4073$\,\kms &                     $V=3661$\,\kms      \\
$T=8739$\,K  &                 $T=5262$\,K  &                 $T=4359$\,K  &                       $T=3533$\,K         \\
He\,{\sc i} NT (36.6\,\%)  &   Mg\,{\sc ii} COL (38.3\,\%)  & Mg\,{\sc ii} COL (48.5\,\%)  &       Ca\,{\sc ii} COL (46.2\,\%)   \\
N\,{\sc ii} COL (25.8\,\%)  &  C\,{\sc ii} COL (35.9\,\%)  &  O\,{\sc i} COL (23.2\,\%)  &         Fe\,{\sc ii} COL (15.2\,\%)   \\
Fe\,{\sc iii} COL (17.2\,\%)  &He\,{\sc i} NT (20.2\,\%)  &   O\,{\sc i} NT (9.2\,\%)  &           Ni\,{\sc ii} COL (13.7\,\%)   \\
C\,{\sc ii} COL (4.8\,\%)  &   C\,{\sc ii} NT (6.3\,\%)  &    Mg\,{\sc i} COL (4.2\,\%)  &         Co\,{\sc ii} COL (12.4\,\%)   \\
He\,{\sc i} COL (4.1\,\%)  &   O\,{\sc i} COL (2.9\,\%)  &    O\,{\sc ii} NT (4.2\,\%)  &          He\,{\sc i} NT (5.4\,\%)         \\
\hline
\end{tabular}
\end{center}
\end{table*}

\begin{table*}
    \caption{Summary of the main coolants in the dominant shells of ejecta model he5p0 in which we enforced a uniform volume filling factor of 20\,\%.
\label{tab_cool_he5p0fvol0p2}
}
\begin{center}
\begin{tabular}{|l|l|l|l|}
\hline
He/N &                             He/C/O &                            O/Ne/Mg &                      Si/S  \\
\hline
$V=8207$\,\kms &                   $V=7620$\,\kms &                    $V=7135$\,\kms &               $V=6140$\,\kms  \\
$T=9352$\,K  &                     $T=5937$\,K  &                      $T=4518$\,K  &                 $T=3356$\,K   \\
He\,{\sc i} NT (32.1\,\%)  &       C\,{\sc ii} COL (54.3\,\%)  &       Mg\,{\sc ii} COL (40.3\,\%)  & Ca\,{\sc ii} COL (65.7\,\%)   \\
N\,{\sc ii} COL (23.2\,\%)  &      He\,{\sc i} NT (22.0\,\%)  &        O\,{\sc i} COL (25.1\,\%)  &   Ni\,{\sc ii} COL (11.0\,\%)   \\
Fe\,{\sc iii} COL (18.8\,\%)  &    Mg\,{\sc ii} COL (9.3\,\%)  &       Ne\,{\sc ii} COL (12.0\,\%)  & Co\,{\sc ii} COL (9.6\,\%)   \\
C\,{\sc ii} COL (6.2\,\%)  &       O\,{\sc ii} COL (4.9\,\%)  &        O\,{\sc i} NT (6.0\,\%)  &     Fe\,{\sc ii} COL (8.1\,\%)   \\
O\,{\sc ii} COL (5.6\,\%)  &       C\,{\sc ii} NT (4.5\,\%)  &         O\,{\sc ii} NT (5.1\,\%)  &    Si\,{\sc ii} NT (7.1\,\%)   \\
\hline
\end{tabular}
\end{center}
\end{table*}

\begin{table*}
    \caption{Summary of the main coolants in the dominant shells of ejecta model he8p0 in which we enforced a uniform volume filling factor of 20\,\%.
\label{tab_cool_he8fvol0p2}
}
\begin{center}
\begin{tabular}{|l|l|l|l|}
\hline
He/N &                            O/C &                             O/Si &                         Fe/He   \\
\hline
$V=6146$\,\kms &                  $V=4929$\,\kms &                  $V=3467$\,\kms &               $V=3390$\,\kms   \\
$T=7489$\,K  &                    $T=4041$\,K  &                    $T=4127$\,K  &                 $T=3364$\,K    \\
He\,{\sc i} NT (44.0\,\%)  &      O\,{\sc i} COL (46.4\,\%)  &      Mg\,{\sc i} COL (20.5\,\%)  &  Fe\,{\sc i} COL (24.3\,\%)    \\
N\,{\sc ii} COL (24.5\,\%)  &     C\,{\sc i} COL (14.9\,\%)  &      O\,{\sc i} NT (18.9\,\%)  &    Ca\,{\sc ii} COL (22.4\,\%)    \\
Fe\,{\sc iii} COL (17.7\,\%)  &   O\,{\sc i} NT (11.0\,\%)  &       Mg\,{\sc ii} COL (17.7\,\%)  & He\,{\sc i} NT (17.1\,\%)    \\
He\,{\sc i} COL (3.7\,\%)  &      C\,{\sc ii} NT (7.4\,\%)  &       O\,{\sc i} COL (17.0\,\%)  &   Co\,{\sc ii} COL (11.4\,\%)    \\
S\,{\sc ii} COL (3.3\,\%)  &      Ca\,{\sc ii} COL (4.3\,\%)  &     Ca\,{\sc ii} COL (11.1\,\%)  & Ni\,{\sc ii} COL (8.3\,\%)    \\
\hline
\end{tabular}
\end{center}
\end{table*}

\end{document}